\documentclass[iop,apj,twocolumn,single,numberedappendix,twocolappendix,tighten]{emulateapj}
\usepackage{enumitem,graphicx,apjfonts,multirow,amsmath,bm}

\shorttitle{Cyclic Convective Dynamo}
\shortauthors{Augustson et al.}

\newcommand{\Rsun}[1]{#1\,\mathrm{R} \,\!\scriptscriptstyle \sun\!}
\newcommand{\Osun}[1]{#1\,\Omega \,\!\scriptscriptstyle \sun\!}
\newcommand{\Prt}{\mathrm{Pr}}
\newcommand{\Prm}{\mathrm{Pm}}

\newcommand{\orho}{\overline{\rho}}
\newcommand{\oent}{\overline{S}}
\newcommand{\oT}{\overline{T}}
\newcommand{\opre}{\overline{P}}

\newcommand{\ddr}[1]{\frac{\partial #1}{\partial r}}
\newcommand{\ddtime}[1]{\frac{\partial #1}{\partial t}}
\newcommand{\sint}{\sin{\theta}}

\newcommand{\avg}[1]{\langle #1 \rangle}
\newcommand{\bigavg}[1]{\big\langle #1 \big\rangle}

\newcommand{\bomega}{\hat{\boldsymbol{\Omega}}}
\newcommand{\grad}{\boldsymbol{\nabla}}
\newcommand{\dvg}{\boldsymbol{\nabla}\bcdot}
\newcommand{\curl}{\boldsymbol{\nabla}\boldsymbol{\times}}
\newcommand{\cnabla}{\boldsymbol{\cdot}\boldsymbol{\nabla}}
\newcommand{\cross}{\boldsymbol{\times}}
\newcommand{\bcdot}{\boldsymbol{\cdot}}

\newcommand{\vv}{\mathbf{v}}
\newcommand{\vB}{\mathbf{B}}
\newcommand{\vJ}{\mathbf{J}}

\newcommand{\Acp}{A_{\varphi}}

\newcommand{\Bcr}{B_{\mathrm{r}}}
\newcommand{\Bct}{B_{\theta}}
\newcommand{\Bcp}{B_{\varphi}}

\newcommand{\rht}{\hat{\boldsymbol{\mathrm{r}}}}
\newcommand{\tht}{\hat{\boldsymbol{\theta}}}
\newcommand{\pht}{\hat{\boldsymbol{\varphi}}}

\newcommand{\vcr}{\mathrm{v_r}}
\newcommand{\vct}{\mathrm{v}_{\theta}}
\newcommand{\vcp}{\mathrm{v}_{\varphi}}

\newcommand{\avgAp}{\langle A_{\varphi} \rangle}
\newcommand{\avgBp}{\langle B_{\varphi} \rangle}
\newcommand{\avgO}{\langle\Omega\rangle}

\newcommand{\TMS}{\mathrm{MS}}

\newcommand{\TMA}{\mathrm{MA}}

\newcommand{\TFS}{\mathrm{FS}}

\newcommand{\TFA}{\mathrm{FA}}

\newcommand{\TCC}{\mathrm{CC}}

\newcommand{\TRD}{\mathrm{RD}}

\newcommand{\MBF}{\langle \vB\rangle}
\newcommand{\MVF}{\langle \vv\rangle}

\newcommand{\MTF}{\langle \Bcp\rangle}
\newcommand{\MPF}{\langle \mathbf{B}_P \rangle}

\newcommand{\mpers}[1]{#1\,\mathrm{m}\,\mathrm{s}^{-1}}
\newcommand{\dgr}[1]{#1^{\circ}}

\begin{document}

\title{Grand Minimum and Equatorward Propagation in a Cycling Stellar Convective Dynamo}
\author{Kyle\ Augustson$^1$, Allan\ Sacha\ Brun$^2$, Mark\ Miesch$^1$ \& Juri\ Toomre$^{3}$}
\affil{$^{1}$High Altitude Observatory, Center Green 1, Boulder, CO 80301, USA}
\affil{$^{2}$Laboratoire AIM Paris-Saclay, CEA/DSM -- CNRS -- Universit\'{e} Paris Diderot, IRFU/SAp, Gif-sur-Yvette, France}
\affil{$^{3}$JILA and Dept. of Astrophysical \& Planetary Sciences, University of Colorado, Boulder, CO 80309, USA}


\begin{abstract}

  The 3-D magnetohydrodynamic (MHD) Anelastic Spherical Harmonic (ASH) code, using slope-limited
  diffusion, is employed to capture convective and dynamo processes achieved in a global-scale
  stellar convection simulation for a model solar-mass star rotating at three times the solar
  rate. The dynamo generated magnetic fields possesses many time scales, with a prominent polarity
  cycle occurring roughly every 6.2~years. The magnetic field forms large-scale toroidal wreaths,
  whose formation is tied to the low Rossby number of the convection in this simulation. The
  polarity reversals are linked to the weakened differential rotation and a resistive collapse of
  the large-scale magnetic field. An equatorial migration of the magnetic field is seen, which is
  due to the strong modulation of the differential rotation rather than a dynamo wave. A poleward
  migration of magnetic flux from the equator eventually leads to the reversal of the polarity of
  the high-latitude magnetic field. This simulation also enters an interval with reduced magnetic
  energy at low latitudes lasting roughly 16~years (about 2.5 polarity cycles), during which the
  polarity cycles are disrupted and after which the dynamo recovers its regular polarity cycles. An
  analysis of this grand minimum reveals that it likely arises through the interplay of symmetric
  and antisymmetric dynamo families. This intermittent dynamo state potentially results from the
  simulations relatively low magnetic Prandtl number. A mean-field-based analysis of this dynamo
  simulation demonstrates that it is of the $\alpha$-$\Omega$ type. The time scales that appear to
  be relevant to the magnetic polarity reversal are also identified.

\end{abstract}

\keywords{stars: magnetic field, rotation -- convection -- magnetohydrodynamics -- turbulence}

\section{Introduction} \label{sec:intro}

The Sun exhibits many time scales, from the ten minute lifetimes of granules to multi-millennial
magnetic activity modulations. One of the most prominent of these scales is the 11-year sunspot
cycle, during which the number of magnetically active regions waxes and wanes. The Sun also
possesses longer-term variability of its magnetic activity such as the 88-year Gleissberg cycle
\citep{gleissberg39}. There are also intermittent and aperiodic phenomenon commonly described as
grand extrema \citep{usoskin13}, such as the Maunder Minimum \citep{eddy76,ribes93}, wherein the
overall magnetic activity of the Sun declines or increases for many polarity cycles relative to a
long-term average.  These longer-term trends in solar activity can be seen both in visual
observations of the number of sunspots as well as in less direct measurements such as radioisotopic
measurements \citep[e.g.,][]{beer98}. During grand minima, the number of sunspots tends to decrease
and sometimes vanish for several polarity cycles. In contrast, their numbers increase over a grand
maximum. Furthermore, during extrema such as the Maunder Minimum and the Modern Maximum,
measurements of cosmogenic radioisotopes suggest that the heliospheric magnetic field strength can
vary by at most a factor of five, but more typically by a factor of two \citep{mccracken07}.  The
simulation presented here shares some of these characteristics, where it shows disrupted cycles and
a decrease in volume-averaged magnetic energy at lower latitudes during an extended interval
covering several magnetic polarity cycles. So, the interval of disrupted magnetic cycles has been
tagged as a ``grand minimum.''

In addition to its large range of time scales, the magnetic field at the solar surface exhibits
complex, hierarchical structures that persist over a vast range of spatial scales. Nevertheless,
large-scale organized spatial patterns of smaller structures such as Maunder's butterfly diagram,
Joy's law, and Hale's polarity law suggest the existence of a structured large-scale magnetic field
within the solar convection zone. On the Sun's surface active regions initially emerge at
mid-latitudes and appear at progressively lower latitudes as the cycle progresses, thus exhibiting
equatorward migration. In contrast, the diffuse field that is comprised of small-scale bipolar
regions migrates toward the poles, with the global-scale reversal of the polar magnetic field
occurring near solar maximum \citep[e.g.,][]{benevolenskaya04,hathaway10}.

Other main-sequence stars also exhibit observable magnetic phenomenon under several measures such as
Ca II, photometric, spectropolarimetric, and X-ray observations \citep[e.g.,][]{baliunas96,
  hempelmann96, favata08, metcalf10, fares13, mathur13}. Such observations have shown that
solar-mass stars younger than the Sun can also possess magnetic activity cycles. These younger stars
tend to rotate more rapidly than the Sun as a consequence of having been born with a relatively high
angular momentum and due to their relatively slow rate of angular momentum loss
\citep[e.g.,][]{barnes07,matt15}.  There are further hints from both observations and from theory
that stellar magnetic cycle periods should be linked to its rotation rate
\citep[e.g.,][]{saar09,jouve10,morgenthaler11}. So, in some senses, the simulation presented here
could be considered to be capturing some of the dynamo behavior of a young Sun-like star. Moreover,
from a theoretical point of view, the ratio of the polarity cycle period to the relevant dynamical
time scales of the rotation period may be of more interest. For the Sun, this ratio is about 287,
and as will be seen later this ratio is about 243 for this model.

\subsection{Cyclic Convective Dynamo Action} \label{sec:cycles}

It has been suspected for at least 60 years that the crucial ingredients for the solar dynamo are
the shear of the differential rotation and the helical nature of the small-scale convective flows
present in the solar convection zone \citep[e.g.,][]{parker55, steenbeck69, parker77}. Though other
models and observations suggest that the surface magnetic fields play a significant role, such as in
the Babcock-Leighton dynamo mechanism \citep[e.g.,][]{babcock61,charbonneau05,miesch12}. Such models
have been fairly successful in capturing aspects of the solar cycles. However, when adopting fully
nonlinear global-scale 3-D MHD simulations \citep[e.g.,][]{gilman83,glatzmaier85,brun04,browning06},
it has been challenging to achieve dynamo action that exhibits a majority of the properties of Sun's
large-scale magnetism.

Recent global-scale simulations of convective dynamos have begun to make substantial contact with
some of the properties of the solar dynamo using a wide variety of numerical methods
\citep[e.g.,][]{ghizaru10,brown11,racine11,kapyla12,augustson13,passos14,fan14}. The simulation
analyzed here fits within this vein of modern stellar dynamo modeling, where it exhibits some
features akin to those observed during solar and stellar cycles. In particular, global-scale
convective dynamo simulations in rotating spherical shells have recently achieved the long-sought
goal of cyclic magnetic polarity reversals with a multi-decadal period. Moreover, some of these
simulations have illustrated that large-scale dynamo action is possible within the bulk of the
convection zone, even in the absence of a tachocline \citep[e.g.,][]{brun04,brown10,kapyla12}.

Global-scale MHD simulations of a more rapidly rotating Sun with the ASH code have produced
polarity-reversing dynamo action that possesses strong toroidal wreaths of magnetism that propagate
poleward as a cycle progresses \citep{brown11}.  These fields are contained within the convection
zone itself, with the majority of the magnetic energy present near the lower boundary. Furthermore,
a recent simulation with ASH employs a dynamic Smagorinski diffusion scheme, wherefore a greater
level of turbulent complexity is achieved for the resolved spatial structures. Those simulations
show that the large-scale toroidal wreaths persist despite the greater pummeling they endure from
the more complex and vigorous convection \citep{nelson13a}. Not only do the toroids of field
persevere, but portions of them can be so amplified that the combination of upward advection and
magnetic buoyancy create loops of magnetic field that rise upward toward the surface
\citep{nelson13b}.

\subsection{Differing Approaches to Sub-Grid-Scale Dissipation} \label{sec:sgs}

Both explicit and implicit large-eddy simulations (LES and ILES) have concurrently paved the road
toward more orderly long-term cycles in a setting that may mimic the solar interior. Indeed, the
first 3-D simulation to produce regular polarity cycles over a long time period utilized the
Eulerian-Lagrangian magnetohydrodynamics code (EULAG-MHD) \citep{ghizaru10}. The polarity cycles in
that simulation occur roughly every 80~years, with the magnetic fields existing primarily at higher
latitudes and within the tachocline at the base of the convection zone
\citep[e.g.,][]{racine11,passos14}. Such dynamo action is likely made possible through two
mechanisms: the first being that the ILES formulation of EULAG attempts to maximize the complexity
of the flows and magnetic fields for a given Eulerian grid resolution, and the second being the
reduction of the enthalpy transport of the largest scales through a relatively simple sub-grid-scale
(SGS) model. The latter mechanism operates through the dissipation of entropy structures by adding a
thermal drag to the entropy equation. This reduces the buoyancy of the resolved convective
structures, and thereby the root-mean-square velocities, which in turn decreases the Rossby number.

The magnetic fields in those EULAG-MHD simulations, however, have primarily shown radial propagation
of structures but little latitudinal variation during a cycle. Though much like prior simulations
using ASH, a recent EULAG-MHD simulation of a Sun-like star rotating at thrice the solar rate can
also produce low-latitude poleward propagating solutions \citep{charbonneau13}. Similarly, 3-D MHD
simulations in spherical segments employing the Pencil code also possess regularly cyclic magnetic
polarity reversals in addition to a rich set of other behavior. In particular, a few of those
polarity reversing solutions were the first to exhibit low-latitude equatorward propagating magnetic
features \citep{kapyla11b,kapyla13}. In those simulations, the stratification and a sufficient level
of turbulence appear to be necessary to achieve the phase alignment between the magnetic field and
the differential rotation required to produce the dynamo wave phenomenon known as the
Parker-Yoshimura effect \citep{warnecke14}.

Inspired by those recent results, a slope-limited diffusion (SLD) scheme was incorporated into ASH
with the express goal of achieving a low effective $\Prt$ and $\Prm$ dynamo, thus attempting to
better mimic the low fluid and magnetic Prandtl numbers present in the solar interior. This effort
minimizes the effects of viscosity, and so extends the inertial range as far as possible for a given
resolution, whereas the thermal and magnetic fields retain their LES-SGS eddy diffusivities.
Consequently, SLD permits more scales to be captured before entering the dissipation range. This in
turn allows more scale separation between the larger magnetic and and smaller kinetic scales
participating in the low $\Prm$ dynamo \citep{ponty05, schekochihin07, brandenburg09}, given that
the ratio of the magnetic to the viscous dissipation scales is greater than unity.  Subsequently,
the kinetic helicity is also greater at small scales than otherwise would be achieved with the
required Newtonian momentum diffusion at the same resolution, which has been shown to have a large
influence on the dynamo efficiency \citep{malyshkin10}.

With the newly implemented SLD scheme, a solution has been found that possesses features similar to
those of the solar dynamo: (i) a regular magnetic polarity cycle, though with a period of 6.2~years,
in which the magnetic polarity reversals occur near the maximum in the magnetic energy, (ii) an
equatorward propagation of magnetic features, (iii) a poleward migration of oppositely-signed flux,
and (iv) the equilibrium of regular cycles is punctuated by an interval where the cycling behavior
is disrupted, the magnetic energy is reduced at low latitudes, and after which the cycle is
recovered. In keeping with the ASH nomenclature for related cases as in \citet{brown10, brown11} and
\citet{nelson13a}, this dynamo solution has been called case K3S.

\subsection{General Layout} \label{sec:layout}

The basic layout of the paper is as follows: \S\ref{sec:methods} contains the details of the
equations solved and of their numerical implementation; \S\ref{sec:overview} provides an overview of
the dynamics of the solution and references to each of the relevant in-depth analysis sections. Then
\S\ref{sec:reversal} assesses the properties of the typical polarity cycles and the mechanisms
contributing to the evolution of the poloidal magnetic field within this dynamo. The grand minimum
seen in this simulation and its properties are discussed in \S\ref{sec:minimum}. The processes
relevant to the equatorward propagation of the magnetic fields during a cycle are covered in
\S\ref{sec:propagate}. An analysis of time scales is given in \S\ref{sec:periods}. Connections to a
mean-field description of the simulation are made in \S\ref{sec:alpha}. Discussion of the
significance of our findings with K3S, and their relation to other studies, is provided in the
concluding \S\ref{sec:conclude}. Appendix A defines the operators used in slope-limited diffusion
and illustrates some of its properties. Appendix B provides a derivation of equations governing the
evolution of the kinetic energy contained in the differential rotation.

\section{Computational Methods} \label{sec:methods}

The 3-D simulation of convective dynamo action presented here as case K3S uses the ASH code to
evolve the Lantz-Braginski-Roberts (LBR) form of the anelastic MHD equations for a conductive plasma
in a rotating spherical shell. ASH solves those equations employing a pseudo-spectral method with
spherical harmonic expansions in the horizontal directions of the entropy, magnetic field, pressure,
and mass flux \citep{clune99,miesch00}. A fourth-order non-uniform finite difference in the radial
direction resolves the radial derivatives. The solenoidality of the mass flux and magnetic vector
fields is maintained through the use of a streamfunction formalism \citep{brun04}. The density,
entropy, pressure, and temperature are linearized about the spherically symmetric background values
$\orho$, $\oent$, $\opre$, and $\oT$ respectively, which are functions of the radial coordinate
only. These linearized thermodynamic variables are denoted $\rho$, $S$, $P$, and $T$. The reduced
pressure $\varomega=P/\orho$ is used in the LBR implementation from which the equivalent
thermodynamic pressure fluctuations can be recovered. The equations solved in ASH retain physical
units, are in spherical coordinates $(r,\theta,\varphi)$, and are evolved in time $t$ as

\vspace{-0.25truein}
\begin{center}
   \begin{align}
     \text{continuity:}  \quad & \displaystyle \dvg{\orho\vv} = 0, \label{eqn:ashcont} \\
     \text{momentum:} \quad & \displaystyle \orho\ddtime{\vv} = -\orho \vv \cnabla \vv -\grad \varomega + \frac{S g}{c_P} \rht \nonumber \\
     \mbox{} & + 2 \orho \vv \cross \bomega_0 + \frac{1}{4\pi}\left(\curl\vB\right)\cross\vB + \dvg{\mathcal{D}}, \label{eqn:ashmom} \\
     \text{energy:} \quad & \displaystyle \orho\oT\ddtime{S} = -\orho\oT\vv \cnabla \left(\oent+S \right) -\nabla \cdot \mathbf{q} + \Phi, \label{eqn:asherg} \\
     \text{flux conservation:}  \quad & \displaystyle \dvg{\vB} = 0, \\
     \text{induction:} \quad & \displaystyle \ddtime{\vB} = \curl\left[\vv\cross\vB-\eta\curl\vB\right], \label{eqn:ashind}
   \end{align}
\end{center}

\noindent with the velocity field being $\vv=\vcr\rht+\vct\tht+\vcp\pht$, and the magnetic field
being $\vB=\Bcr\rht+\Bct\tht+\Bcp\pht$. $\bomega_0=\Omega_0 \hat{\mathbf{z}}$ is the angular
velocity of the rotating frame, $\hat{\mathbf{z}}$ is the direction along the rotation axis, and the
magnitude of the gravitational acceleration is $g$. The diffusion tensor $\mathcal{D}$, which
includes both viscous and slope-limited components, and the dissipative term $\Phi$ are

\vspace{-0.25truein}
\begin{center}
   \begin{align}
     \displaystyle \mathcal{D}_{ij} &= 2 \orho \nu \left[ e_{ij} - \frac{1}{3} \dvg{\vv} \delta_{ij} \right] + \mathcal{F}_{\vv, ij}^{\mathrm{sld}}, \label{eqn:vsstress} \\
     \displaystyle \Phi &= 2\orho\nu\left[e_{ij} e_{ij} - \frac{1}{3} \left(\dvg{\vv}\right)^2\right] + \frac{4\pi\eta}{c^2}\vJ^2 + \dvg{\mathbf{F}_{\mathrm{ke}}^{\mathrm{sld}}}-\orho\vv\cdot\dvg{\mathcal{F}_{\vv}^{\mathrm{sld}}}, \label{eqn:heating}
   \end{align}
\end{center}

\begin{figure}[t!]
   \begin{center}
   \includegraphics[width=0.495\textwidth]{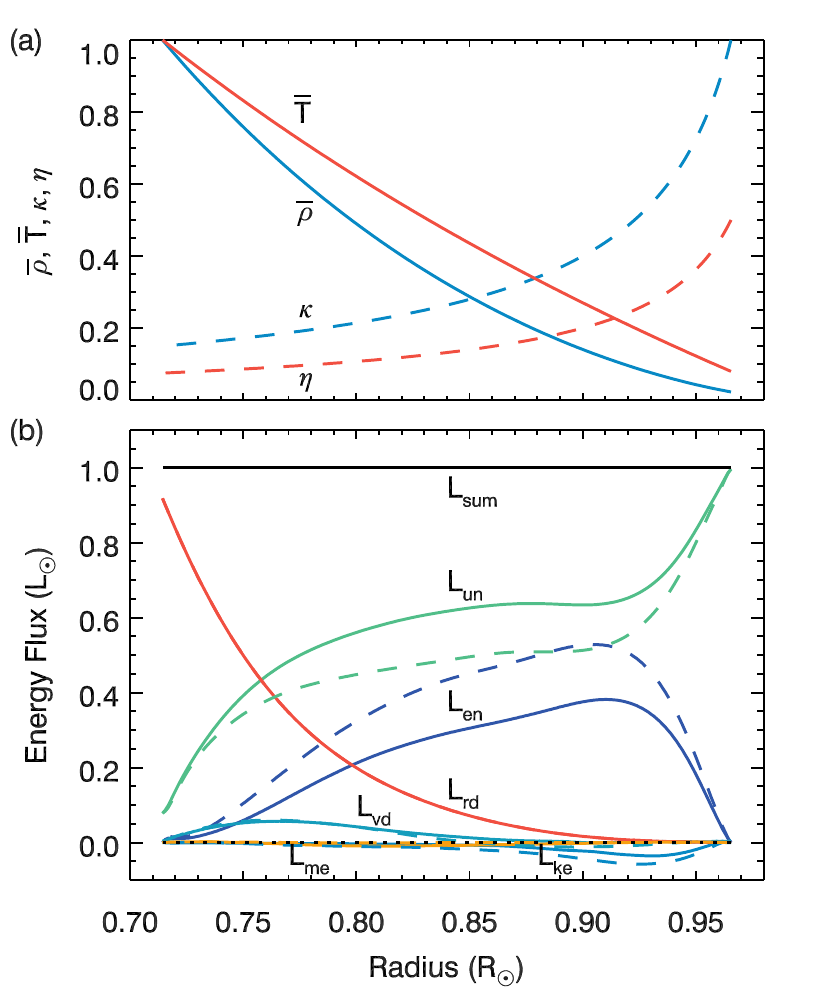} \figcaption{Background stratification and
     energy flux balance in K3S as a function of fractional solar radius ($r/\Rsun{}$). (a) The
     isentropic background state, showing $\orho$ (solid blue) and $\oT$ (solid red). The entropy
     diffusion coefficient $\kappa$ and magnetic diffusion coefficient $\eta$ are shown as the
     dashed-blue and dashed-red lines respectively. All quantities have been normalized by their
     maximum value. (b) The time and horizontally-averaged radial energy fluxes represented as
     luminosities (e.g., the flux in a quantity $x$ is $L_x = 4\pi r^2 F_x$) in units of the solar
     luminosity ($L_{\sun}=3.86\times10^{33} \, \mathrm{erg\,s^{-1}}$). The fluxes are shown
     averaged over two intervals, with solid lines averaged around a magnetic energy maximum and the
     dashed lines near a magnetic energy minimum. The lines are total flux ($L_{\mathrm{sum}}$) in
     black, radiative flux ($L_{\mathrm{rd}}$) in red, enthalpy flux ($L_{\mathrm{en}}$) in blue,
     conductive entropy flux ($L_{\mathrm{un}}$) in green, kinetic energy ($L_{\mathrm{ke}}$) in
     light blue, slope-limited diffusion flux ($L_{\mathrm{vd}}$) in teal, and Poynting flux in
     ($L_{\mathrm{me}}$) orange. \label{fig1}}
   \end{center}
\end{figure}

\noindent involving the stress tensor $e_{ij}$, the effective kinematic eddy viscosity $\nu$, the
magnetic eddy resistivity $\eta$, and the current density $\vJ = c/4\pi\curl\vB$. The slope-limited
velocity diffusion tensor is $\mathcal{F}_{\vv}^{\mathrm{sld}}$, the kinetic energy slope-limited
diffusion flux vector is $\mathbf{F}_{\mathrm{ke}}^{\mathrm{sld}}$, and they are computed using the
algorithm shown in Appendix A. The difference of the divergence of the two fluxes accounts for the
change in entropy due to the SLD operator acting on the velocity field. The energy flux $\mathbf{q}$
is comprised of a radiation flux (in the diffusion approximation) and an inhomogeneous turbulent
entropy diffusion flux,

\vspace{-0.25truein}
\begin{center}
  \begin{equation}
     \displaystyle \mathbf{q} = \kappa_r \orho \mathrm{c_P} \grad \left(\oT+T\right)+\kappa \orho
     \oT \grad S, \label{eqn:ediff}
  \end{equation}
\end{center}

\begin{figure}[t!]
   \begin{center}
     \includegraphics[width=0.475\textwidth]{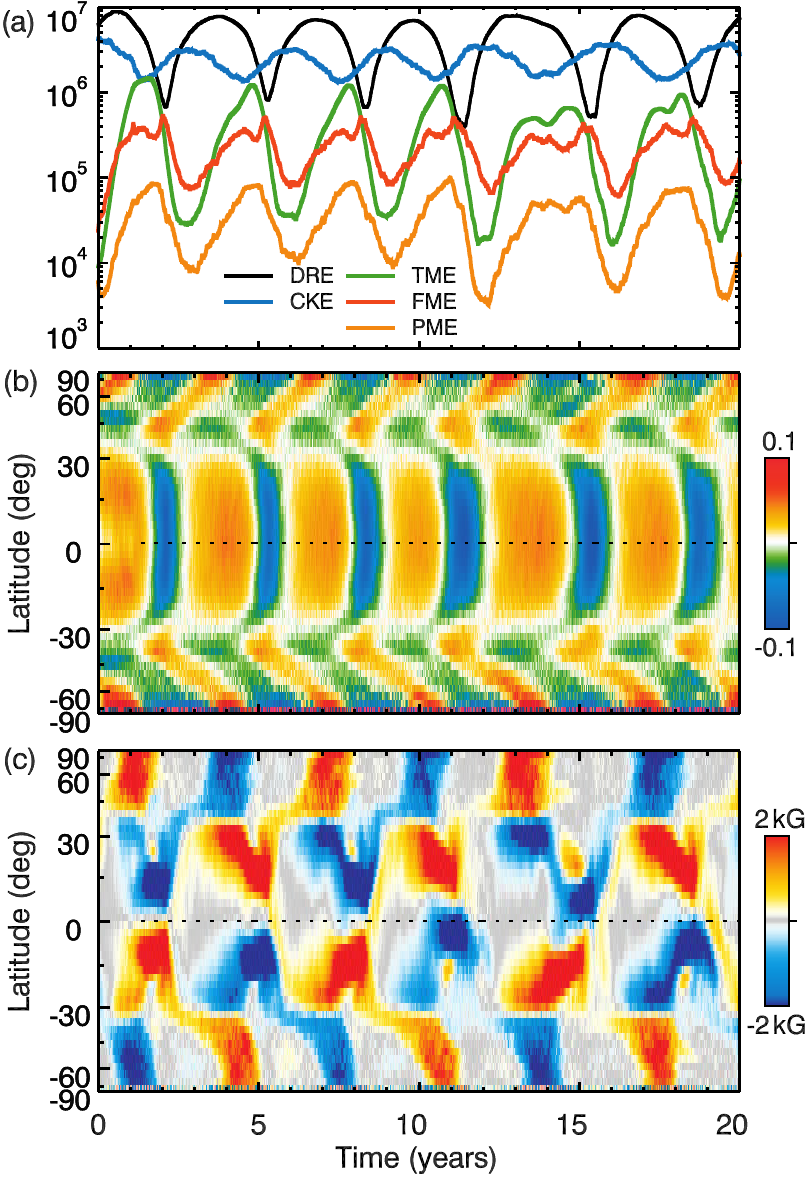} \figcaption{Evolution of the energy
       densities as well as the angular velocity variations and mean (longitudinally-averaged)
       toroidal magnetic field $\avgBp$ at $\Rsun{0.92}$ over the first 20~years of the
       simulation. (a) Time variation of the volume-averaged energy density of the differential
       rotation (DRE, black), nonaxisymmetric flows (CKE, blue), axisymmetric toroidal magnetic
       energy (TME, green), axisymmetric poloidal magnetic energy (PME, orange), and nonaxisymmetric
       magnetic energy (FME, red) in units of $\mathrm{erg}\, \mathrm{cm}^{-3}$ (b) Time-latitude
       diagram of angular velocity variations
       $\avg{\Delta\Omega}/\Omega_0=\left(\avgO-\{\Omega\}\right)/\Omega_0$ in cylindrical
       projection, elucidating the propagation of equatorial and polar branches of torsional
       oscillations. The color indicates faster rotation in red and slower rotation in blue, with
       departures of up to $\pm 10$\% of the bulk rotation rate. (c) Time-latitude diagram of
       $\avgBp$ in cylindrical projection, exhibiting the equatorward migration of the wreaths from
       the tangent cylinder, and the poleward propagation of the higher latitude field, with the
       polarity of the field such that red (blue) tones indicate positive (negative) toroidal
       field. \label{fig2}}
   \end{center}
\end{figure}

\noindent with $\kappa_r$ the molecular radiation diffusion coefficient, and $c_{\mathrm{P}}$ the
specific heat at constant pressure. The entropy diffusion flux has the thermal eddy diffusivity
$\kappa$ acting on the entropy fluctuations. A calorically-perfect ideal gas equation of state is
used for the mean state, about which the fluctuations are linearized as

\vspace{-0.25truein}
\begin{center}
  \begin{align}
     \displaystyle \opre &= (\gamma-1)\mathrm{c_P}\orho\oT/\gamma,  \label{eqn:idealgas} \\ 
     \displaystyle \rho/\orho &= P/\opre - T/\oT = P/\gamma \opre - S/\mathrm{c_P}, \label{eqn:asheos}
  \end{align}
\end{center}

\noindent with $\gamma=5/3$ the adiabatic exponent. The anelastic system of MHD equations requires
12 boundary conditions in order to be well posed. One of the primary goals of this work is to assess
the generation of magnetic field and how it impacts the organization of angular momentum and energy
in the simulation. Thus, the following impenetrable, torque-free, and flux transmitting boundary
conditions are employed

\vspace{-0.25truein}
\begin{center}
  \begin{equation}
    \displaystyle \vcr = \ddr{}\left(\frac{\vct}{r}\right) = \ddr{}\left(\frac{\vcp}{r}\right) = \ddr{S} = 0, \quad \mathrm{on} \; r=r_1 \; \mathrm{and} \; r_2. \label{eqn:bdrycond}
  \end{equation}
\end{center}

\noindent The magnetic boundary conditions are perfectly conducting at the lower radial boundary
($r_1$) and matching to a potential field at the upper radial boundary ($r_2$), implying that

\vspace{-0.25truein}
\begin{center}
  \begin{equation}
    \left. \displaystyle \Bcr\right|_{r_1} = 0 \quad \mathrm{and} \quad \left. \vB\right|_{r_2} = \nabla\Psi \Rightarrow \triangle \Psi = 0, \label{eqn:magbdry}
  \end{equation}
\end{center}

\noindent with $\Psi$ the magnetic potential. The solution of Laplace's equation defines the three
components of $\vB$ at the upper boundary. Further details of the implementation and formulation of
the ASH code can be found in \citet{clune99} and \citet{brun04}. 

Here a one solar mass star, with a solar luminosity, is considered that is rotating at three times
the solar rate. An isentropic background stratification is employed that closely resembles the
helioseismically-constrained Model S stratification \citep{christdals96}, with its normalized
spherically-symmetric profiles of density ($\orho$) and temperature ($\oT$) shown in Figure
\ref{fig1}(a). The simulated domain stretches from the base of the convection zone at
$r_1=\Rsun{0.72}$ to the upper boundary of the simulation at $r_2=\Rsun{0.97}$, where $\Rsun{}$~$=
6.96\times 10^{10} \, \mathrm{cm}$. This approximation omits the near-surface region and any regions
below the convection zone, such as a tachocline. The simulation K3S has a resolution of $N_r\times
N_{\theta} \times N_{\varphi} = 200\times 256\times 512$, corresponding to a horizontal resolution
with a maximum spherical harmonic degree of $\ell_{\mathrm{max}} = 170$. In what follows, the
operator $\avg{}$ indicates a longitudinal average (or mean) of a quantity, whereas the operator
$\{\}$ indicates a longitudinal and temporal average. The extent of the temporal average varies
depending upon the context of its use, so that interval will be indicated when the operator is
invoked.

The SLD mechanism implemented in the ASH code, and used in case K3S, is similar to the schemes
presented in \citet{rempel09} and \citet{fan13}, though it has been modified to compensate for the
grid convergence at the poles. This diffusive operator is detailed in Appendix A. SLD acts locally
to achieve a monotonic solution by limiting the slope in each coordinate direction of a piecewise
linear reconstruction of the unfiltered solution. The scheme minimizes the steepest gradient, while
the rate of diffusion is regulated by the local velocity. It is further reduced through a function
$\varphi$ that depends on the eighth power of the ratio of the cell-edge difference $\delta_i q$ and
the cell-center difference $\Delta_i q$ in a given direction $i$ for the quantity $q$. This limits
the action of the diffusion to regions with large differences in the reconstructed solutions at
cell-edges. Since SLD is computed in physical space, it incurs the cost of smaller time steps due to
the convergence of the grid at the poles, which is largely mitigated by introducing a filtering
operator that depends upon latitude. The resulting diffusion fields are projected back into spectral
space and added to the solution with a forward Euler time step. 

\begin{figure*}[t!]
   \begin{center}
     \includegraphics[width=\textwidth]{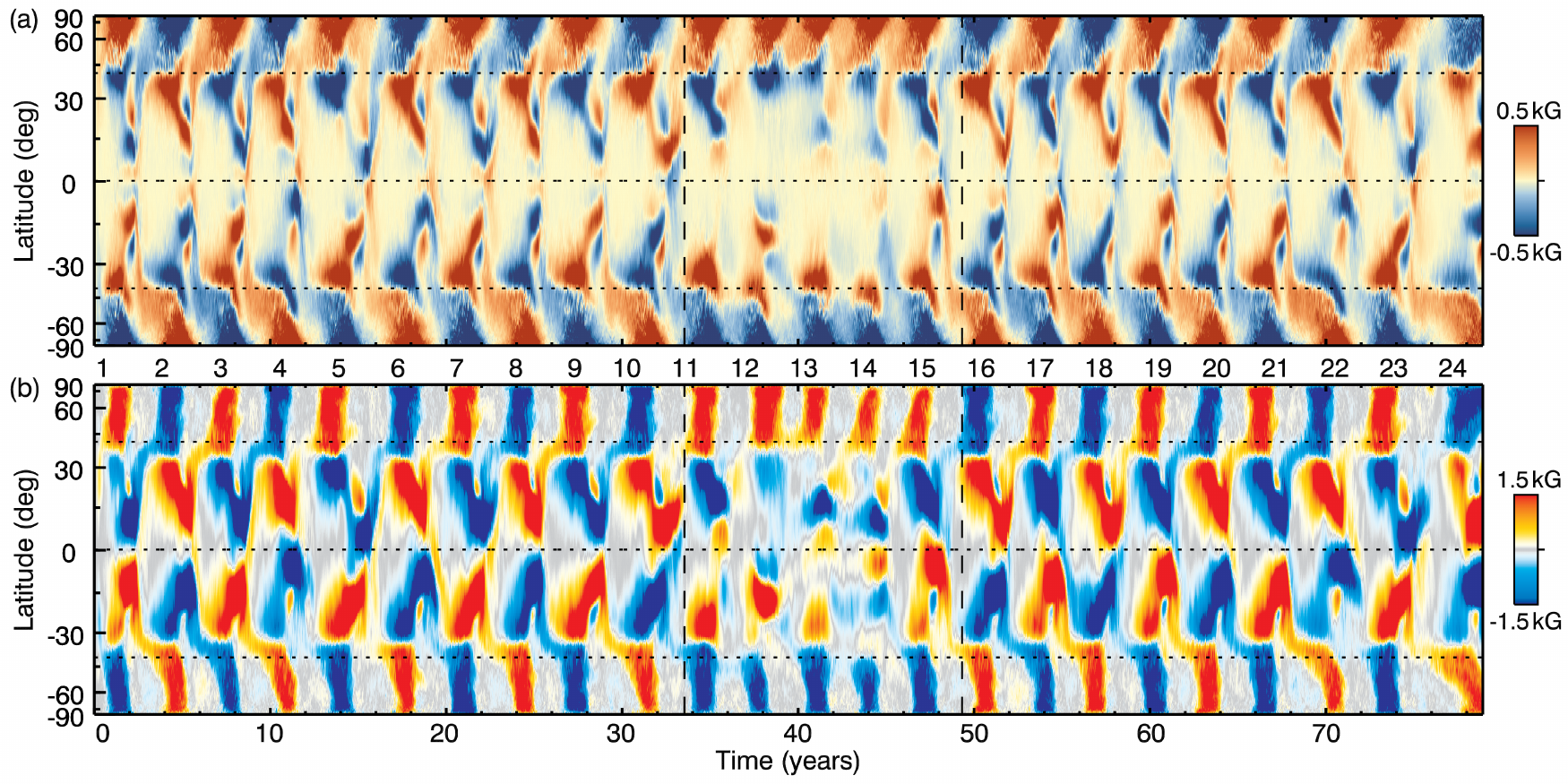} \figcaption{Evolution of the mean
       (longitudinally-averaged) radial $\avg{\Bcr}$ and toroidal $\avg{\Bcp}$ magnetic fields over
       an extended interval of the K3S simulation, with its regular cycling interrupted by a grand
       minimum during the interval roughly between 33 to 49 years. (a) Time-latitude diagram of
       $\avg{\Bcr}$ at $\Rsun{0.92}$ in cylindrical projection, elucidating the poleward propagation
       of mid and high-latitude magnetic field and the equatorward propagation of lower latitude
       field. (b) Time-latitude diagram of $\avg{\Bcp}$ at $\Rsun{0.92}$ in cylindrical projection,
       exhibiting the equatorward migration of the wreaths from the tangent cylinder (horizontal
       dotted lines at $\pm\dgr{43}$) and the poleward propagation of the higher latitude field. The
       polarity of the fields are such that red (blue) tones indicate positive (negative) field. The
       interval containing the grand minimum is marked by vertical dashed lines. Each magnetic
       energy cycle is labeled starting at unity.
     \label{fig3}}
   \end{center}
\end{figure*}

The SLD has been restricted to act only on the velocity field in this simulation. This mimics a
lower thermal and magnetic Prandtl number ($\Prt$, $\Prm$) than otherwise attainable through a
purely Newtonian diffusion operator with the spatial resolution used in this simulation. Yet a weak
viscous eddy diffusion is retained in addition to the SLD operator in order to reduce the condition
number of the matrices used in ASH for the implicit Crank-Nicholson time stepping method. In
contrast, the entropy and magnetic fields remain fully under the influence of an eddy diffusion,
with both a radially-dependent entropy diffusion $\kappa$ and resistivity $\eta$. The eddy diffusion
coefficients are roughly similar in form to those of case D3 from \citet{brown10} and case D3a
\citet{nelson13a}, with $\kappa$, $\nu$, and $\eta \propto \orho^{\; -1/2}$ and where the profiles
of $\kappa$ and $\eta$ are shown in Figure \ref{fig1}(a). The value of these diffusion coefficients
at the upper boundary are $\kappa(r_2) = 1.6\times 10^{13}$, $\nu(r_2) = 4\times 10^{8}$, and
$\eta(r_2) = 8\times 10^{12}$ with the units of each coefficient being $\mathrm{cm}^2\,
\mathrm{s}^{-1}$.

Since the majority of the viscous diffusion and dissipation is handled with the SLD scheme, it is
somewhat involved to estimate standard fluid parameters such as the Reynolds number. However, a
detailed analysis carried out in Appendix A.3 provides an estimate for the effective SLD
viscosity. This also permits the estimation of the Reynolds number as $\mathrm{Re}\approx 350$, as
well as thermal Prandtl number as $\mathrm{Pr}\approx 0.115$ and a magnetic Prandtl number of
$\mathrm{Pr}_\mathrm{m} \approx 0.23$.  This is about a factor of two lower than in previous ASH
simulations, which typically are carried out with $\mathrm{Pr}=1/4$ and
$\mathrm{Pr}_{\mathrm{m}}=1/2$. The effective magnetic Reynolds number is then
$\mathrm{Re}_{\mathrm{m}}=\mathrm{Pr}_{\mathrm{m}}\mathrm{Re}_{\mathrm{eff}} \approx 90$.  Further,
the Rayleigh number can be characterized at mid-convection zone as $\mathrm{Ra}=\Delta\overline{S} g
d^3/c_P \nu \kappa \approx 6.3\times 10^5$ and the Taylor number as $\mathrm{Ta}=4\Omega_0^2
d^4/\nu^2 \approx 9.1\times 10^7$, where $\Omega_0=\Osun{3}$ or $7.8\times 10^{-6} \mathrm{rad
  s^{-1}}$. The Rossby number, when defined with the enstrophy as
$\mathrm{Ro}=|\nabla\times\vv|/2\Omega_0$, varies with the magnetic cycle between 0.12 at magnetic
maximum to 0.33 at magnetic minimum. Thus, some of these parameters differ significantly from other
simulations that present features similar to this dynamo, namely those with equatorward propagating
fields such as in \citet{racine11} and \citet{warnecke14} where the effective thermal and magnetic
Prandtl numbers are between two and ten times larger. However, other relevant parameters are quite
similar such as the effective Reynolds numbers as well as the Rayleigh, Taylor, and Rossby numbers.

\begin{figure*}[ht!]
   \begin{center}
   \includegraphics[width=0.95\textwidth]{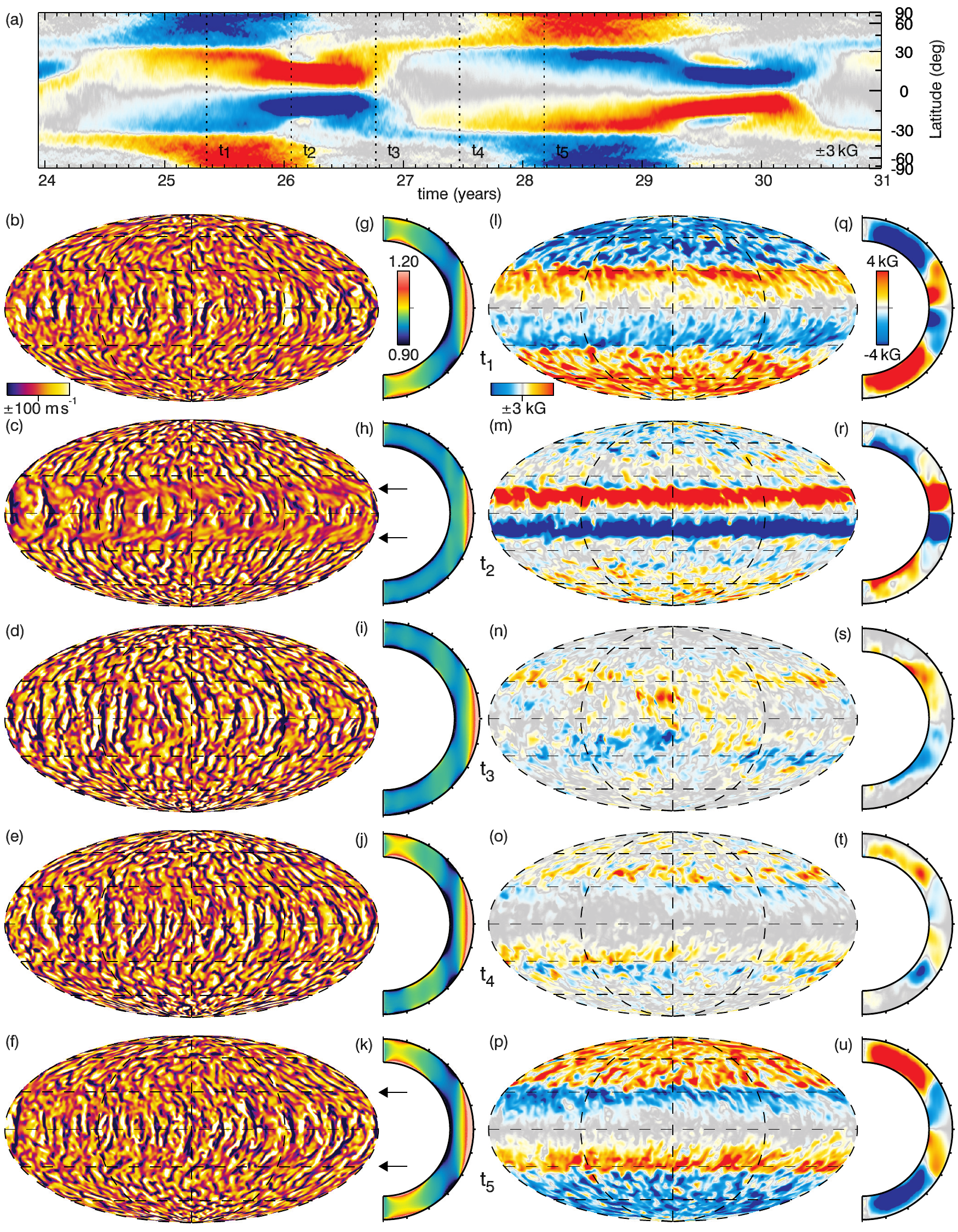} \figcaption{Changing velocity and magnetic
     fields during magnetic cycling. (a) The cylindrical projection of the mean toroidal magnetic
     field $\avgBp$ is shown with time and latitude at $\Rsun{0.90}$ through the course of a full
     polarity cycle (capturing cycles 8 and 9), with colorbar as in (l). The times $t_1$ through
     $t_5$ are indicated with vertical dashed lines. (b)-(f) The horizontal structure of the radial
     velocity at $\Rsun{0.90}$ is displayed in a global Mollweide projection (with dashed lines
     marking the equator and latitudes every $\dgr{30}$). The colorbar is given in (b).  Downflows
     are darker tones and upflows lighter tones. Latitudes that are influenced by strong magnetic
     fields are indicated with arrows. (g)-(k) A 20-day-time and longitude-averaged angular velocity
     $\{\Omega\}/\Omega_0$ is shown in the meridional plane, with colorbar as in (g) where faster
     rotation is in red tones and slower in blue tones. (l)-(p) The longitudinal field $\Bcp$ at
     $\Rsun{0.90}$ is illustrated, with the shared colorbar given in (l). (q)-(u) A 20-day-time and
     longitude-average of the toroidal magnetic field $\{\Bcp\}$ is shown, with colorbar as in
     (q). \label{fig4}}
   \end{center}
\end{figure*}

\section{Overview of the Cycling Dynamics} \label{sec:overview}

 With the formulation of the problem established, the discussion now turns to an overview of the
 dynamics occurring within the K3S simulation.  A first diagnostic of the K3S simulation is to
 assess its cycles in a global sense.  This is easily achieved by considering the evolution of
 volume-averaged energy densities of various components of the flow and magnetic field. The
 definition of the various energy densities are

\vspace{-0.25truein}
\begin{center}
  \begin{align}
   & \mathrm{DRE} = \frac{1}{2}\orho\avg{\vcp}^2,  \; \mathrm{MCE} = \frac{1}{2}\orho\left(\avg{\vcr}^2 + \avg{\vct}^2\right), \nonumber \\
   &\mathrm{CKE} = \frac{1}{2}\orho\left(\vv-\avg{\vv}\right)^2, \; \mathrm{TME} = \frac{1}{8\pi}\avg{\Bcp}^2, \nonumber \\ 
   &\mathrm{PME} = \frac{1}{8\pi}\left(\avg{\Bcr}^2 + \avg{\Bct}^2\right), \; \mathrm{FME} = \frac{1}{8\pi}\left(\vB-\avg{\vB}\right)^2, 
  \end{align}
\end{center}

\noindent where the total kinetic energy is $\mathrm{DRE}+\mathrm{MCE}+\mathrm{CKE}$ and the total
magnetic energy is $\mathrm{TME}+\mathrm{PME}+\mathrm{FME}$.

The periodic modulation of the kinetic energy in both the convection (CKE) and the differential
rotation (DRE) can be seen in Figure \ref{fig2}(a), which covers the first 20~years of evolution of
the simulation. Those changes are accompanied by the wax and wane of the energy contained in the
magnetic field, though with a different temporal shift. The mean (longitudinally-averaged) toroidal
magnetic fields (TME) contain the most magnetic energy, being formed by the action of the
differential rotation on the poloidal magnetic field (e.g., \S\ref{sec:tme}). The nonaxisymmetric
magnetic fields (FME) have the second greatest magnetic energy, whereas the mean
(longitudinally-averaged) poloidal magnetic fields (PME) contain the least amount of magnetic
energy. A double-peaked structure can be seen in the FME as well as in the TME. These variations in
turn are largely due to the modulation of the differential rotation over the course of a cycle,
which becomes readily apparent in Figures \ref{fig2}(b). Particularly, the pole and equator are
accelerated as the system recovers from the quenching of the differential rotation that occurs
during the magnetic maxima, leading to the first peak. There is also a phase difference in the peak
in the magnetic energy between the deep and upper convection zones, which results in the second
peak. The specific correlations and mechanisms that are behind this oscillatory behavior are covered
below and in later sections.

The polarity reversals of the magnetic field are illustrated for the first 20~years of the
simulation in Figure \ref{fig2}(c) and over an extended interval of the simulation in Figure
\ref{fig3}. The magnetic field begins to regularly oscillate roughly every 3.1~years between
positive and negative polarity states. Such regular cycling behavior arises shortly after the
roughly 2~year kinematic growth phase of the magnetic fields, which began at year zero when this MHD
simulation was initialized by inserting a dipolar magnetic field into a preexisting but mature
hydrodynamic simulation. That initial magnetic field had a strength of about 100~G at the base of
the convection zone. The initial energy in that magnetic field is about $10^5$ times smaller than
the total kinetic energy. The oscillations in the magnetic energy then continue throughout the
entire evolution of the system. Throughout this paper two cycle periods will be cited. There is a
3.1~year magnetic energy cycle measured between maxima in the magnetic energy (or half-polarity
cycle), which could be considered to be akin to the 11~year sunspot cycle. There is also a 6.2~year
polarity cycle measured as the interval between magnetic maxima that have the same polarity as seen
in Figure \ref{fig3} for instance. This polarity cycle is akin to the 22~year solar polarity
cycle.

The overall structure of the magnetic fields during a magnetic cycle is readily apparent in both
Figures \ref{fig3}(a) and (b). The mean radial magnetic field ($\avg{\Bcr}$) is largely
confined to higher latitudes, whereas the mean toroidal magnetic field field ($\avg{\Bcp}$) has both
prominent polar and low-latitude branches. At the radius where the magnetic fields in Figure
\ref{fig3} are sampled, the mean radial and toroidal magnetic fields differ by about a factor
of three in magnitude. This ratio is approximately maintained throughout the domain, leading to the
roughly order of magnitude difference between the toroidal and poloidal magnetic energies seen in
Figure \ref{fig2}(a). There is an interval roughly between years 33 and 49 as seen in Figure
\ref{fig3} during which the system fails to fully reverse its polarity for five magnetic
cycles. This interval will be referred to as a ``grand minimum.''  While the choice of the beginning
and end of this interval is somewhat arbitrary, this interval was chosen to be between the minimum
in the magnetic energy near the upper boundary of the last ``normal'' cycle (near year 33, or cycle
11) and the similar minimum of the last abnormal cycle (near year 49, or cycle 16). The magnetic
energy cycle is still operating during that interval, with the polar field waxing and waning but not
fully reversing. However, the polarity cycles are disrupted at lower latitudes and the magnetic
energy there is significantly reduced. This will be further explored in \S\ref{sec:minimum}.

\subsection{Magnetic Energy Cycle in Detail} \label{sec:magcyc}

Figure \ref{fig4} illustrates the morphology of the convection, differential rotation, and the
longitudinal magnetic fields in space and time over the course of a polarity reversal. Particularly,
Figures \ref{fig4}(b)-(f) shows the convective patterns represented in radial velocities that are
prevalent during different phases of the cycle (labeled as times $t_1$-$t_5$ in Figure
(\ref{fig4})), with elongated and north-south aligned flows at low latitudes (banana cells) and
apparently smaller scales at higher latitudes. Such flows are typical in the
rotationally-constrained convection captured in global-scale large-eddy MHD simulations
\citep[e.g.,][]{miesch00,kapyla11a,guerrero13,augustson12,augustson13}. In aggregate, the velocity
field of those rotationally-aligned convective cells produce correlations in the velocity field that
yield strong Reynolds stresses that act to accelerate the equator and slow the poles. In concert
with turbulent heat transport, such stresses serve to rebuild and maintain the differential rotation
during each cycle. Indeed, when combined with the longitudinally-averaged Lorentz force and Maxwell
stresses, those stresses induce the modulation in the angular velocity seen in Figures \ref{fig2}(b)
and \ref{fig4}(g)-(k). The values of the shear do not quite reach zero, but are reduced by about
60\% relative to their maximum value at lower latitudes. The elements of angular momentum transport
that give rise to such modulation are further discussed in \S\ref{sec:dre}.

The presence of large-scale and longitudinally-connected magnetic structures is evident in $\Bcp$ as
shown in Figures \ref{fig4}(l)-(p). Such toroidal structures have been dubbed wreaths
\citep{brown10}. In this simulation, there are two evolving counter-polarized, lower-latitude
wreaths that form in the region near the tangent cylinder at nearly all depths, meaning that the
latitude of formation decreases with depth. This region is also where the peak in the latitudinal
gradient of the differential rotation exists for much of a magnetic energy cycle (Figures
\ref{fig2}(b) and \ref{fig4}(g)-(k)). In the latter set of figures, it is clear that the radial
shear is roughly proportional to $r\sin{\theta}$ at low latitudes, for the differential rotation is
largely cylindrical. Though, it tends decrease near the upper boundary. So for figures showing
radial cuts, a radius of $0.92 R_{\odot}$ was chosen to emphasize the region of greatest shear as it
corresponds to the depth where $d\Omega/dr$ is largest.

There are also polar caps of magnetism that possess a magnetic polarity that is reversed compared to
that of the low-latitude wreaths. These caps act to moderate the polar differential rotation, which
would otherwise tend to accelerate and hence establish fast polar vortices. The average structure of
the wreaths and caps at each point in the cycle is apparent in $\{\Bcp\}$ exhibited in Figures
\ref{fig4}(q)-(u), which is averaged over 20~days at each time $t_i$. The wreaths appear rooted at
the base of the convection zone, whereas the caps have the bulk of their energy in the lower
convection zone above its base. As will be seen in \S\ref{sec:evodyno}, the wreaths are initially
generated higher in the convection zone while the wreath generation mechanism (primarily the
$\Omega$-effect) migrates equatorward and toward the base of the convection zone over the course of
the cycle. The equatorward migration takes place largely in the upper convection zone nearer the
beginning of the cycle at times $t_1$ and $t_2$. At later times, the mean toroidal magnetic field
near the base of the convection zone migrates poleward and begins to build up the polar magnetic
caps, which have a polarity opposite to the flux generated at lower latitudes. This represents a
very different dynamo mechanism relative to a typical flux-transport dynamo.

The changes in the structure of the convection seen in Figure \ref{fig4} plays a role in the
dynamo, for they induce changes in the Reynolds stress and the electromotive force (EMF) that
generates the magnetic field. As a cycle proceeds, the magnetic fields disrupt the alignment and
correlations of the convective cells through Lorentz forces, which is particularly evident in Figure
\ref{fig4}(c). The presence of the magnetic fields, in addition to modifying the structure of
the low-latitude convection, modulates the global convective amplitudes as might be ascertained by
comparing Figures \ref{fig4}(b)-(f).  Particularly, while the magnetic field gathers strength
during a cycle, the strong longitudinally-connected magnetic fields also create a thermal shadow,
weakening the thermal driving of the equatorial cells, as indicated with arrows.

Such influences of the magnetic fields on the convection and its ability to transport heat are also
apparent, albeit less directly, in Figure \ref{fig1}(b) where the enthalpy, entropy diffusion, and
kinetic energy fluxes are modulated by about 30\% throughout a cycle. These fluxes vary largely in
phase with the cycle, where at magnetic maximum the fluxes are smallest and near minimum they are
largest. The reduction of the convective amplitudes also leads to the angular momentum transport of
the flows being diminished as the magnetic fields become stronger (see \S\ref{sec:dre}). The effects
of the magnetic fields on the convection are also captured in the ebb and flow of the kinetic energy
contained in the nonaxisymmetric velocity field, which here varies by about 50\% over the magnetic
energy cycle (CKE in Figure \ref{fig2}(a)). Indeed, signatures of such in-phase
magnetically-modulated convection have also been detected in EULAG-MHD simulations
\citep{cossette13,charbonneau14}. These magnetic feedback mechanisms are in keeping with the
predicted impacts of strong longitudinal fields in the convection zone suggested by
\citet{parker87}.

\begin{figure}[t!]
   \begin{center}
     \includegraphics[width=0.45\textwidth]{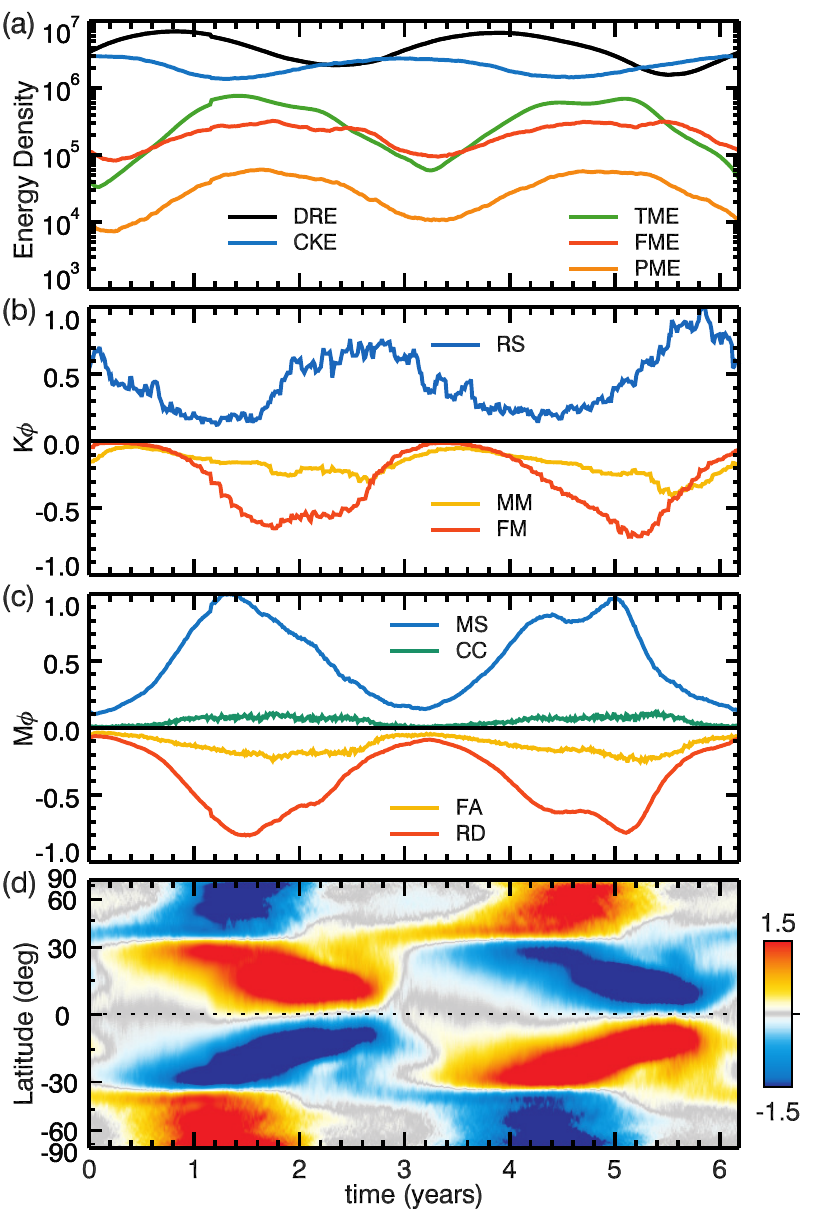} \figcaption{Energy densities and energy
       generation rates over the average polarity cycle. (a) Log-linear plot of the energy density
       of the differential rotation (DRE, black), nonaxisymmetric flow (CKE, blue), toroidal
       magnetic field (TME, green), nonaxisymmetric magnetic field (FME, red), poloidal magnetic
       energy (PME, orange) in units of $\mathrm{erg\, cm}^{-3}$. (b) Volume-integrated differential
       rotation energy generation rate ($K_{\varphi}$, as in Equation \ref{eqn:evokdr}), with the
       rate due to Reynolds stresses (RS, blue), mean magnetic stresses (MM, yellow), and
       fluctuating Maxwell stresses (FM, orange). (c) Volume-integrated mean toroidal magnetic
       energy generation rate ($M_{\varphi}$, as in Equation \ref{eqn:tme}), with the rate due mean
       shear (MS, blue), compressive motions (CC, green), fluctuating advection (FA, yellow), and
       resistive diffusion (RD, red). In (b) and (c), the values are normalized by the maximum
       absolute value of the generation rates. (d) The cycle and longitude-averaged toroidal
       magnetic field $\avgBp$ evaluated over the average magnetic polarity cycle shown at
       $\Rsun{0.92}$, with a color bar given in units of kG. \label{fig5}}
   \end{center}
\end{figure}

There is also the direct impact of the large-scale Lorentz forces on the differential rotation
(e.g., the \citet{malkus75} effect). This process and the magnetic influences on the convection
described above combine to explain why the differential rotation seen in Figure \ref{fig2}(b) cannot
be fully maintained during the cycle. Rather, the angular velocity has substantial variations
throughout the cycle (Figure \ref{fig2}(a)), which are largely driven by the strong feedback of the
magnetic fields (see \S\ref{sec:dre} and Figure \ref{fig5}). Such strong nonlinear Lorentz force
feedbacks are not without precedent, as they have been seen in previous convective dynamo
simulations as well \citep[e.g.,][]{gilman83,brun04,browning08,brown11}.

\section{Anatomy of a Reversal} \label{sec:reversal}

The analysis of dynamical elements contributing to regular magnetic polarity cycles is aided by
building a composite of a number of our regular cycles. This composite omits the interval designated
as the grand minimum.

\subsection{Average Polarity Cycle} \label{sec:avgcyc}

The average polarity cycle is shown in Figure \ref{fig5}, which has been formed by identifying the
common structures in each polarity cycle, obtaining the times of the beginning and end of each
polarity cycle as defined through these structures, and then stretching each polarity cycle to be
the same length in time and co-adding them. The statistical significance of this process is greatly
aided by the regularity of the magnetic polarity cycle period, which typically varies by only about
10\% of the average polarity cycle period (see \S\ref{sec:periods}). In particular, Figure
\ref{fig5}(a) shows the time evolution of the relevant volume-integrated components of the total
energy density. The differential rotation energy (DRE) is the largest component of the total kinetic
energy followed by the kinetic energy in the convection (CKE). The energy contained in the
meridional circulation is the smallest component, being roughly three orders of magnitude smaller
than the DRE and is omitted from Figure \ref{fig5}(a). Such a small contribution of the meridional
flow to the overall kinetic energy is typical of global-scale convection simulations
\citep[e.g.,][]{brown08,augustson12,brun14}.

The total magnetic energy at its peak is about 30\% of the total kinetic energy, or about 50\% of
the CKE, placing the K3S dynamo close to equipartition when averaged over the cycles and the
domain. There are certainly some cycles and most certainly some regions in the computational domain
in which the kinetic and magnetic energies are near equipartition. The bulk of the magnetic energy
resides in the low-latitude magnetic wreaths and polar caps (TME), with the energy in the
nonaxisymmetric field (FME) a close second. The energy contained in the mean poloidal field (PME) is
about an order of magnitude smaller.

Figure \ref{fig5}(a) further elucidates three prominent phase shifts, one between the differential
rotation and the convection, a second between the magnetic field and the differential rotation, and
a third between the convection and the magnetic field. The first phase shift is apparent when
comparing peak values of CKE and DRE, where the CKE precedes the DRE by about a year. The second
shift has the peak in DRE preceding that of TME by about 0.6~years. Finally, there is an additional
phase shift between maxima of CKE and TME, with CKE preceding TME by about 1.5~years. Each of these
phase shifts are related to the nonlinear coupling of the convective Reynolds stresses, the
differential rotation, the Lorentz forces and the Maxwell stresses as will be further studied in
\S\ref{sec:dre} and \ref{sec:tme}.

The magnetic energy cycles visible in Figure \ref{fig3} can be numbered starting with unity
at year zero and extending up to cycle 24 at year 80. There are quantifiable differences between the
even and odd-numbered magnetic energy cycles over that extended interval that become more apparent
in the averaged polarity cycle shown in Figure \ref{fig5}. Such a temporal parity is similar to the
observed behavior of the sunspots known as the Gnevyshev-Ohl rule \citep[e.g.,][]{charbonneau10}.
Those asymmetries are reflected in the behaviors exhibited by the energy densities and their
transport mechanisms in the average polarity cycle shown in Figure \ref{fig5}.  The average of the
odd-numbered cycles are captured there as the first 3.1~year interval, and the average even cycle is
shown over the succeeding 3.1~year interval. In particular, there is a deeper minimum in the
magnetic energy densities as the odd-numbered cycles are entered as compared to the the entrance of
the even-numbered cycles. When TME and PME are assessed alone, the deeper minima become more
apparent, with their energy densities being about 50\% and 25\% lower respectively (Figure
\ref{fig5}(a)). Such temporal differences in the properties of the odd and even cycles are also
evident in the transport mechanisms. For instance the Reynolds stresses and Maxwell stresses of the
odd cycles have a broader variation in time than the even cycles, where they are more sharply peaked
in both of those quantities (Figure \ref{fig5}(b)). In contrast, the magnetic energy production
terms exhibit the opposite correlation, with the production terms being more peaked during the
even-numbered cycles (Figure \ref{fig5}(c)). Yet this simulation must be run longer to better assess
the statistical significance of this temporal asymmetry.  However, similar signatures of such
differences between even and odd magnetic cycles have been found within the long-running cycles of
an EULAG-MHD simulation \citep{passos14}.

\subsection{Maintaining the Differential Rotation}\label{sec:dre}

The evolution of the energy contained in the differential rotation is critical to the behavior of
the K3S dynamo, and its evolution is considered explicitly here. A detailed derivation of Equation
\ref{eqn:evokdr} is given in Appendix B. In particular, it is shown in Appendix B that the boundary
fluxes of mean kinetic energy of the longitudinal flows are zero as is the volume integrated energy
arising from the advection of the angular velocity.  Therefore, the evolution of the
volume-integrated differential rotation kinetic energy (DRE) is

\vspace{-0.35truein}
\begin{center}
   \begin{align}
      K_{\varphi} &= \frac{d\mathrm{DRE}}{dt} = \displaystyle \int_V dV \ddtime{}\frac{1}{2}\orho\avg{\vcp}^2 = \int_V dV \bigg[\overbrace{\orho\lambda\avg{\vcp'\vv'}\cnabla\avg{\Omega}}^{\mathrm{RS}} \nonumber \\ 
        & -\overbrace{\frac{\lambda}{4\pi}\avg{\Bcp}\avg{\vB}\cnabla\avg{\Omega}}^{\mathrm{MM}} -\overbrace{\frac{\lambda}{4\pi}\avg{\Bcp'\vB'}\cnabla\avg{\Omega}}^{\mathrm{FM}} \bigg]. \label{eqn:evokdr}
   \end{align}
\end{center}

\noindent with $\vv'=\vv-\avg{\vv}$ the nonaxisymmetric velocity, $\vB'=\vB-\avg{\vB}$ the
nonaxisymmetric magnetic field, as well as $\avg{\vv}$ and $\avg{\vB}$ the axisymmetric velocity and
magnetic field respectively.

Figure \ref{fig5}(b) shows the evolution of the primary components contributing to the dissipation
and production of differential rotation kinetic energy (DRE) given in Equation (\ref{eqn:evokdr}).
Clearly, the Reynolds stresses (RS) are the only significant means of producing DRE. The
contribution of the SLD viscous stresses are very small when integrated over the volume, being over
two orders of magnitude less than the RS. Thus they are not shown. The contribution of the RS to the
DRE varies substantially throughout a cycle, which is a reflection of the Lorentz force impacting
the morphology of convective structures that can be formed and thus their capacity to generate
DRE. The magnetic fields do not play just a passive role either, for they actively dissipate and
transfer energy as well. Indeed, both the mean magnetic stresses (MM) and the fluctuating Maxwell
stresses (FM) contribute to the global transfer of DRE to the magnetic energy reservoir where some
of this energy will be dissipated via a resistive channel. More importantly, the FM can act to
inhibit local turbulence and vortical motions, acting much like an anisotropic and inhomogeneous
viscous dissipation, whereas the MM act primarily on the large-scale flows such as the differential
rotation. The FM dominate throughout much of the cycle, though the MM play a larger role during
minima. Nevertheless, the amplitudes of the FM and MM during a magnetic energy cycle are tightly
correlated with the magnetic energy densities, as expected.

The Reynolds stresses (RS) reach a peak about 0.4~years before a magnetic minimum and then begin to
decrease through the minimum and the rest of the cycle (Figure \ref{fig5}(b)). If it were primarily
the magnetic fields that modify the RS, one might expect that the RS terms would be maximum at the
minimum of the magnetic energy. Instead, the RS are maximum when the differential rotation is at a
minimum. There are likely two reasons for this: one is that the energy in the convection is growing
at that time, leading to an increase in the RS, and the other is that the shear of the differential
rotation itself modifies the velocity correlations of the convective structures. The shear of the
differential rotation will radially and longitudinally stretch the equatorial columns of convection
(banana cells) that are primarily responsible for building it. This tends to diminish the velocity
correlations responsible for generating the Reynolds stresses.

\begin{figure*}[t!]
   \begin{center}
     \includegraphics[width=\textwidth]{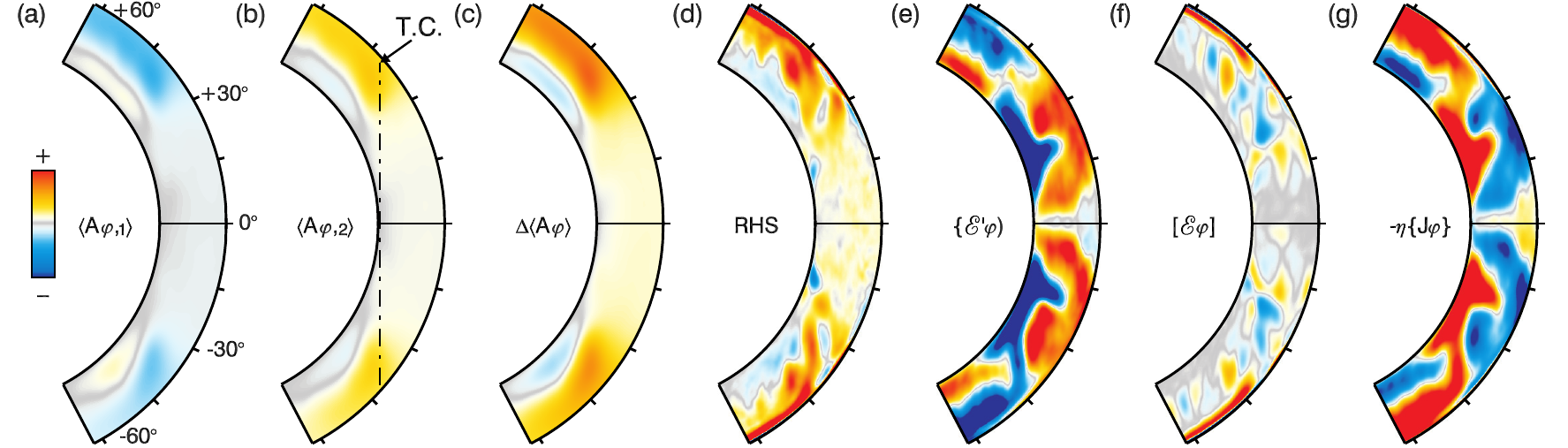} \figcaption{Time evolution of poloidal
       magnetic potential $\avgAp$ through a typical magnetic energy cycle. Longitudinal component
       of the magnetic potential (a) at $t_1$ called $\avg{A_{\varphi,1}}$ and (b) $t_2$ called
       $\avg{A_{\varphi,2}}$, and their difference (c) $\avg{\Delta A_\varphi}=\avg{A_{\varphi,2}} -
       \avg{A_{\varphi,1}}$. The sum of the right-hand-side terms in Equation (\ref{eqn:daphi}) is
       shown as panel (d). The components of the sum are shown individually as: (e) the turbulent
       EMF $\{\mathcal{E}'_\varphi\}$, (f) the mean EMF $\left[\mathcal{E}_\varphi\right]$, and (g)
       the resistive diffusion $-\eta \{J_\varphi\}$. The location of the tangent cylinder (T.~C.) is
       shown in (b). \label{fig6}}
   \end{center}
\end{figure*}

\subsection{Building Toroidal Magnetic Structures} \label{sec:tme}

The coherent large-scale wreath-like magnetic structures have been realized in many stellar
convective dynamo simulations utilizing very different codes such as in
\citet{browning06,brown08,ghizaru10,kapyla12,augustson13}, and \citet{nelson13a}. The common feature
shared by all those simulations is that the regions in which the wreaths form in the convection zone
is typically one where the Rossby number is low. For the ASH simulations, another feature that
appears to promote the formation of longitudinal magnetic structures is a perfectly conducting lower
boundary condition, which requires the field to be horizontal there. Thus the formation of magnetic
wreaths in the K3S simulation is generally promoted both through its relatively low Rossby number as
well as through the use of perfectly conducting lower boundary condition. These mean toroidal
magnetic fields $\avgBp$ are shown in Figures \ref{fig2}, \ref{fig3}, and
\ref{fig4}. Such magnetic fields are initially generated and subsequently maintained by
similar processes. During the growth phase of the magnetic field, the shear of the differential
rotation acts to fold and wind the initial poloidal field into large-scale longitudinal magnetic
structures. In this kinematic phase, the shear and meridional flows are largely unaffected and can
be considered stationary relative to the time scales of the growing field. However, once the
magnetic fields are strong enough, they begin to impact the convective flows that cross them through
Lorentz forces.  Hence, the magnetic field strength becomes saturated as the back-reaction of the
Lorentz forces increases the alignment of the velocity field and the magnetic field, which reduces
both its generation and can lead to its destruction. To quantify these processes, consider the time
evolution of the toroidal magnetic energy (TME), which can be represented as

\vspace{-0.25truein}
\begin{center}
   \begin{align}
      M_\varphi &= \frac{d\mathrm{TME}}{dt} = \displaystyle \int_V dV\ddtime{}\frac{\avg{\Bcp}^2}{8\pi} \label{eqn:tme} \\
               &= \int_V dV\frac{\MTF}{4\pi}\pht\cdot\bigg[\overbrace{\MBF\cnabla\MVF}^{\TMS}
               + \overbrace{\bigavg{\vB'\cnabla\vv'}}^{\TFS} - \overbrace{\MVF\cnabla\MBF}^{\TMA} \nonumber \\ 
               &- \overbrace{\bigavg{\vv'\cnabla\vB'}}^{\TFA} + \overbrace{\left(\avg{\vB}\avg{\vcr} 
                 + \avg{\vB'\vcr'}\right) \ddr{\ln{\orho}}}^{\TCC} - \overbrace{\curl\left(\eta\avg{\vJ}\right)}^{\TRD} \bigg]. \nonumber
   \end{align}
\end{center}

\noindent A detailed derivation of the production terms for the mean magnetic fields in spherical
coordinates is provided in Appendix A of \citet{brown10}. The terms in Equation (\ref{eqn:tme}) are
the production of magnetic energy by mean shear ($\mathrm{MS}$), fluctuating shear ($\mathrm{FS}$),
mean advection ($\mathrm{MA}$), fluctuating advection ($\mathrm{FA}$), compressive correlations
($\mathrm{CC}$), and resistive diffusion ($\mathrm{RD}$).

The significant volume-integrated components of Equation (\ref{eqn:tme}) are shown in Figure
\ref{fig5}(c). As suggested above, the $\Omega$-effect or mean shear (MS) here is the dominant means
of producing magnetic energy in the toroidal fields, which is accompanied by a weak contribution
from the compressive terms (CC). In contrast, resistive dissipation (RD) and fluctuating advection
(FA) dissipate TME. The other terms comprise less than 5\% of the total production or dissipation of
TME. While there is a generation of TME when all the terms are summed during much of the magnetic
energy cycle, it is clear that much of the temporally local generation through mean shearing effects
(or the $\Omega$-effect) is counter-balanced by dissipative processes. As with the poloidal
generation mechanisms as seen in \S\ref{sec:genpol}, the generation of field is greater than its
rate of dissipation during the growth phase of the energy cycle, as to be expected of a convective
dynamo whose magnetic Reynolds number is supercritical. Whereas during the declining phase of the
cycle, dissipation dominates these processes and so the magnetic energy declines. There is also a
strong correlation between the generation of field through the compressive mechanism and dissipation
by fluctuating advection.  Their amplitudes are, however, not perfectly matched. Instead, the energy
dissipated through the FA term is energy that is converted into either mechanical energy or magnetic
energy such as the poloidal and nonaxisymmetric magnetic fields. Note that the magnetic energy
dissipation through the FA term provides a first indication that the dynamo operating in K3S is of
an $\alpha$-$\Omega$ type rather than $\alpha^2$-$\Omega$ within the context of mean-field dynamo
theory \citep[e.g.,][]{krause80}. Indeed, the dissipative character of the FA term is more
reminiscent of the diffusive mean-field $\beta$ effect, which is defined in \S\ref{sec:alpha}.

\subsection{Generating Poloidal Fields} \label{sec:genpol}

The time evolution of the magnetic field can be recovered from the magnetic vector potential, where
for instance the mean toroidal magnetic vector potential $\avg{\Acp}$ captures the poloidal magnetic
field as $\MPF = \curl(\avgAp\pht)$. In particular, the behavior of $\avg{\Acp}$ is governed by the
following form of the induction equation

\vspace{-0.25truein}
\begin{center}
   \begin{align}
     \displaystyle \ddtime{\avgAp} = \pht\cdot\left[\avg{\vv'\times\vB'} + \avg{\vv}\times\avg{\vB} 
                                      - \eta\avg{\vJ}\right]. & \label{eqn:daphidt}
   \end{align}
\end{center}

\noindent In what follows, $\boldsymbol{\mathcal{E}}=\vv\times\vB$ is the electromotive force
(EMF). Thus the turbulent electromotive force (EMF, $\boldsymbol{\mathcal{E}}'$) is defined as
$\avg{\boldsymbol{\mathcal{E}}'}=\avg{\vv'\times\vB'}$.  The diffusion is proportional to the
product of the current $\vJ=c/4\pi\curl\vB$ and the magnetic diffusion coefficient $\eta$. As noted
in \citet{nelson13a}, the definite time integral of this equation subsequently yields

\vspace{-0.25truein}
\begin{center}
   \begin{align}
     &\displaystyle \Delta\avgAp = \avg{A_{\varphi,2}} - \avg{A_{\varphi,1}}
                                 = \{\mathcal{E}'_\varphi\} + \left[\mathcal{E}_\varphi\right] - \eta \{J_{\varphi}\} \nonumber \\
     &=\!\!\!\int_{t_1}^{t_2}\!\!\!\!\!\! dt \pht\cdot\avg{\vv'\times\vB'} + \int_{t_1}^{t_2}\!\!\!\!\!\! dt \pht\cdot\left(\avg{\vv}\times\avg{\vB}\right)
                                 - \int_{t_1}^{t_2}\!\!\!\!\!\! dt \eta\avg{J_\varphi}. \label{eqn:daphi}
   \end{align}
\end{center}

\noindent where $\left[\mathcal{E}_\varphi\right]$ denotes the time integral of the mean component
of the EMF. This can be interpreted as the difference between two snapshots of the longitudinal
vector potential being proportional to three time-integrated terms: the longitudinal-average of the
turbulent EMF, the mean EMF, and the magnetic diffusion. Since only $\avg{\Acp}$ is being
considered, Equations (\ref{eqn:daphidt}) and (\ref{eqn:daphi}) are rendered gauge invariant since
$\partial_\varphi \avg{\Acp} = 0$.

The mechanisms that set the time scales relevant to the reversal of the poloidal field are difficult
to assess. Namely, these mechanisms require information about the collective action of the turbulent
convection upon existing magnetic structures as well as the complex self-interaction of convection
to produce differential rotation. These processes are inherently nonlocal in space as magnetic
energy from the local and small-scale action of helical motions upon a large-scale toroidal magnetic
structure leads to a large-scale poloidal field, and thus is also nonlocal in time as the
large-scale structures evolve on longer time scale than the convection.  However, these processes
can be individually assessed, beginning with an illustration of the several components of the
production of poloidal magnetic field. To help further disentangle the various influences of the
convection on the turbulent production of magnetic field, a mean-field analysis of the K3S dynamo is
provided in \S\ref{sec:alpha}.

\begin{figure}[t!]
   \begin{center}
   \includegraphics[width=0.485\textwidth]{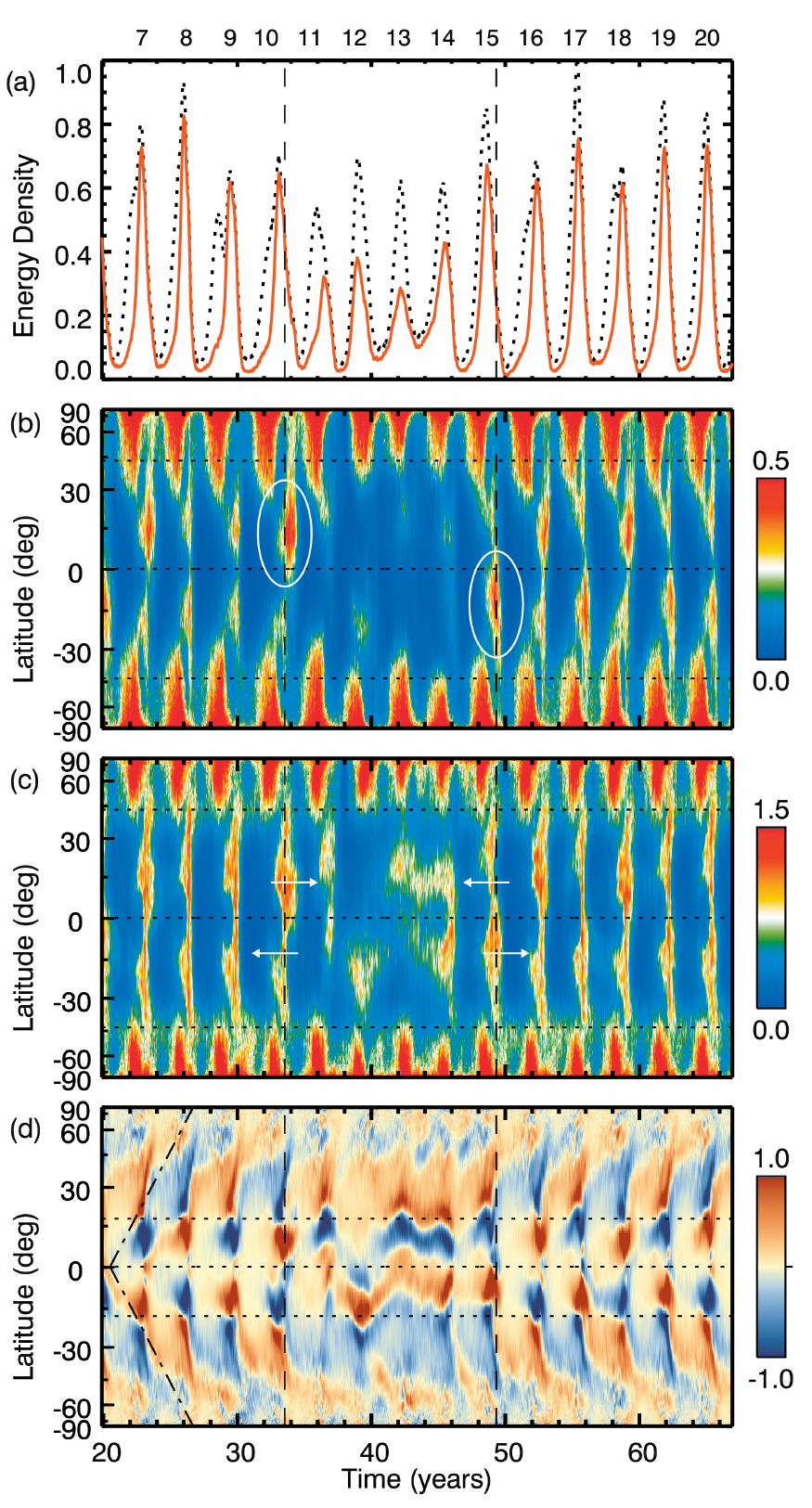} \figcaption{Evolution of total magnetic
     energy, nonaxisymmetric magnetic energy, and radial magnetic field through the grand minimum.
     (a) Time variation of the magnetic energy through the grand minimum. The normalized and
     volume-averaged total magnetic energy is depicted with integrals taken over the total volume
     (dotted black line) and at lower latitudes between $\pm\dgr{40}$ (red line). (b) Time advance
     of the nonaxisymmetric magnetic field magnitude ($|B_r|$ for $m>0$) at $\Rsun{0.95}$, with
     regions of strong equatorially antisymmetry encircled. (c) Time and latitude dependence of the
     nonaxisymmetric magnetic field magnitude ($|B_r|$ for $m>0$) at $\Rsun{0.75}$, with hemispheric
     temporal lags indicated with arrows. (d) Time-latitude diagram of $\avg{\Bcr}$ at a depth of
     $\Rsun{0.75}$ with dot-dashed lines helping to illustrate the primary evolution of the polarity
     cycle. Units in (b), (c), and (d) are kG. The tangent cylinder is indicated with horizontal
     dotted lines and the grand minimum by vertical dashed lines.\label{fig7}}
   \end{center}
\end{figure}

The terms in Equation (\ref{eqn:daphi}) are shown in Figure \ref{fig6}, where the times $t_1$ and
$t_2$ in Equation (\ref{eqn:daphi}) are taken at the peak values of the magnetic energy on either
side of a minimum in magnetic energy. Thus a magnetic energy cycle and a magnetic polarity reversal
are captured. In order, Figure \ref{fig6} shows the two instances of the vector potential (Figures
\ref{fig6}(a), (b)) whose difference (Figure \ref{fig6}(c)) closely corresponds to the sum of the
time-integrated components of the EMF and the magnetic diffusion, which are shown individually in
the last three panels. Comparing Figures \ref{fig6}(e, g), the predominant competition is between
the fluctuating EMF and the resistive diffusion, with a modest contribution from the mean EMF
(Figure \ref{fig6}(f)) contributing to the full sum (Figure \ref{fig6}(d)).

Despite the large degree of cancellation between the fluctuating EMF and the resistive dissipation,
the two act together to reverse the polarity of the poloidal field. In particular, the fluctuating
EMF provides the dominant means of reversing the vector potential at lower latitudes (outside the
tangent cylinder), whereas the resistive dissipation dominates at the higher latitudes (inside the
tangent cylinder). Such an arrangement is largely due to the disparate spatial scales of the
convection present in the magnetic field inside and outside the tangent cylinder, with the more
easily dissipated smaller scales being prevalent at higher latitudes. There is also a temporal lag
with the generation of the turbulent EMF and the later action of the magnetic diffusion.
Nonetheless, at high latitudes and close to the upper boundary, the mean EMF makes a significant
contribution to reversing the magnetic vector potential. This portion of the mean EMF is
predominantly due to the poleward meridional flow in that region, namely $\avg{\vct}\avg{\Bcr}$.
This term must dominate as the longitudinal average of the radial velocity is quite small in this
near-surface high-latitude region due to the cancellation of small-scale convective flows and due to
the impenetrable boundary condition. This influence of the meridional flow suggests that some
aspects of a flux-transport dynamo could be operating in the near-surface region of this
simulation. However, the direction of the transport is reversed relative to a typical flux-transport
dynamo.

\section{Characterizing the Grand Minimum} \label{sec:minimum}

Some 3-D convective dynamo simulations have attained magnetic cycles that also show a longer-term
modulation in the amplitude of their peak magnetic energy
\citep[e.g.,][]{brown11,augustson13,charbonneau13}. The K3S simulation shows similar properties,
though with the additional features of a significant disruption and later recovery of the magnetic
polarity cycles. Indeed, Figures \ref{fig3}, \ref{fig7}, and \ref{fig8} show different
aspects of the dynamo during the 16~year interval in the evolution of case K3S in which the polarity
cycles are substantially disrupted and the magnetic energy is reduced. The volume-averaged magnetic
energy density at lower-latitudes is decreased by a factor of two (Figure \ref{fig7}(a)),
which is reflected in the larger decrease in the magnetic energy in the longitudinal fields relative
to the other energy components. During this ``grand minimum,'' the quite regular and self-similar
cycles seen prior to it are nearly lost. In particular, the mean radial magnetic field in the deep
convection zone does not exhibit a polarity cycle at low-latitudes, whereas there is a semblance of
a polarity cycle at higher latitudes (Figure \ref{fig7}(d)). In the upper convection zone,
the higher-latitudes do not reverse their polarity, though the lower-latitudes do retain something
akin to a polarity cycle (Figure \ref{fig3}(a)).

Despite those disruptions, both the spatial and temporal coherency of the cycles are recovered after
this interval and persist for the last 30~years of the 80~year-long simulation. Due to the decrease
in the volume-averaged magnetic energy density during this interval and because it retains it
magnetic energy cycle of 3.1~years, it is fairly similar to observed characteristics of the Sun
during an average grand minimum.  The largest difference between what is being called a ``grand
minimum'' here and the characteristics of the grand minima seen in cosmogenic isotope data is that
the heliospheric magnetic field appears to have maintained the reversals of its dipolar mode and its
magnetic energy apparently was reduced by roughly a factor of four during an average grand minimum
\citep{mccracken07}.

\subsection{Entering and Exiting the Grand Minimum} \label{sec:spectra}

\begin{figure}[t!]
   \begin{center}
   \includegraphics[width=0.45\textwidth]{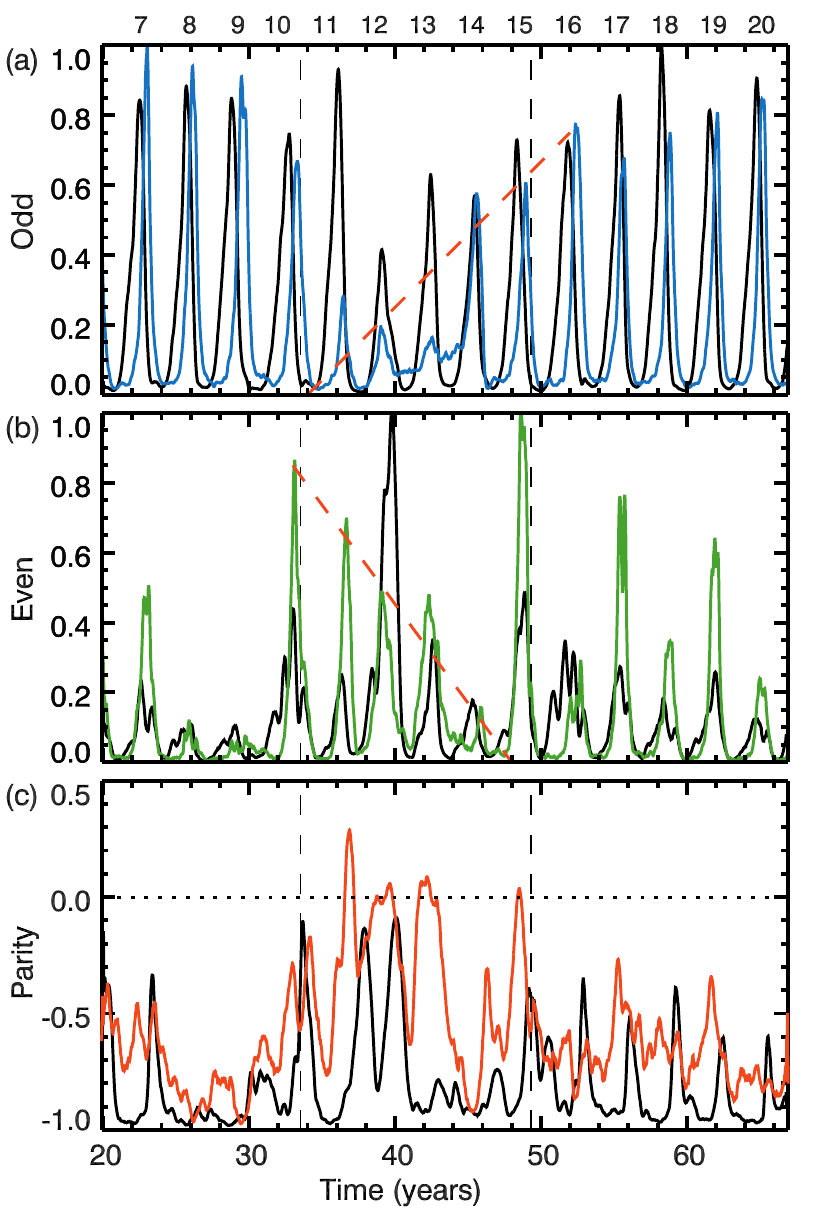} \figcaption{Time variation of the radial
     magnetic energy through the grand minimum at two depths. (a) Power in odd $\ell$, $m=0$ modes
     of $\Bcr^2/8\pi$ at a depth of $\Rsun{0.75}$ (blue) and $\Rsun{0.95}$ (black). (b) Power in
     even $\ell$, $m=0$ modes of $\Bcr^2/8\pi$ at a depth of $\Rsun{0.75}$ (green) and $\Rsun{0.95}$
     (black). The red dashed lines in (a) and (b) indicate the general trends of the magnetic
     energies. (c) Magnetic parity, showing the dominance of odd parity during normal cycles and
     more even parity during the minimum at two depths ($\Rsun{0.75}$ orange, $\Rsun{0.95}$ black),
     indicating that even parity becomes prominent at depth entering and during the grand
     minimum. \label{fig8}}
   \end{center}
\end{figure}

As the interval of reduced magnetic energy and disrupted cycles is entered, there is an anomalous
excitation of low-$m$ modes, where $m$ is the longitudinal wavenumber. This event appears to be
precipitated by an asymmetry of the magnetic field in time and relative to the equator (Figures
\ref{fig7}(b), (c), (d)). The regions of strong magnetic field in the nonaxisymmetric modes
are circled in Figure \ref{fig7}(b), and the direction of the hemispheric temporal lags are
indicated by arrows in Figure \ref{fig7}(c). The atypical excitation of the nonaxisymmetric
modes occurs near the end of cycle 10, during what should normally be a minimum in the magnetic
energy. This likely disrupts the normally clean polarity reversals and may permit the longer-term
excitation of the axisymmetric even-$\ell$ modes. Those even-$\ell$ modes are equatorially symmetric
with $m=0$, where $\ell$ is the spherical harmonic degree. Cycle 10, which begins near year 30, is
an atypical magnetic energy cycle during which there is a strong cross-equatorial filament of radial
magnetic field (Figure \ref{fig3}(a)) and where only the northern hemisphere exhibits a
significant equatorward propagation (Figure \ref{fig3}(b), (d)).  This led to a substantial
temporal lag between the northern and southern hemispheres throughout the magnetic energy
cycle. While the precise physical mechanisms that yield such a state are ambiguous, the symmetric
modes of the radial magnetic field were strongly excited as the grand minimum is entered at year 33
(Figures \ref{fig7}(c) and \ref{fig8}(b)). Furthermore, the subsequent four energy cycles of
the grand minimum do not fully reverse the odd-$\ell$ axisymmetric modes, whereas some of the
even-$\ell$ axisymmetric modes do begin to reverse (particularly at depth).

Given the larger contribution of the even modes during the grand minimum, the $\Omega$-effect (or
$\MPF\cdot\nabla\avg{\Omega}$) is less efficient at building and maintaining strong longitudinal
magnetic fields through latitudinal shear. However, the symmetric modes of the radial magnetic field
can influence the dynamo at low latitudes, where the differential rotation has a strong radial
gradient due to the cylindrical rotation profile. Hence during the grand minimum, the longitudinal
magnetic field is largely generated by radial shear rather than by latitudinal shear. Both the
radial magnetic field and the radial gradient of the angular velocity tend to be weaker than their
latitudinal counterparts, leading to the weaker longitudinal magnetic fields seen during this grand
minimum.  Moreover, as was shown in \citet{strugarek13}, the primary influence on the dipolar mode
is the differential rotation, whereas the quadrupolar mode was fed energy through coupling to
small-scale convection. Such nonlocal couplings may also be at work in K3S as well. Such differences
in the primary inverse energy cascades of the dynamo are indicative of the sensitivity of the dynamo
to the symmetry of the convection as well as the magnetic field.

There is a decay of the even modes and an increase in the energy of the odd modes throughout the
magnetic energy cycles of the grand minimum, as indicated with dashed lines in Figures \ref{fig8}(a)
and (b). This likely permits the exit from the grand minimum near year 49 into another interval of
regular equatorially antisymmetric cycles. These regular magnetic energy cycles involve a prominent
alternation in the peak energy of the even modes between successive magnetic energy cycles. Cycle 15
possesses the first such high peak after the grand minimum. Indeed, it appears that the entrance
into and the exit from the grand minimum are heralded by the excitation of the even modes of the
poloidal magnetic field (Figure \ref{fig8}(b)). Such findings are not without precedent, as similar
issues regarding the relative influence of higher-order multipole modes and their interactions
within magnetic dynamos have been discussed
\citep[e.g.,][]{tobias97,brandenburg08,nishikawa08,gallet09,derosa12,karak13}. What is unique here
are the strongly excited low-$m$ modes and the temporal lag between the hemispheres, during the
cycles that are active as the grand minimum is entered and exited. These atypical events appear to
excite the symmetric dynamo modes and diminish the influence of the antisymmetric modes throughout
the grand minimum.

\subsection{Dynamo Families and Magnetic Field Parity} \label{sec:parity}

The interplay of the antisymmetric (primary) and symmetric (secondary) dynamo families and their
parity has some precedent within the context of the solar dynamo as explored in \citet{derosa12}.
Rather than attempting to assess the behavior of each mode to quantify how these two dynamo families
interact within the K3S simulation, it is useful to construct a more encompassing measure of all the
modes involved in the dynamo.  One such measure is the parity of the radial magnetic field, which
provides a scalar indication of the relative importance of the symmetric and antisymmetric modes
for each longitudinal wavenumber $m$ and radius $r$. It is defined as

\vspace{-0.25truein}
\begin{center}
  \begin{align}
    &\mathcal{P}\left(r,m\right) = \frac{B_{\mathrm{even}}^2\left(r,m\right) - B_{\mathrm{odd}}^2\left(m\right)}{B_{\mathrm{even}}^2\left(r,m\right) + B_{\mathrm{odd}}^2\left(r,m\right)}, \label{eqn:parity} \\
    &B_{\mathrm{even}}^2\left(r,m\right) = \sum_{\underset{\mathrm{even}}{\ell + m}} \left| \Bcr\left(r,\ell, m\right) \right|^2, \;
     B_{\mathrm{odd}}^2\left(r,m\right) = \sum_{\underset{\mathrm{odd}}{\ell + m}} \left| \Bcr\left(r,\ell, m\right) \right|^2, \nonumber
  \end{align}
\end{center}

\begin{figure*}[t!]
   \begin{center}
     \includegraphics[width=\textwidth]{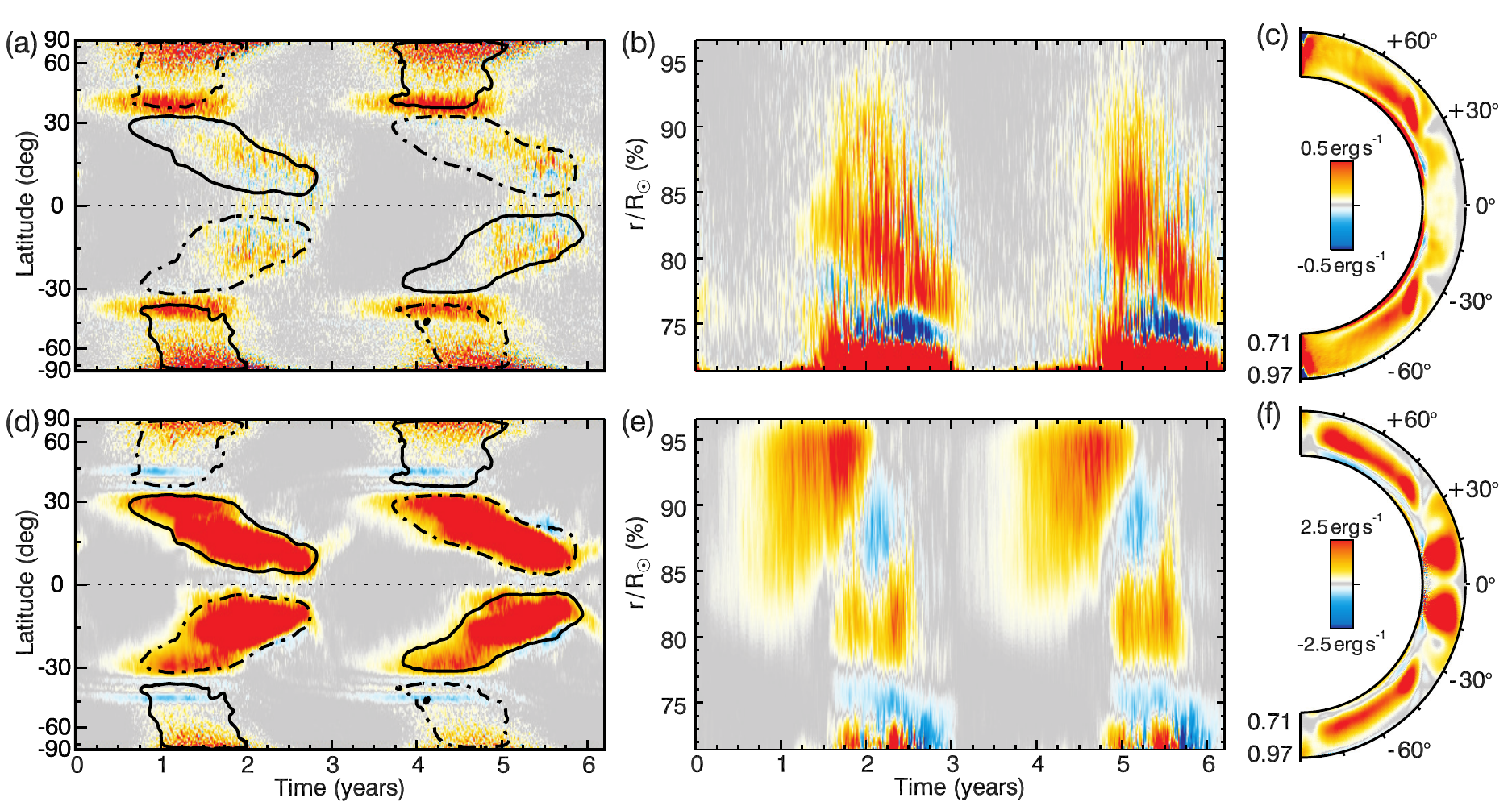} \figcaption{Comparison of the evolution of the
       mean poloidal magnetic energy production through the fluctuating EMF ($P_{FL}$) and mean
       toroidal magnetic energy production by mean shear ($T_{MS}$) over the average polarity
       cycle. An overlay of the contours $\avg{\Bcp}$ at 500~G is also shown, with solid contours
       being of positive polarity and dash-dotted of negative polarity. (a) Mean poloidal magnetic
       energy generation $P_{FL}$ at $\Rsun{0.92}$ shown with latitude and time in cylindrical
       projection, illustrating the strong field generation above about $\pm 30^{\circ}$ and the
       weak generation that accompanies the equatorward propagation of the toroidal wreaths. (b)
       $P_{FL}$ plotted with depth $r/\Rsun{}$ at a latitude of $25^{\circ}$. The tapering with
       depth arises from the strong radial dependence of $\vB_P$, with a downward direction of
       propagation. (c) $P_{FL}$ averaged over the magnetic energy cycle rendered in the meridional
       plane. (d) Mean toroidal magnetic energy generation $T_{MS}$ exhibited over the average cycle
       and with latitude in cylindrical projection at $\Rsun{0.92}$, showing the equatorward
       migration of field generation. (e) $T_{MS}$ in depth and time at a latitude of $25^{\circ}$,
       with two zones of migration evident. (f) $T_{MS}$ averaged over a magnetic energy cycle,
       where the equatorial and polar regions represent spatially separated zones of generation. The
       color bars are shared between (a)-(c) and (d)-(f).
       \label{fig9}}
   \end{center}
\end{figure*}

\noindent with $\ell$ the spherical harmonic degree. When this parity measure is negative, the
radial magnetic field favors an equatorially antisymmetric state, whereas a positive parity
indicates that the radial magnetic field is in a symmetric state. The axisymmetric ($m=0$) parity of
the radial magnetic field in case K3S is usually large and negative as is illustrated by Figure
\ref{fig8}(c), meaning that the system strongly prefers most of the energy being in the
antisymmetric (odd-$\ell$) modes during typical cycles. In the near-surface region this is
especially true, as it possesses an average magnetic parity of $-0.9$. The magnetic parity in the
deeper convection zone is also quite negative being $-0.7$ on average, when the grand minimum is
omitted from that average. In contrast, the nonaxisymmetric modes (those with $m>0$) show no parity
preference, where their near-zero parity is maintained during both typical cycles and during the
grand minimum.

During regular polarity cycles, there is also a distinct phase delay between the peak in magnetic
energy in the odd-ordered modes at depth and those near the surface, as seen in Figure
\ref{fig8}(a).  Furthermore, it shows that the reversal in the odd-ordered modes usually occurs
later at depth than it does near the surface, indicating the top-down nature of the poloidal
magnetic field reversal. In contrast, the even-ordered modes show no such phase discrepancies.  Yet
in Figure \ref{fig8}(b), there is a prominent alternation in the peak energy of the even modes
between successive magnetic energy cycles during the regular cycles after the grand minimum. Such
behavior accounts for the similarity to the Gnevyshev-Ohl rule that those cycles seem to
obey. However, prior to entering the grand minimum this pattern is lost, and it is only regained as
the grand minimum is exited.

The three magnetic energy cycles prior to entering the grand minimum were all atypical.  In
particular, magnetic energy cycles 8 and 9 were more equatorially antisymmetric than a typical
cycle, as indicated by the low magnitude of their even-$\ell$ axisymmetric modes (Figure
\ref{fig8}(b)). The surface layers also reversed earlier than normal relative to when the magnetic
fields in the deep convection zone reversed, being almost 2.5~years earlier instead of the average
of one year (Figure \ref{fig8}(a)). This preceded cycle 10, where the lengthened minimum and
reversal phase discrepancy were especially pronounced in the southern hemisphere. This breaks the
pronounced equatorial antisymmetry of the magnetic field. Such a change in symmetry is largely
maintained throughout the grand minimum, when the parity remains near zero. This indicates the
stronger coupling of the symmetric and antisymmetric dynamo families within that interval. Indeed,
throughout the grand minimum, both the deep convection zone and the upper convection zone have a
significant contribution to the amplitude of the radial magnetic field from both the symmetric
(even-$\ell$) and antisymmetric (odd-$\ell$) modes. In that interval, the ratio of the power in
those modes remains close to unity within the convection zone, which is particularly evident in the
deep convection zone as visible in the magnetic parity shown in \ref{fig8}(c).

The parity variations similar to those seen in Figure \ref{fig8}(c) have also been found and
studied in nonlinear mean-field dynamo models \citep{tobias97,moss00,brooke02}. In those studies,
there is a clear variability of the parity of the dynamo models between the antisymmetric and
symmetric modes, with the dynamo solution being often accompanied by a strong quadrupolar
component. Such variability, which also leads to intervals of deep minima, confirms that the
interplay between the dynamo families is critical to achieve intermittent and aperiodic dynamo
states. These studies have further shown that low magnetic Prandtl numbers favor more irregular and
time-dependent solutions with extended intervals of minimal activity. Similarly the SLD treatment
used for the K3S simulation, by favoring lower magnetic Prandtl number dynamos, is likely at the
origin of the appearance of the deep minimum described above. In the \citet{moss00} and
\citet{brooke02} studies, the symmetric dynamo modes can become completely dominant. In contrast,
the symmetric dynamo modes do not fully become dominant in the K3S simulation. Rather, it is likely
that the interplay of the two dynamo families leads to the grand minimum and the return to the
regular magnetic cycles captured in the low magnetic Prandtl number regime realized here.

\section{Equatorward Propagation of Magnetic Wreaths} \label{sec:propagate}

Having now studied how the regular cycles could be disrupted, it is appropriate to analyze in
further detail how the magnetic energy is generated in both space and time. This will also show how
the equatorward propagation may arise in K3S.

\subsection{Spatio-Temporal Evolution of Dynamo Mechanisms} \label{sec:evodyno}

As will be seen in \S\ref{sec:alpha}, the K3S simulation is most akin to an $\alpha$-$\Omega$
mean-field dynamo, requiring an assessment of the two dominant modes of magnetic energy
generation. These modes are the generation of poloidal magnetic energy through the fluctuating EMF
(the $\alpha$-effect, e.g., \citet{moffatt78}) and the generation of mean toroidal magnetic energy
through the differential rotation ($\Omega$-effect) acting on the mean poloidal magnetic field. The
first mechanism is denoted $\avg{P_{\mathrm{FL}}} =
\avg{\mathbf{B_P}\cdot\nabla\times\mathcal{E}'_{\varphi}\pht}$ and the second $\avg{T_{\mathrm{MS}}}
= \lambda\avg{\Bcp}\MPF\cdot\nabla\Omega$. What is also of note, had the generation of the magnetic
field been instead shown, nearly identical propagating patterns and correlations are seen.

In Figure \ref{fig9}, two primary regions of production of poloidal magnetic field are visible: one
at low latitudes and another at high latitudes. The low-latitude regions of poloidal energy
production are associated with convective cells acting upon the equatorially migrating magnetic
wreaths. While the mechanism is somewhat similar for the high-latitude poloidal field, it primarily
originates in the helical action of convection on the polar caps. The primary source of magnetic
energy for those caps of toroidal magnetic field is the more rapidly rotating poles, which are
established during the magnetic minimum as seen in Figures \ref{fig2} and \ref{fig4}.

The spatial and temporal separation of mean poloidal and toroidal field generation is particularly
evident when comparing Figures \ref{fig9}(a) and \ref{fig9}(d). The greatest generation of poloidal
field is concentrated in the polar regions and near the tangent cylinder, with the low latitudes
playing much less of a role. In contrast, the generation of mean toroidal magnetic field is greatest
at low latitudes throughout the bulk of the cycle. In Figures \ref{fig9}(b) and \ref{fig9}(e), there
is a relatively thin region in radius where the sign of the poloidal magnetic energy generation rate
reverses. This sign reversal is largely due to the change in the sign of the kinetic helicity of the
convection that occurs near the base of the convection zone. Such a kinetic helicity reversal is
expected due to the influence of the lower boundary \citep[e.g.,][]{miesch00}, which has an impact
on the $\alpha$-effect as will be discussed in \S\ref{sec:alpha}.

The beginning of each magnetic energy cycle in Figure \ref{fig9} occurs at years 0.0 and 3.1, when
the magnetic fields are weakest. At those times, the toroidal and poloidal magnetic fields begin to
grow at roughly $\pm\dgr{30}$ (Figures \ref{fig2}(c), \ref{fig3}, \ref{fig9}(a),
(d)). The turbulent action of the convection on this newly generated wreath sustains the poloidal
field generation through $P_{FL}$ on the high-latitude edge of the wreath, which is near the tangent
cylinder. In combination with the polar EMF that emerges from the action of convection on the
longitudinal fields there, this sustains the poloidal field that in turn allows the wreaths to be
maintained through the shearing action of the differential rotation. However, once the polar
differential rotation has been quenched, the toroidal magnetic fields begin to decay. Subsequently,
the poloidal field generation that had been quite prominent at the tangent cylinder vanishes. The
remaining generation of the poloidal field then moves equatorward, advancing with the migration of
the wreaths. Yet that poloidal field generation is still largely on the high-latitude edge of the
low-latitude wreaths. At this point, the strong longitudinal magnetic field at low latitudes has
begun to significantly feedback on the equatorial differential rotation, modifying the structure of
the convection and diminishing the differential rotation. Indeed, the centroid for the greatest
dynamo action propagates equatorward and downward in radius as the magnetic energy cycle progresses,
which is evident in the time-latitude diagram (Figure \ref{fig9}(a), (d)) and which is suggested in
the time-radius diagram (Figure \ref{fig9}(b)). Hence, the equatorial migration begun at the surface
makes its way deeper into the domain as the magnetic energy cycle advances.

The low-latitude wreaths of field eventually lose their coherence and energy through the lack of
sufficient differential rotation to sustain them (e.g., Figure \ref{fig2}(b)), the destructive
influence of the convection (Figure \ref{fig7}), and also due to cross-equatorial flux
cancellation. Once those magnetic field structures have been sufficiently diminished, the diffusion
and convection serve to rapidly redistribute the remaining magnetic flux. This is evident in Figures
\ref{fig3} and \ref{fig4}. As the end of each cycle is approached, the wreaths converge on the
equator and their resulting destruction changes the morphology of the convective cells. The modified
convective patterns better permit the poleward migration and diffusion of the surviving low-latitude
magnetic field polarity. Such actions lead to the topological reconnection of the large-scale
magnetic field. This migrating field is of the opposite sense compared to the previous cycle's polar
cap.  Being of greater amplitude compared to the remaining polar magnetic field, those fields
establish the sense of the subsequent cycle's polar field. Thus the polarity of the subsequent
magnetic field seems to be determined by the EMF generated at the equator, as was also seen in
\citet{augustson13} and \citet{nelson13a}. This source of poloidal magnetic field begins to be
generated once the toroidal magnetic fields are sufficiently close to the equator to enable a strong
cross-equatorial interaction.  It is sustained throughout the rest of the cycle. During this period,
the origin of this poloidal field generation is the action of convection on the low-latitude edge of
the wreaths.

\begin{figure}[t!]
   \begin{center}
   \includegraphics[width=0.485\textwidth]{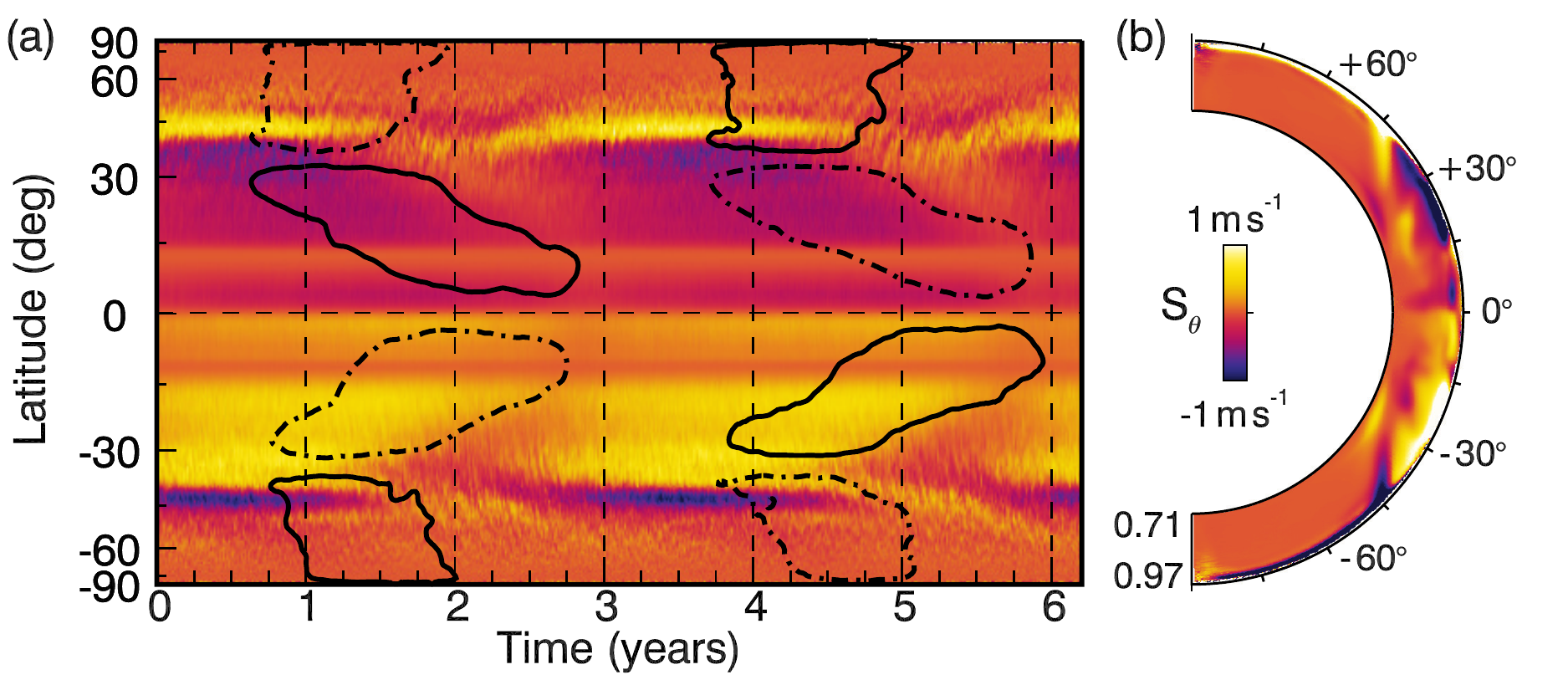} \figcaption{The latitudinal component of
     the propagation direction of a dynamo wave, corresponding to Equation (\ref{eqn:stheta}) with
     $\overline{\alpha}=\alpha_{(\varphi\varphi)}$. (a) Latitudinal propagation velocity $S_\theta$
     shown over an average polarity cycle at $\Rsun{0.92}$ with time and latitude. An overlay of the
     contours $\avg{\Bcp}$ at 500~G is also shown, with solid contours being of positive polarity
     and dash-dotted of negative polarity. (b) The latitudinal propagation $\{S_\theta\}$ averaged
     over the polarity cycle and shown in the meridional plane. Dark tones indicate negative
     latitudinal propagation, light tones positive latitudinal propagation.
     \label{fig10}}
   \end{center}
\end{figure}

\subsection{Kinematic Versus Nonlinear Dynamo Waves} \label{sec:illusions}

The equatorward propagation of magnetic features observed in this simulation, which is visible in
Figure \ref{fig2}(b) and the broad panorama of Figure \ref{fig3}(b), will now be assessed to
disentangle which mechanisms permit such behavior. The equatorward propagation in kinematic
$\alpha$-$\Omega$ dynamo models is traditionally attributed to the propagation of a dynamo wave. In
kinematic theory the propagation direction of such a wave is given by the Parker-Yoshimura rule
\citep[e.g.,][]{parker55,yoshimura75} as

\vspace{-0.25truein}
\begin{center}
  \begin{equation}
    \mathbf{S} = -\lambda\overline{\alpha}\pht\cross\grad\frac{\Omega}{\Omega_0}, \label{eqn:stheta}
  \end{equation}
\end{center}

\noindent where $\lambda = r\sint$ and
$\overline{\alpha}=-\tau_o\avg{\vv'\boldsymbol{\cdot\omega'}}/3$, with
$\boldsymbol{\omega'}=\curl{\vv'}$. Thus $\overline{\alpha}$ depends on the convective overturning
time $\tau_o$ and the kinetic helicity. When Lorentz-force back-reactions are taken into account,
there is also a current helicity contribution to $\overline{\alpha}$
\citep[e.g.,][]{pouquet76,gruzinov94}. \citet{augustson14} demonstrated that the kinematic
expression in Equation \ref{eqn:stheta} with $\overline{\alpha}$ defined using only the kinetic
helicity fails to explain the equatorward propagation seen in this dynamo simulation. The current
helicity can in principle reverse the sign of the scalar $\alpha$-effect \citep{warnecke14}.  For
instance, it has recently been shown that the direction of the propagation of the magnetic field
through a cycle can depend upon the Parker-Yoshimura mechanism
\citep[e.g.,][]{racine11,kapyla13,warnecke14}.

\begin{figure}[t!]
   \begin{center}
   \includegraphics[width=0.485\textwidth]{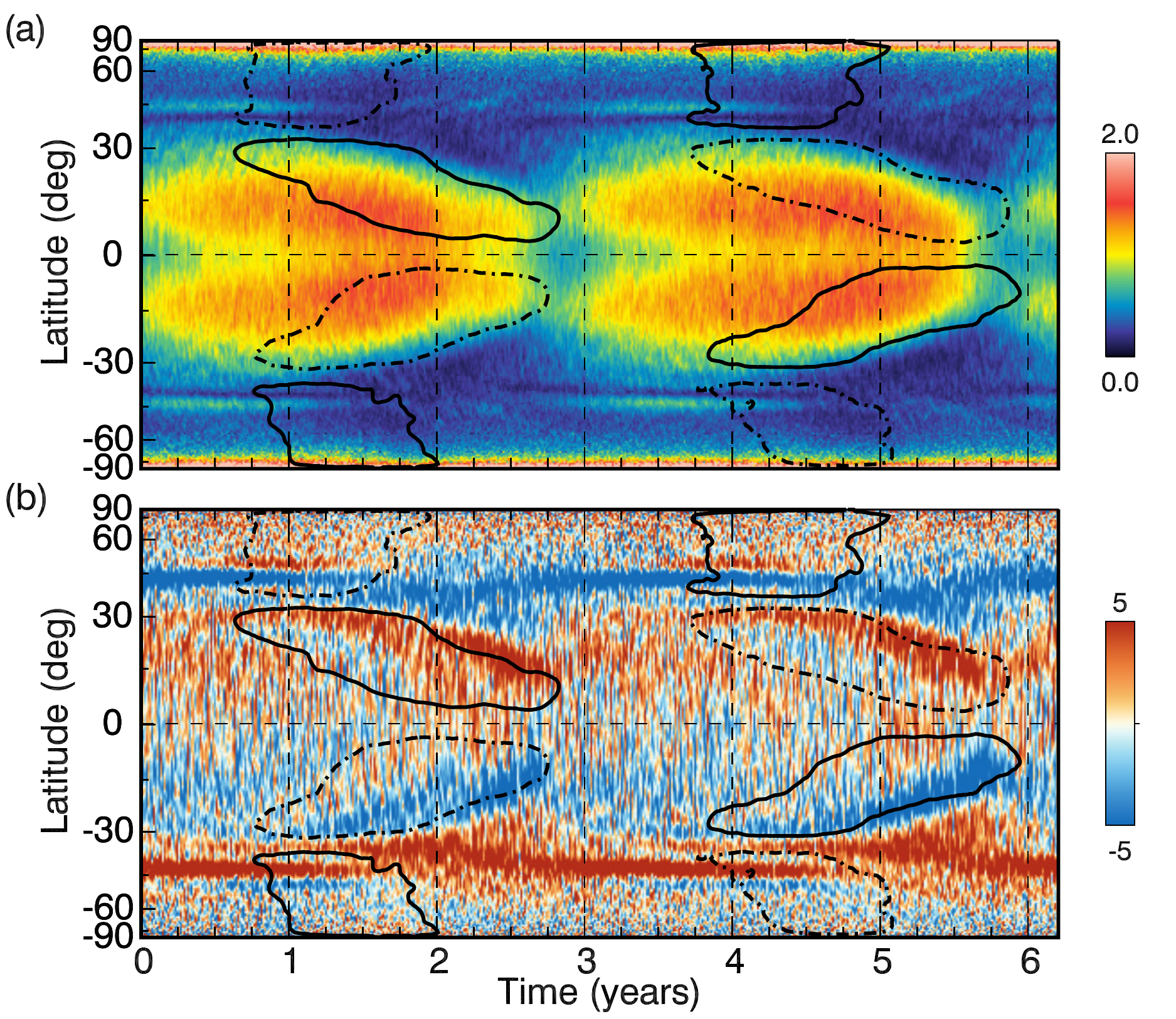} \figcaption{Coevolution of the mean
     toroidal magnetic field $\avg{\Bcp}$ at $\Rsun{0.92}$ over the average magnetic polarity cycle
     with (a) the magnitude of the mean angular velocity gradient
     $\Rsun{}$~$|\nabla\Omega|/\Omega_0$ and (b) latitudinal velocity $\avg{\vct}$ of the evolving
     meridional circulation in units of $\mathrm{m\, s^{-1}}$. Here $\avgBp$ is overlain with
     positive magnetic field as solid lines and negative field as dashed lines, with the contours
     corresponding to a 500~G strength field. \label{fig11}}
   \end{center}
\end{figure}

Rather than explicitly including the current helicity here, $\overline{\alpha}$ can be procured from
the simulation itself. As described in \S\ref{sec:alpha}, the full $\alpha$ tensor is obtained from
an SVD decomposition. When the scalar $\alpha$-effect utilizes information from that derived
$\alpha$ tensor, it will automatically include information about both the current and kinetic
helicities. Hence, the $\alpha_{(\varphi\varphi)}$ component of the $\alpha$ tensor is used in place
of $\overline{\alpha}$ in Equation \ref{eqn:stheta} to determine the propagation direction of the
dynamo wave predicted by mean-field theory. This is shown in Figure \ref{fig10}. With such a
definition of $\overline{\alpha}$, the Parker-Yoshimura sign rule still does not hold for this
simulation. More precisely, the sign of the latitudinal propagation of a dynamo wave is of the
opposite sign required for both the equatorward migration of low-latitude magnetic structures and
the poleward propagation of magnetic field as seen in Figure \ref{fig10}. So, although the
Parker-Yoshimura mechanism seems to be at work in other convective dynamo simulations
\citep{racine11,warnecke14}, it does not appear to be operating here.

Thus, another mechanism must be sought to explain the equatorward propagation in the K3S simulation.
In contrast to the intuition derived from mean-field theory, the dominant mechanism appears to be
the nonlinear feedback of the magnetic fields upon the differential rotation. The tight correlation
between the presence of $\Bcp$ and the angular velocity gradient $|\nabla\Omega|$ is demonstrated in
Figure \ref{fig11}(a). Since the latitudinal shear serves to build and maintain the magnetic wreaths
(as in \S\ref{sec:tme}), the latitude of peak magnetic energy corresponds to that of the greatest
shear. So the region with available shear moves progressively closer to the equator as the Lorentz
forces of the wreaths locally weaken the shear (Figure \ref{fig11}(a)). Hence the appearance of
equatorward motion in K3S is attributed to a nonlinear dynamo wave. This interpretation is
consistent with the substantial modulations of the differential rotation seen in Figure \ref{fig2}
and the equatorward migration of the toroidal source term shown in Figure \ref{fig9}. In a
mean-field dynamo, the equatorward propagation of the toroidal source term arises from the
equatorward propagation of the poloidal magnetic field. Here, the equatorward propagation of
$\nabla\Omega$ also contributes.

Accompanying the local weakening of the gradient of the differential rotation is a meridional flow
that is gyroscopically induced at the poleward edge of the low-latitude magnetic wreaths as seen in
Figure \ref{fig11}(b), with gyroscopic pumping defined as in \citet{mcintyre98} and
\citet{miesch11}. In particular, the torque provided by the divergence of the Lorentz force and the
Maxwell stresses produce a change in the local shear which in turn induces a change in the
meridional flow. Indeed the spatio-temporal correlation between the changing differential rotation,
the mean toroidal magnetic field, and the meridional flow seen in Figure \ref{fig11} appears
to support a nonlinear dynamo wave. This mechanism for producing an equatorward migration of
magnetic field relies upon the complex dynamical coupling of the differential rotation, meridional
flows, and the magnetic field.

\section{Assessing the Cycle Periods} \label{sec:periods}

A correlation analysis of the dominant processes in this simulation is useful to quantitatively
ascertain the magnetic and polarity cycle periods and the correlations. Such analysis also permits
the estimation of the variance of those cycles. First consider the dynamical coupling of the mean
magnetic fields $\avg{\vB}$ and the mean angular velocity $\avgO$, which plays a crucial role in
regulating the magnetic energy cycle. The significant spatial and temporal correlation between
$\avgBp$ and angular velocity variations $\avg{\Delta\Omega}$ during reversals is apparent when
comparing Figures \ref{fig2}(a) and \ref{fig2}(b), and it is readily seen in Figure \ref{fig11}(a),
revealing the strong nonlinear coupling of the magnetic field and the differential rotation.

\subsection{Correlation Analysis} \label{sec:corr}

\begin{figure}[t!]
   \begin{center}
     \includegraphics[width=0.45\textwidth]{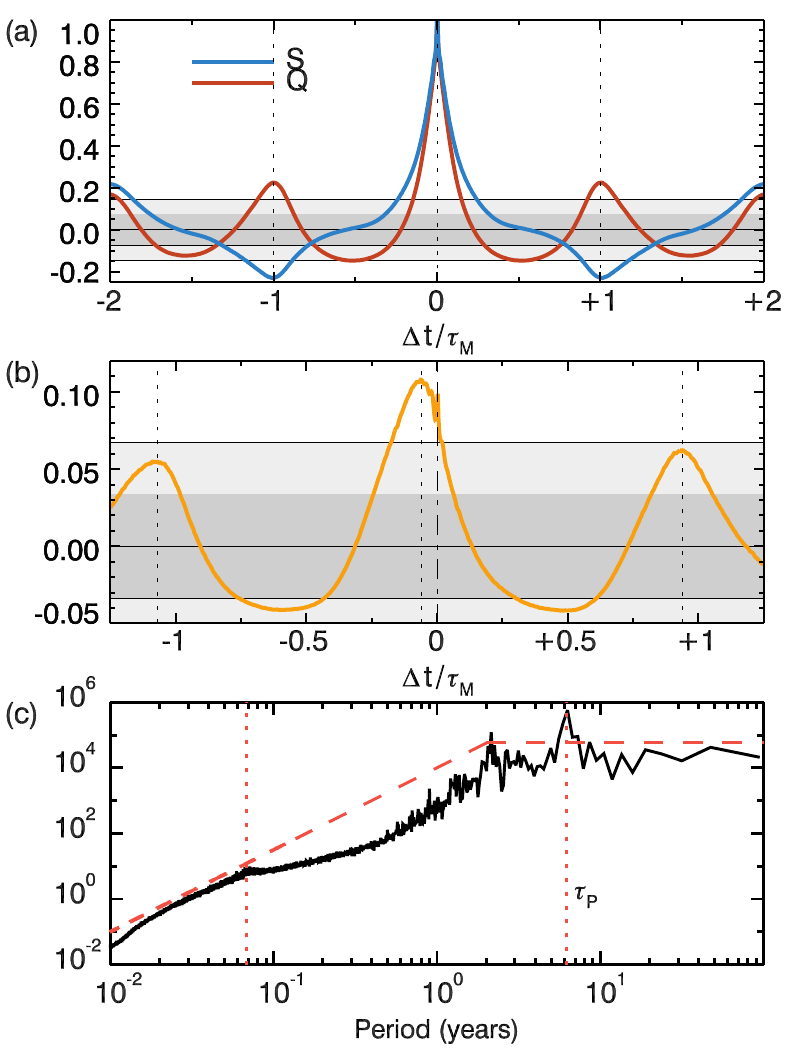} \figcaption{Auto and cross-correlation
       of Lorentz-force and mean toroidal magnetic field production by mean-shear. (a)
       Volume-averaged temporal auto-correlation of mean toroidal magnetic field generation by mean
       shear ($S$, blue curve) and the same for the mean Lorentz force impacting the mean angular
       velocity ($Q$, red curve) plotted against temporal lags $\Delta \mathrm{t}$ normalized by the
       magnetic energy cycle period $\tau_\mathrm{M}$. Confidence intervals are shown as shaded gray
       regions, with the 67\% interval in darker gray and 95\% in lighter gray. (b)
       Cross-correlation of the mean poloidal energy production through the fluctuating EMF
       ($P_{\mathrm{FL}}$) and the mean toroidal magnetic energy production due to the mean shear
       ($T_{\mathrm{MS}}$). (c) The volume-averaged temporal power spectrum for $\avg{\Bcp}$. The
       red vertical dotted lines indicate the convective overturning time of 0.07~years and the
       polarity reversal time scale of 6.2~years and the red dashed line indicates a $1/f^2$ pink
       noise spectrum. \label{fig12}}
   \end{center}
\end{figure}

The dynamics that couples the differential rotation and the mean toroidal magnetic field is captured
in two terms: the mean toroidal magnetic field generation due to mean shear ($S = \lambda
\avg{\vB_{P}} \cnabla \avgO$, with $\avg{\vB_{P}}$ the mean poloidal field) and the azimuthal
component of the mean Lorentz-force $Q = \boldsymbol{\hat{\varphi}\cdot} \avg{\curl{\avg{\vB}}}
\cross \avg{\vB}$, which acts on the longitudinal component of the momentum equation (Equation
(\ref{eqn:ashmom})). As shown in Figure \ref{fig12}(a), the auto-correlation of each of these
components of the MHD system reveals that $Q$ varies with a period corresponding to the magnetic
energy cycle, whereas $S$ is self-correlated over the polarity cycle period. Similarly, the
cross-correlation between $P_{\mathrm{FL}}$ and $T_{\mathrm{MS}}$ indicates their close temporal
correlation.  There is, however, a phase lag of about $0.05\tau_{\mathrm{M}}$ or 2~months between
the two primary energy generation mechanisms for the mean toroidal and poloidal magnetic fields
(Figure \ref{fig12}(b)). Yet there is also a high degree of temporal regularity between cycles, with
the auto-correlation of both quantities remaining significant with 95\% confidence for a single
polarity cycle and with 67\% confidence for three such cycles as indicated by the shaded areas of
Figures \ref{fig12}(a), (b). To further quantify the regularity of the cycles one could consider the
full-width half-maximum of the auto-correlation of $S$ and $Q$ as well as the cross-correlation of
$P_{\mathrm{FL}}$, and $T_{\mathrm{MS}}$ (Figure \ref{fig12}(b)). Indeed, an average of the FWHM of
those correlated quantities yield a standard variance of the cycle period of about 4~months or about
10\% of the magnetic energy cycle period of 3.1~years.

This analysis has been undertaken primarily to emphasize two basic features of the K3S dynamo.
First, the time scale for the Lorentz-force feedback on the differential rotation, and hence the
energy cycle time scale, is $\tau_M$. This is the basic mechanism that sets the clock for the
magnetic energy cycles. Second, the polarity cycle time $\tau_P$ is then twice the energy cycle time
scale, which simply follows from the mathematical behavior of the square of a sinusoid being
directly related to a sinusoid of twice of its period. To further illustrate the time scales that
arise in this simulation, the volume-averaged frequency power spectrum of $\avg{\Bcp}$ is shown in
Figure \ref{fig12}(c). There is clear peak in the power spectrum around the average polarity cycle
time $\tau_P$ of 6.2~years. The width of this peak gives a hint as to the variance of the magnetic
polarity cycle period, which appears to be about one year. If the peak in power at $\tau_P$ is
subtracted, then there is a nearly uniform distribution of time scales longer than about 2~years,
potentially reflecting the aperiodic character of those time scales. The power at shorter time
scales decreases roughly as the inverse square of the frequency, which is common in the pink noise
of highly complex and nonlinear systems. There is also a broad peak at time scales corresponding to
the convective overturning time of about 25~days, or 0.07~years.

\subsection{Potential Time Scales}

In addition to the primary identification of these time scales, the processes assessed in previous
sections suggest that there at least two processes that lead to those time scales. One is the
effective rate of conversion of differential rotation kinetic energy into mean toroidal magnetic
energy ($\tau_\Omega$). The second is the time required to diffuse magnetic field from the equator
to the pole ($\tau_\eta$). A third potentially relevant time scale is related to the rate at which
the magnetic field converges toward the equator may also be important. This in turn can be related
to the gyroscopically-pumped meridional circulation induced by the changes in the magnetic fields as
they approach the equator. The first time scale $\tau_\Omega$ can be measured as

\vspace{-0.25truein}
\begin{center}
  \begin{align}
    \tau_\Omega = \frac{2\pi}{\Delta\Omega}\frac{\left|\avg{\Bcp}\right|}{\left|\MPF\right|} \approx 3.6\, \mathrm{years}.
  \end{align}
\end{center}

\noindent The diffusion time scale from the equator to the pole is

\vspace{-0.25truein}
\begin{center}
  \begin{align}
    \tau_\eta = \frac{\pi}{2\left(r_2-r_1\right)}\int_{r_1}^{r_2} \frac{dr r^2}{\eta} \approx 6.7\, \mathrm{years}.
  \end{align}
\end{center}

\noindent Finally, the meridional circulation time scale follows from the average time it takes for
an ensemble of particles to traverse a path of the meridional circulation in K3S or

\vspace{-0.25truein}
\begin{center}
  \begin{align}
    \tau_{\mathrm{MC}} = \frac{1}{N_p}\sum_{\mathbf{p}_i}\oint_{\mathbf{p}_i} \frac{d\boldsymbol{\ell}\cdot\mathbf{v}_{\mathrm{MC}}}{\mathrm{v}_{\mathrm{MC}}^2} \approx 1.2\, \mathrm{years},
  \end{align}
\end{center}

\noindent where each particle path is denoted as $\mathbf{p}_i$. With these simple estimates of
these relevant time scales, it is seen that the shear time scale and the diffusion time scale are
close to the magnetic energy cycle period and to the magnetic polarity cycle respectively. However,
since all three of these mechanisms are operating concurrently, a single time scale might emerge
from their combined influence. One such way to accomplish this is to take the geometric mean of
these time scales, which yields 3.1~years, which reflects the observed magnetic energy cycle period
$\tau_M$. This, however, only loosely suggests that these three processes may share in setting the
magnetic energy cycle period as well as the polarity cycle period. Indeed, the shearing and the
diffusion time scales alone are nearly adequate to explain the magnetic and polarity cycle time
scales.

\section{Mean-Field Analysis} \label{sec:alpha}

The complex set of processes at work in this dynamo solution, as just assessed in
\S\ref{sec:reversal} and \ref{sec:evodyno}, yield a dynamo that falls outside of the broad classes
of $\alpha$-$\Omega$ dynamos that underpin much of the mean-field theory (MFT) of MHD
\citep[e.g.,][]{steenbeck66,moffatt78,krause80,brandenburg05}, which typically assume a temporally
constant differential rotation and fluctuating EMF. Rather, the feedback of the magnetic field on
both the nonaxisymmetric and axisymmetric components of the convective flows is critical to the
operation of the dynamo running in this simulation. This suggests that, if one were to attempt to
fully model these dynamics in the context of MFT, one would need to include an $\alpha$-quenching
mechanism and the Malkus-Proctor effect. Nevertheless, helical turbulent convection is largely
responsible for the generation of poloidal field. As such, MFT provides a route to assess and
quantify the various zeroth-order influences of the turbulent velocity field upon the generation of
the turbulent electromotive force (EMF, $\boldsymbol{\mathcal{E}}'$). Thus a spatially varying, but
temporally constant, and $\delta$-correlated $\alpha$-effect is now examined.

\subsection{Examining the Turbulent Electromotive Force} \label{sec:mfemf}

As seen in \S\ref{sec:genpol} and \S\ref{sec:evodyno}, $\boldsymbol{\mathcal{E}}'$ is largely
responsible for the generation of poloidal magnetic field in this simulation.  Therefore, the
generation of poloidal field will be characterized through the mean-field evolution of the mean
toroidal vector potential $\avg{A_{\varphi}}$ as in Equation (\ref{eqn:daphi}), which is gauge
independent since it only considers longitudinally-averaged quantities.  The connection between MFT
and the EMF achieved in this simulation will be examined by noting that the first-order expansion of
$\boldsymbol{\mathcal{E}}'$ around the mean magnetic field and its gradient is

\vspace{-0.25truein}
\begin{center}
  \begin{equation}
    \avg{\boldsymbol{\mathcal{E}}'} = \alpha \avg{\vB} + \beta \nabla \avg{\vB} + \mathcal{O}\left(\partial\avg{\vB}/\partial t,\nabla^2\avg{\vB}\right),
  \end{equation}
\end{center}

\noindent where $\alpha$ is a rank two pseudo-tensor and $\beta$ is a rank three tensor. In the
following the $\beta$ term will be neglected for simplicity. However, this does increase the
systematic error in estimating the $\alpha$ tensor. A SVD decomposition that includes the
$\beta$-effect has been undertaken in order to provide a lower bound on this systematic error. It is
21\% when averaged over all components of the $\alpha$ tensor. In this analysis $\alpha$ has been
expanded as $\alpha \langle\vB\rangle = \alpha_S\avg{\vB} + \boldsymbol{\gamma}\times\avg{\vB}$,
with $\alpha_S$ being the symmetric portion of $\alpha$ and $\boldsymbol{\gamma}$ the antisymmetric
portion. The latter is also known as the turbulent pumping velocity
\citep[e.g.,][]{krause80,kapyla06}. The diagonal components of $\alpha$ are automatically
symmetric. Yet the symmetrized elements of $\alpha$ are all denoted as $(ij)$.

To reconstruct the $\alpha$ tensor from the simulation data, its individual components are
determined from a temporal sequence of data at each radial and latitudinal grid point using a method
similar to the least-squares singular value decomposition (SVD) methodology described in
\citet{racine11}. Such a local fitting technique assumes that each point may be treated
independently, which precludes the capture of temporal and spatial correlations that can influence
the dynamo action \citep{brown11,nelson13a,augustson13}. Yet it has the advantage that the magnetic
fields and electric currents from the simulation constrain the components of $\alpha$. The
reconstruction of $\alpha$ is carried out using the data from the extended interval of 80~years, as
shown in Figure \ref{fig3}.

\begin{figure}[t!]
   \begin{center}
   \includegraphics[width=0.44\textwidth]{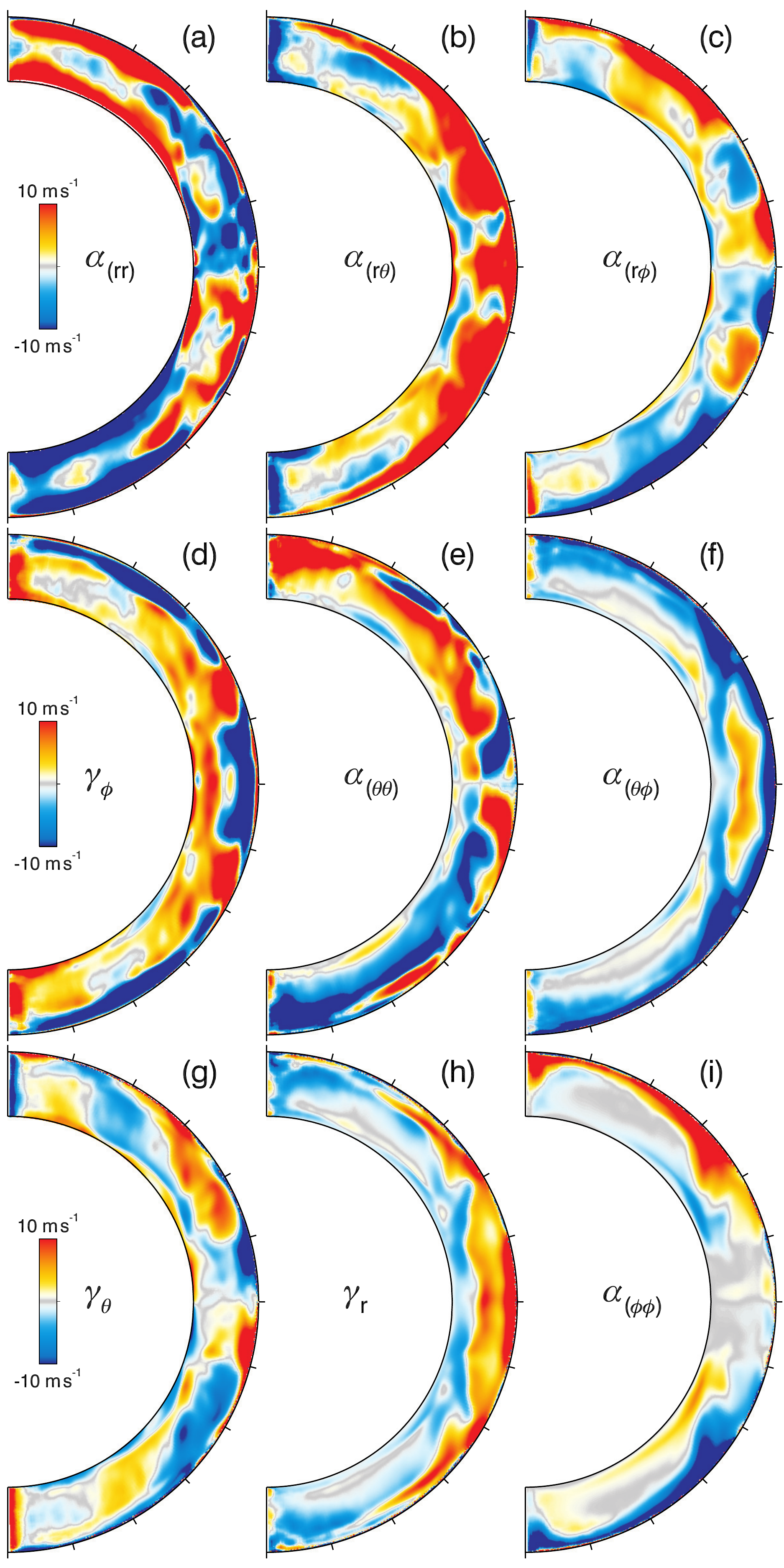} \figcaption{A mean-field theoretic
     interpretation of the dynamics. The nine components of the $\alpha$ tensor are shown, as
     computed using a SVD technique. The diagonal components of $\alpha$ are shown in (a), (e), and
     (i). The symmetrized components of the tensor are shown in (b), (c), and (f).  The
     antisymmetric turbulent pumping components are shown in (d), (g), and (h). The colorbar and
     scaling are uniform across all panels. \label{fig13}}
   \end{center}
\end{figure}

Figure \ref{fig13} shows the nine components of $\alpha$ that contribute to
$\avg{\boldsymbol{\mathcal{E}}'}$. The magnitude of the components of the $\alpha$ tensor are on par
with other solar dynamo simulations \citep{racine11, nelson13a, simard13}.  All of the $\alpha$
tensor components contribute to the generation of the magnetic vector potential through the
turbulent EMF and thus to the poloidal and toroidal magnetic fields. The spatial structures of
$\alpha_{(r\varphi)}$, $\alpha_{(\varphi\varphi)}$, and $\gamma_{\theta}$ visible in Figures
\ref{fig13}(c), (g), and (i) are roughly antisymmetric between the two hemispheres, with the other
two components $\alpha_{(\theta\varphi)}$ and $\gamma_r$ being nearly north-south symmetric.  The
antisymmetry arises from the influence of Coriolis forces on the flows, with the cyclonic turbulence
present at higher latitudes being an important contributor to the sign and magnitude of these
fields. Such an arrangement is also indicative of the dipolar nature of the poloidal magnetic field,
which has a symmetric latitudinal magnetic field and an antisymmetric radial magnetic field. Thus,
it is not a surprise that the equatorially symmetric components of $\alpha$ are those that act on
the latitudinal magnetic field.  All of the components shown also have structures that reflect the
global-scale properties of the convection zone (CZ), where there is a radial symmetry or
antisymmetry about the middle of the CZ. This is particularly evident in $\alpha_{(\theta\varphi)}$
and $\alpha_{(\varphi\varphi)}$, where the divergence of vortical downflows and the convergence of
swirling upflows gives rise to this feature. As might be expected, there is a latitudinal dependence
to these structures, where a distinct transition in flow morphology occurs across the tangent
cylinder.

Most of the features seen here are also present in the $\alpha$ tensor yielded in the analysis of an
EULAG-MHD simulation shown in \citet{racine11}. In particular, the components of the tensor possess
a radial antisymmetry with respect to a radius related to the point at which the vorticity of the
bulk of the flows reverses its sign, which is above the base of the convection zone. Moreover, the
radial and longitudinal elements of $\alpha$ are latitudinally antisymmetric, much like the
simulation here. The latitudinal components are symmetric reflecting the importance of the dipolar
component of the magnetic field. Some of the tensor elements of the EULAG-MHD simulation may appear
to differ by a sign relative to the $\alpha$ tensor seen in Figure \ref{fig13}. However, this is
due to the use of latitude rather than co-latitude in displaying the results of \citet{racine11}.

\subsection{The Efficiency of the $\alpha$ Effect} \label{sec:aeff}

An interesting measure of the dynamo is how efficiently the convective flows can regenerate existing
mean magnetic fields. The dynamo efficiency of these flows can be surmised by finding the average
magnitude of an estimated $\alpha$-effect relative to the rms value of the nonaxisymmetric velocity
field. One such measure of the dynamo efficiency $E$ is

\vspace{-0.25truein}
\begin{center}
  \begin{align}
    \langle\frac{\alpha}{v_{\mathrm{rms}}}\rangle &\sim E = \frac{3}{2\left(r_2^3-r_1^3\right)}\!\!\sum_{a,b}\iint{\!\!\!\mathrm{d}r \mathrm{d}\theta r^2 \sin{\theta} \sqrt{\frac{\alpha_{ab}\alpha^{ab}}{\{\vv'\cdot\vv'\}}}}, \label{eqn:dyneff}
  \end{align}
\end{center}

\noindent where $\{\vv'\cdot\vv'\}$ is the sum of the diagonal elements of the Reynolds stress
tensor averaged over the duration of the simulation and over all longitudes. For the K3S simulation,
this measure yields a dynamo efficiency of 70\%. If one considers only the generation of
$\mathcal{E}'_\varphi$ by summing over only the components of $\alpha_{i\varphi}$, this efficiency
measure yields 15\%, which is half of the 30\% value found for simulations of F-type stars
\citep{augustson13}. Such a level of efficiency is what one might expect given the factor of five
between the rate of mean toroidal magnetic energy generation and poloidal energy generation seen in
Figures \ref{fig9}(c) and (f). One can utilize Equation \ref{eqn:dyneff} to provide a measure of the
importance of each of the relative components of $\alpha$ as

\vspace{-0.25truein}
\begin{center}
  \begin{align}
    &\langle\frac{\alpha_{ij}}{v_{\mathrm{rms}}}\rangle \sim \epsilon_{ij} = \frac{3}{2 E\left(r_2^3-r_1^3\right)}\!\!\iint{\!\!\!\mathrm{d}r \mathrm{d}\theta r^2 \sin{\theta} \sqrt{\frac{\alpha_{ij}\alpha^{ij}}{\{\vv'\cdot\vv'\}}}} \nonumber \\
    &= \begin{bmatrix} \epsilon_{(rr)} & \epsilon_{(r\theta)} & \epsilon_{(r\varphi)} \\ 
                              \epsilon_{\gamma_\varphi} & \epsilon_{(\theta\theta)} & \epsilon_{(\theta\varphi)} \\ 
                              \epsilon_{\gamma_\theta} & \epsilon_{\gamma_r} & \epsilon_{(\varphi\varphi)} 
       \end{bmatrix} 
     = \begin{bmatrix} 0.355 & 0.124 & 0.073 \\
                       0.110 & 0.103 & 0.066 \\
                       0.062 & 0.054 & 0.053
     \end{bmatrix}.\label{eqn:dynnorm}
  \end{align}
\end{center}

Equation \ref{eqn:dynnorm} clearly indicates that the $\alpha_{(rr)}$ component is dominant, with
those processes contributing to $\alpha_{(rr)}$ being about 2.9 times more efficient than the next
largest component $\alpha_{(r\theta)}$ and 6.7 times more efficient than the smallest component
$\alpha_{(\varphi\varphi)}$. Indeed, the upper two by two matrix formed by $\alpha_{(rr)}$,
$\alpha_{(r\theta)}$, $\alpha_{(\theta\theta)}$ and $\gamma_\varphi$ possesses the terms that make
the largest contribution to the $\alpha$-effect. Specifically, these terms encapsulate those
turbulent processes that are the most efficient at converting mean poloidal magnetic field into
toroidal magnetic field. However, an assessment of the magnetic energy generating capacity of the
$\alpha$ and $\Omega$-effects requires a proper tensor norm of the relevant energy generation
terms. Such an assessment also permits the characterization of the dynamo in the context of
mean-field dynamo theory.

\subsection{Mean-Field Characterization of the Dynamo} \label{sec:chardyn}

To assess which category of mean-field dynamos that the K3S dynamo falls into, one can measure the
relative influence of the $\alpha$ effect to that of the $\Omega$ effect. In particular, the
following ratio quantifies this

\vspace{-0.25truein}
\begin{center}
  \begin{align}
    \frac{\alpha_\phi}{\Omega} = \frac{3}{2\left(r_2^3-r_1^3\right)}\!\!\iint\!\!\! dr d\theta r^2 \sint \left|\frac{\avg{\Bcp}\pht\boldsymbol{\cdot\nabla\times}\avg{\boldsymbol{\mathcal{E}'}}}{\lambda\avg{\Bcp}\MPF\boldsymbol{\cdot\nabla}\avg{\Omega}}\right|. \label{eqn:alpvomg}
  \end{align}
\end{center}

\noindent With this measure one can also quantify the impact of individual components of the
$\alpha$-effect that generates mean toroidal magnetic field. Overall, it is found that the $\alpha$
effect is 11.9 times smaller than the $\Omega$ effect. If each of the six terms in the numerator of
Equation \ref{eqn:alpvomg} are considered individually, it is seen that the term
$\alpha_{(\theta\varphi)}$ is about 50\% larger than the average of the other terms. Moreover, the
terms in the longitudinal component of the $\alpha$ effect arising from $\mathcal{E}'_\theta$ are
43\% larger than those from $\mathcal{E}'_r$. This implies first that the generation of toroidal
magnetic field through the $\alpha$-effect is weak and that those terms that contribute to that
effect are dominated by the horizontal components of $\alpha$.

The component $\alpha_{(\varphi\varphi)}$ effectively represents the action of helical convection
on the mean toroidal magnetic field. In the context of the classical $\alpha$-$\Omega$ dynamo, that
term is usually taken to be the largest if not the only term in the $\alpha$ tensor. In the K3S
dynamo this term has the smallest contribution, being about three times smaller than the components
with the largest efficiency norm. However, it is also acting on the mean toroidal field, which is
the largest component of the magnetic field. To quantify the efficacy of the poloidal field
generation by the turbulent EMF relative to that of the toroidal magnetic field, consider the
following measure

\vspace{-0.25truein}
\begin{center}
  \begin{align}
    \frac{\alpha_P}{\alpha_\varphi} = \frac{3}{2\left(r_2^3-r_1^3\right)}\iint dr d\theta r^2 \sint \left|\frac{\MPF\boldsymbol{\cdot\nabla\times}\avg{\boldsymbol{\mathcal{E}'}}}{\avg{\Bcp}\pht\boldsymbol{\cdot\nabla\times}\avg{\boldsymbol{\mathcal{E}'}}}\right|. \label{eqn:alpvalp}
  \end{align}
\end{center}

\noindent This average yields a ratio of 4.4 for the relative generation of poloidal and toroidal
magnetic field through the $\alpha$ effect. This is consistent with the above argument that the
$\alpha$ effect is relatively unimportant for the generation of the toroidal magnetic field.

It is evident that the impact of the convection upon the mean magnetic fields provides the primary
regenerative mechanism for the mean poloidal field at low latitudes, while the diffusion of poloidal
field is the primary mechanism operating at higher latitudes (see \S\ref{sec:genpol}). In contrast,
the mean toroidal magnetic fields are primarily built through the interaction of the mean poloidal
magnetic field and the rotational shear. Therefore, given the presence of the differential rotation
to sustain the mean toroidal magnetic field and the convective regeneration and diffusion of the
poloidal magnetic field, this simulation could be characterized as an $\alpha$-$\Omega$
dynamo. However, the nonlocal spatio-temporal correlations contained in the generation terms of the
magnetic field in this simulation makes an exact characterization of the dynamo within the context
of mean-field theory difficult. For instance, if this dynamo were simply a kinematic
$\alpha$-$\Omega$ dynamo, the Parker-Yoshimura mechanism should be sufficient to explain the
equatorward propagation of magnetic field structures.

\section{Conclusions and Discussion} \label{sec:conclude}

The 3-D simulation K3S self-consistently exhibits five prominent features: (i) regular magnetic
energy cycles during which the magnetic polarity reverses near the magnetic energy minimum; (ii)
magnetic polarity cycles with a period of $\tau_P = 6.2$~years, where the equatorially antisymmetric
modes of the poloidal magnetic field returns to the polarity of the initial condition; (iii) the
equatorward migration of longitudinal field structures during these cycles; (iv) the poleward
migration of oppositely-signed flux; and (v) a ``grand minimum,'' where there is a period of
relative magnetic quiescence at low-latitudes and disrupted polarity cycles after which the previous
polarity cycle is recovered. These aspects bear resemblance to some of the behavior of solar
magnetism. It, however, does possess different time scales. Further, it does not have explicit
surface flux emergence, and there is a significant modulation of the differential rotation.

The most prevalent properties of K3S involve both a prominent solar-like differential rotation and
distinctive wreaths of magnetism. Similar properties have been realized in a broad range of
simulations carried out previously with ASH \citep[e.g.,][]{brown10,brown11,augustson13,nelson13a}.
The primary characteristic shared among these simulations is that they typically have a low Rossby
number, leading to the formation of large-scale toroidal magnetic wreaths. Another common feature in
global-scale convective dynamo simulations is that the interplay between the mean toroidal magnetic
field and the angular velocity leads to a significant modulation of the differential rotation. In
K3S this contributes to the waxing and waning of the magnetic energy since the production of mean
toroidal magnetic energy relies upon the shear of the differential rotation though the $\Omega$
effect. In particular, as the shear reaches a minimum due to the Lorentz forces, the magnetic fields
subsequently weaken. Yet once the magnetic fields are sufficiently diminished, the convective
patterns at low latitudes regain structures that can more efficiently regenerate the differential
rotation. Thus the magnetic field generation starts anew. The strong feedback of the Lorentz forces
on the differential rotation leads to the equatorward propagation of the magnetic fields. As the
wreaths approach the equator, there is an increase in the cross-equatorial magnetic flux that
permits the low latitude convection to generate poloidal magnetic fields with the opposite polarity
of the dominant magnetic wreaths. This oppositely-signed magnetic flux is then advected and
diffused, eventually reaching the poles and completing a magnetic polarity reversal. This dynamo
regime is distinct from mean-field models of cyclic dynamos that do not involve Lorentz-force
feedbacks, and in some ways it operates in contrast to typical flux-transport dynamos. Such robust
properties appear to be unaffected by the new SLD treatment for diffusion of vorticity. What SLD
admits are noticeably lower magnetic Prandtl numbers $\mathrm{Pm}$, which allow a richer suite of
temporal variability. The lower $\mathrm{Pm}$, combined with a strong stratification, has resulted
in K3S possessing more regular cycles and reversals than obtained in previous wreath-building ASH
simulations.

The substantial advances in supercomputing capabilities has led to the recent development of a
series of global-scale 3-D MHD simulations that all possess the interplay of convection, rotation,
and magnetism. The K3S simulation builds upon this contemporary work
\citep[e.g.,][]{ghizaru10,brown10,racine11,brown11,kapyla12,augustson13,nelson13a,fan14}. The two
common threads in those studies is the generation of large-scale coherent magnetic wreaths within
the convection zone and the presence of a solar-like differential rotation achieved due to the low
Rossby number of the convective flows. Some of those simulations exhibit regular magnetic polarity
reversals. Indeed, both K3S and the Millennium simulation \citep{charbonneau14}, which also includes
a tachocline at the base of the convection zone, have attained many such reversals over an extended
interval. Furthermore, K3S and the Millennium simulation both show that the amplitude of the
magnetic cycles appears to be regulated by the feedback of the Lorentz forces on the differential
rotation \citep{racine11}. They do, however, appear to operate differently in the context of
mean-field theory. Namely, the Millennium simulation seems to be akin to an $\alpha^2$-$\Omega$
dynamo \citep{racine11}, whereas in K3S the $\alpha$ effect generating toroidal magnetic fields is
quite weak leading it to be more similar to the classical $\alpha$-$\Omega$ dynamo. In a smaller
subset of those simulations including K3S, equatorward propagation of the large-scale magnetic
structures is another shared feature during their magnetic cycles. In the simulations of
\citet{kapyla13} and \citet{warnecke14}, a sufficiently large density stratification and the linear
dynamo waves of mean-field theory appear to provide an explanation for the equatorward propagation
of the magnetic fields. Similarly, K3S shares a large density stratification.  However, unlike a
kinematic $\alpha$-$\Omega$ dynamo, the equatorward propagation arises from the nonlinear
interaction of the magnetic fields and the differential rotation.

The K3S simulation exhibits an interval of 16~years during which the magnetic cycles are disrupted
and the magnetic energy is reduced at low latitudes, after which the regular cycles resume. This is
somewhat reminiscent of solar grand minima, when the observed solar magnetic energy is substantially
reduced and sunspot emergence is largely interrupted \citep[e.g.,][]{ribes93,mccracken07}. As such,
the disruption of regular cycling in K3S has been loosely identified as a grand minimum, since this
is the first appearance of such long-sought behavior in a 3-D global-scale simulation. The likely
mechanism that leads to intermittency such as this grand minimum in K3S is the interplay of
symmetric and antisymmetric dynamo families. During the typical magnetic energy cycles, the
antisymmetric dynamo family is dominant throughout the majority of the cycle in which the magnitude
of the odd-ordered modes of the poloidal magnetic field is much larger than those of the
even-ordered modes. In contrast during the grand minimum, the even modes can be equal to, and at
times greater than, the magnitude of the odd modes. The increased symmetry about the equator of the
poloidal field disrupted the ability of the dynamo to fully reverse the antisymmetric modes and led
to a weaker dynamo state during the grand minimum, in which the low-latitude volume-integrated
magnetic energy decreased by about 50\%.

What is particularly striking is that during the magnetic energy cycle preceding the grand minimum,
magnetic structures in the northern and the southern hemispheres become highly asynchronous, leading
to the strong excitation of many symmetric modes in the poloidal magnetic field. It also appears
that the typical phasing of the magnetic field between the deep convection zone (CZ) and that of the
upper CZ is disrupted. Together those losses of phase coherence admit dynamo action from both the
symmetric and the antisymmetric modes, which heralds the entrance of the grand minimum. Eventually,
the symmetric modes decay in amplitude, allowing the antisymmetric family and thus the dipole mode
to reassert its dominance. In earlier studies of simpler though nonlinear mean-field dynamo models,
there is a clear variability between the dominance of antisymmetric or symmetric dynamo modes in
those models \citep[e.g.,][]{tobias97,moss00,brooke02,bushby06}. Such variability can lead to
intervals of deep minima.  Moreover, it was seen that the interplay between the dynamo families is
critical to achieve intermittent and aperiodic dynamo states. Those studies have further shown that
low magnetic Prandtl numbers favor solutions with extended intervals of minimal activity. This
provides some background for how the lower $\mathrm{Pm}$ achieved with the SLD treatment in K3S may
lead to the appearance of deep minima and the coupling of the symmetric and antisymmetric dynamo
families.

Though the K3S simulation does share robust features with other global-scale convective dynamos, it
must be stated that the results do depend on the dissipation through the effective values of the
Reynolds and magnetic Reynolds numbers, Re and Rm. Yet there are currently no global convective
dynamo simulations in the known literature that demonstrate convergence with increasing Re and
Rm. Since the SLD diffusion is linked with the spatial resolution, this sensitivity to Re and Rm
translates to a dependence on resolution. Particularly, reducing the effective viscous diffusion
using the SLD method likely permitted this solution to enter an interesting parameter
regime. Further simulations are being carried out to characterize parameter sensitivities, including
the identification of the factors that set the cycle period and the search for asymptotic behavior.
These results will be presented in future papers. The SLD method has been shown to converge for
solar surface magnetoconvection simulations \citep{rempel12}, but large-scale dynamos may be more
subtle. Nevertheless, the K3S simulation lies in an interesting parameter regime that exhibits a
novel, self-consistent, cyclic convective dynamo with intermittent cycle modulation that may share
some dynamic features with dynamos in more extreme parameter regimes and indeed, in stars.

Despite rotating three times faster than the Sun and parameterizing large portions of its vast range
of scales, some of the features of the dynamo that may be active within the Sun's interior have been
realized in this global-scale ASH simulation. For instance, the period of the magnetic polarity
cycle in K3S is 6.2 years. This period is about 243 times the rotation period, which can be compared
to the Sun's ratio of about 287. Under a linear scaling of the rotation rate, a comparable solar
simulation could have a half-period of 9.3 years, close to the sunspot cycle period of the Sun. So,
although this model star rotates more rapidly than the Sun and has a shorter cycle period, the ratio
of the cycle period to the rotation period is not much different.  Particularly, its rapid rotation
helped to put it into an interesting Rossby number regime that permits a solar-like differential
rotation (fast equator, slow poles), and once there it produced a cycle period that may be
non-dimensionally comparable to that of the Sun. Indeed, the increasingly frequent emergence of 3-D
convective dynamo simulations that exhibit solar-like dynamo features such as this one are reshaping
the understanding of the physics of convective dynamos. Thus, future global-scale dynamo simulations
promise new insights into solar and stellar dynamos as supercomputing resources continue to advance.

\section*{Acknowledgments}

The authors wish to thank an anonymous referee for extensive and helpful comments. A singular
thanks is due to Nicholas Featherstone for his effort in greatly improving the computational
efficiency and scaling of the ASH code. The authors also thank Bradley Hindman, Mark Rast, Matthias
Rempel, and Regner Trampedach for useful conversations. This research was supported by NASA through
the Heliophysics Theory Program grant NNX11AJ36G, with additional support for Augustson first
through the NASA NESSF program by award NNX10AM74H and second through the NCAR Advanced Study
Program. NCAR is supported by the NSF. A.S. Brun acknowledges financial support through ANR TOUPIES,
CNES Solar Orbiter grant and INSU/PNST. The computations were primarily carried out on Pleiades at
NASA Ames with SMD grants g26133 and s0943, and also used XSEDE resources for analysis. This work
further utilized the Janus supercomputer, which is supported by the NSF award CNS-0821794 and the
University of Colorado Boulder.

\bibliography{apj-jour,k3s}

\begin{thebibliography}{91}
\expandafter\ifx\csname natexlab\endcsname\relax\def\natexlab#1{#1}\fi

\bibitem[{{Augustson} {et~al.}(2012){Augustson}, {Brown}, {Brun}, {Miesch}, \&
  {Toomre}}]{augustson12}
{Augustson}, K.~C., {Brown}, B.~P., {Brun}, A.~S., {Miesch}, M.~S., \&
  {Toomre}, J. 2012, \apj, 756, 169

\bibitem[{{Augustson} {et~al.}(2014){Augustson}, {Brun}, {Miesch}, \&
  {Toomre}}]{augustson14}
{Augustson}, K.~C., {Brun}, A.~S., {Miesch}, M.~S., \& {Toomre}, J. 2014, Cool
  Stars 18 Electronic Proc., 1, 437

\bibitem[{{Augustson} {et~al.}(2013){Augustson}, {Brun}, \&
  {Toomre}}]{augustson13}
{Augustson}, K.~C., {Brun}, A.~S., \& {Toomre}, J. 2013, \apj, 777, 153

\bibitem[{{Babcock}(1961)}]{babcock61}
{Babcock}, H.~W. 1961, \apj, 133, 572

\bibitem[{{Baliunas} {et~al.}(1996){Baliunas}, {Nesme-Ribes}, {Sokoloff}, \&
  {Soon}}]{baliunas96}
{Baliunas}, S.~L., {Nesme-Ribes}, E., {Sokoloff}, D., \& {Soon}, W.~H. 1996,
  \apj, 460, 848

\bibitem[{{Barnes}(2007)}]{barnes07}
{Barnes}, S.~A. 2007, \apj, 669, 1167

\bibitem[{{Beer} {et~al.}(1998){Beer}, {Tobias}, \& {Weiss}}]{beer98}
{Beer}, J., {Tobias}, S., \& {Weiss}, N. 1998, Solar Phys., 181, 237

\bibitem[{{Benevolenskaya}(2004)}]{benevolenskaya04}
{Benevolenskaya}, E.~E. 2004, \aap, 428, L5

\bibitem[{{Borges} {et~al.}(2008){Borges}, {Carmona}, {Costa}, \&
  {Don}}]{borges08}
{Borges}, R., {Carmona}, M., {Costa}, B., \& {Don}, W.~S. 2008, J. Comp. Phys.,
  227, 3191

\bibitem[{{Brandenburg}(2009)}]{brandenburg09}
{Brandenburg}, A. 2009, \apj, 697, 1206

\bibitem[{{Brandenburg} \& {Spiegel}(2008)}]{brandenburg08}
{Brandenburg}, A., \& {Spiegel}, E.~A. 2008, Astro. Nachr., 329, 351

\bibitem[{{Brandenburg} \& {Subramanian}(2005)}]{brandenburg05}
{Brandenburg}, A., \& {Subramanian}, K. 2005, \physrep, 417, 1

\bibitem[{{Brooke} {et~al.}(2002){Brooke}, {Moss}, \& {Phillips}}]{brooke02}
{Brooke}, J., {Moss}, D., \& {Phillips}, A. 2002, \aap, 395, 1013

\bibitem[{{Brown} {et~al.}(2008){Brown}, {Browning}, {Brun}, {Miesch}, \&
  {Toomre}}]{brown08}
{Brown}, B.~P., {Browning}, M.~K., {Brun}, A.~S., {Miesch}, M.~S., \& {Toomre},
  J. 2008, \apj, 689, 1354

\bibitem[{{Brown} {et~al.}(2010){Brown}, {Browning}, {Brun}, {Miesch}, \&
  {Toomre}}]{brown10}
---. 2010, \apj, 711, 424

\bibitem[{{Brown} {et~al.}(2011){Brown}, {Miesch}, {Browning}, {Brun}, \&
  {Toomre}}]{brown11}
{Brown}, B.~P., {Miesch}, M.~S., {Browning}, M.~K., {Brun}, A.~S., \& {Toomre},
  J. 2011, \apj, 731, 69

\bibitem[{{Browning}(2008)}]{browning08}
{Browning}, M.~K. 2008, \apj, 676, 1262

\bibitem[{{Browning} {et~al.}(2006){Browning}, {Miesch}, {Brun}, \&
  {Toomre}}]{browning06}
{Browning}, M.~K., {Miesch}, M.~S., {Brun}, A.~S., \& {Toomre}, J. 2006, \apjl,
  648, L157

\bibitem[{{Brun} {et~al.}(2014){Brun}, {Garcia}, {Houdek}, {Nandy}, \&
  {Pinsonneault}}]{brun14}
{Brun}, A.~S., {Garcia}, R.~A., {Houdek}, G., {Nandy}, D., \& {Pinsonneault},
  M. 2014, \ssr, 1, 1

\bibitem[{{Brun} {et~al.}(2004){Brun}, {Miesch}, \& {Toomre}}]{brun04}
{Brun}, A.~S., {Miesch}, M.~S., \& {Toomre}, J. 2004, \apj, 614, 1073

\bibitem[{{Bushby}(2006)}]{bushby06}
{Bushby}, P.~J. 2006, \mnras, 371, 772

\bibitem[{{Charbonneau}(2010)}]{charbonneau10}
{Charbonneau}, P. 2010, Liv. Rev. Sol. Phys., 7, 3

\bibitem[{{Charbonneau}(2013)}]{charbonneau13}
---. 2013, J. Phys. Conf. Ser., 440, 012014

\bibitem[{{Charbonneau}(2014)}]{charbonneau14}
---. 2014, \araa, 52, 251

\bibitem[{{Charbonneau} {et~al.}(2005){Charbonneau}, {St-Jean}, \&
  {Zacharias}}]{charbonneau05}
{Charbonneau}, P., {St-Jean}, C., \& {Zacharias}, P. 2005, \apj, 619, 613

\bibitem[{{Christensen-Dalsgaard} {et~al.}(1996){Christensen-Dalsgaard},
  {Dappen}, {Ajukov}, {Anderson}, {Antia}, {Basu}, {Baturin}, {Berthomieu},
  {Chaboyer}, {Chitre}, {Cox}, {Demarque}, {Donatowicz}, {Dziembowski},
  {Gabriel}, {Gough}, {Guenther}, {Guzik}, {Harvey}, {Hill}, {Houdek},
  {Iglesias}, {Kosovichev}, {Leibacher}, {Morel}, {Proffitt}, {Provost},
  {Reiter}, {Rhodes}, {Rogers}, {Roxburgh}, {Thompson}, \&
  {Ulrich}}]{christdals96}
{Christensen-Dalsgaard}, J., {Dappen}, W., {Ajukov}, S.~V., {et~al.} 1996,
  Science, 272, 1286

\bibitem[{{Clune} {et~al.}(1999){Clune}, {Elliott}, {Miesch}, {Toomre}, \&
  {Glatzmaier}}]{clune99}
{Clune}, T.~L., {Elliott}, J.~R., {Miesch}, M.~S., {Toomre}, J., \&
  {Glatzmaier}, G.~A. 1999, Para. Comp., 25, 361

\bibitem[{{Cossette} {et~al.}(2013){Cossette}, {Charbonneau}, \&
  {Smolarkiewicz}}]{cossette13}
{Cossette}, J.-F., {Charbonneau}, P., \& {Smolarkiewicz}, P.~K. 2013, \apjl,
  777, L29

\bibitem[{{DeRosa} {et~al.}(2012){DeRosa}, {Brun}, \& {Hoeksema}}]{derosa12}
{DeRosa}, M.~L., {Brun}, A.~S., \& {Hoeksema}, J.~T. 2012, \apj, 757, 96

\bibitem[{{Eddy}(1976)}]{eddy76}
{Eddy}, J.~A. 1976, Science, 192, 1189

\bibitem[{{Fan} \& {Fang}(2014)}]{fan14}
{Fan}, Y., \& {Fang}, F. 2014, \apj, 789, 35

\bibitem[{{Fan} {et~al.}(2013){Fan}, {Featherstone}, \& {Fang}}]{fan13}
{Fan}, Y., {Featherstone}, N., \& {Fang}, F. 2013, ArXiv e-prints

\bibitem[{{Fares} {et~al.}(2013){Fares}, {Moutou}, {Donati}, {Catala},
  {Shkolnik}, {Jardine}, {Cameron}, \& {Deleuil}}]{fares13}
{Fares}, R., {Moutou}, C., {Donati}, J.-F., {et~al.} 2013, \mnras

\bibitem[{{Favata} {et~al.}(2008){Favata}, {Micela}, {Orlando}, {Schmitt},
  {Sciortino}, \& {Hall}}]{favata08}
{Favata}, F., {Micela}, G., {Orlando}, S., {et~al.} 2008, \aap, 490, 1121

\bibitem[{{Gallet} \& {P{\'e}tr{\'e}lis}(2009)}]{gallet09}
{Gallet}, B., \& {P{\'e}tr{\'e}lis}, F. 2009, \pre, 80, 035302

\bibitem[{{Ghizaru} {et~al.}(2010){Ghizaru}, {Charbonneau}, \&
  {Smolarkiewicz}}]{ghizaru10}
{Ghizaru}, M., {Charbonneau}, P., \& {Smolarkiewicz}, P.~K. 2010, \apjl, 715,
  L133

\bibitem[{{Gilman}(1983)}]{gilman83}
{Gilman}, P.~A. 1983, \apjs, 53, 243

\bibitem[{{Glatzmaier}(1985)}]{glatzmaier85}
{Glatzmaier}, G.~A. 1985, \apj, 291, 300

\bibitem[{{Gleissberg}(1939)}]{gleissberg39}
{Gleissberg}, W. 1939, The Observatory, 62, 158

\bibitem[{{Gruzinov} \& {Diamond}(1994)}]{gruzinov94}
{Gruzinov}, A.~V., \& {Diamond}, P.~H. 1994, Phys. Rev. Lett., 72, 1651

\bibitem[{{Guerrero} {et~al.}(2013){Guerrero}, {Smolarkiewicz}, {Kosovichev},
  \& {Mansour}}]{guerrero13}
{Guerrero}, G., {Smolarkiewicz}, P.~K., {Kosovichev}, A.~G., \& {Mansour},
  N.~N. 2013, \apj, 779, 176

\bibitem[{{Hathaway}(2010)}]{hathaway10}
{Hathaway}, D.~H. 2010, Liv. Rev. Sol. Phys., 7, 1

\bibitem[{{Hempelmann} {et~al.}(1996){Hempelmann}, {Schmitt}, \& {St{\c
  e}pie{\'n}}}]{hempelmann96}
{Hempelmann}, A., {Schmitt}, J.~H.~M.~M., \& {St{\c e}pie{\'n}}, K. 1996, \aap,
  305, 284

\bibitem[{{Jiang}(1996)}]{jiang96}
{Jiang}, G. 1996, J. Comp. Phys., 126, 202

\bibitem[{{Jouve} {et~al.}(2010){Jouve}, {Brown}, \& {Brun}}]{jouve10}
{Jouve}, L., {Brown}, B.~P., \& {Brun}, A.~S. 2010, \aap, 509, A32

\bibitem[{{K{\"a}pyl{\"a}} {et~al.}(2006){K{\"a}pyl{\"a}}, {Korpi},
  {Ossendrijver}, \& {Stix}}]{kapyla06}
{K{\"a}pyl{\"a}}, P.~J., {Korpi}, M.~J., {Ossendrijver}, M., \& {Stix}, M.
  2006, \aap, 455, 401

\bibitem[{{K{\"a}pyl{\"a}} {et~al.}(2011{\natexlab{a}}){K{\"a}pyl{\"a}},
  {Mantere}, \& {Brandenburg}}]{kapyla11b}
{K{\"a}pyl{\"a}}, P.~J., {Mantere}, M.~J., \& {Brandenburg}, A.
  2011{\natexlab{a}}, Astro. Nachr., 332, 883

\bibitem[{{K{\"a}pyl{\"a}} {et~al.}(2012){K{\"a}pyl{\"a}}, {Mantere}, \&
  {Brandenburg}}]{kapyla12}
---. 2012, \apjl, 755, L22

\bibitem[{{K{\"a}pyl{\"a}} {et~al.}(2013){K{\"a}pyl{\"a}}, {Mantere}, {Cole},
  {Warnecke}, \& {Brandenburg}}]{kapyla13}
{K{\"a}pyl{\"a}}, P.~J., {Mantere}, M.~J., {Cole}, E., {Warnecke}, J., \&
  {Brandenburg}, A. 2013, \apj, 778, 41

\bibitem[{{K{\"a}pyl{\"a}} {et~al.}(2011{\natexlab{b}}){K{\"a}pyl{\"a}},
  {Mantere}, {Guerrero}, {Brandenburg}, \& {Chatterjee}}]{kapyla11a}
{K{\"a}pyl{\"a}}, P.~J., {Mantere}, M.~J., {Guerrero}, G., {Brandenburg}, A.,
  \& {Chatterjee}, P. 2011{\natexlab{b}}, \aap, 531, A162

\bibitem[{{Karak} \& {Choudhuri}(2013)}]{karak13}
{Karak}, B.~B., \& {Choudhuri}, A.~R. 2013, Res. Astron. Astrophys., 13, 1339

\bibitem[{{Krause} \& {Raedler}(1980)}]{krause80}
{Krause}, F., \& {Raedler}, K.-H. 1980, {Mean-field magnetohydrodynamics and
  dynamo theory}

\bibitem[{{Malkus} \& {Proctor}(1975)}]{malkus75}
{Malkus}, W.~V.~R., \& {Proctor}, M.~R.~E. 1975, J. Fluid Mech., 67, 417

\bibitem[{{Malyshkin} \& {Boldyrev}(2010)}]{malyshkin10}
{Malyshkin}, L.~M., \& {Boldyrev}, S. 2010, Phys. Rev. Lett., 105, 215002

\bibitem[{{Mathur} {et~al.}(2013){Mathur}, {Garc{\'{\i}}a}, {Morgenthaler},
  {Salabert}, {Petit}, {Ballot}, {R{\'e}gulo}, \& {Catala}}]{mathur13}
{Mathur}, S., {Garc{\'{\i}}a}, R.~A., {Morgenthaler}, A., {et~al.} 2013, \aap,
  550, A32

\bibitem[{{Matt} {et~al.}(2015){Matt}, {Brun}, {Baraffe}, {Bouvier}, \&
  {Chabrier}}]{matt15}
{Matt}, S.~P., {Brun}, A.~S., {Baraffe}, I., {Bouvier}, J., \& {Chabrier}, G.
  2015, \apjl, 799, L23

\bibitem[{{McCracken}(2007)}]{mccracken07}
{McCracken}, K.~G. 2007, J. Geophys. Res. (Space Physics), 112, 9106

\bibitem[{{McIntyre}(1998)}]{mcintyre98}
{McIntyre}, M.~E. 1998, Progress of Theoretical Physics Supplement, 130, 137

\bibitem[{{Metcalfe} {et~al.}(2010){Metcalfe}, {Basu}, {Henry}, {Soderblom},
  {Judge}, {Kn{\"o}lker}, {Mathur}, \& {Rempel}}]{metcalf10}
{Metcalfe}, T.~S., {Basu}, S., {Henry}, T.~J., {et~al.} 2010, \apjl, 723, L213

\bibitem[{{Miesch} \& {Brown}(2012)}]{miesch12}
{Miesch}, M.~S., \& {Brown}, B.~P. 2012, \apjl, 746, L26

\bibitem[{{Miesch} {et~al.}(2000){Miesch}, {Elliott}, {Toomre}, {Clune},
  {Glatzmaier}, \& {Gilman}}]{miesch00}
{Miesch}, M.~S., {Elliott}, J.~R., {Toomre}, J., {et~al.} 2000, \apj, 532, 593

\bibitem[{{Miesch} \& {Hindman}(2011)}]{miesch11}
{Miesch}, M.~S., \& {Hindman}, B.~W. 2011, \apj, 743, 79

\bibitem[{{Mignone} {et~al.}(2010){Mignone}, {Tzeferacos}, \&
  {Bodo}}]{mignone10}
{Mignone}, A., {Tzeferacos}, P., \& {Bodo}, G. 2010, J. Comp. Phys., 229, 5896

\bibitem[{{Moffatt}(1978)}]{moffatt78}
{Moffatt}, H.~K. 1978, {Magnetic field generation in electrically conducting
  fluids}

\bibitem[{{Morgenthaler} {et~al.}(2011){Morgenthaler}, {Petit}, {Morin},
  {Auri{\`e}re}, {Dintrans}, {Konstantinova-Antova}, \&
  {Marsden}}]{morgenthaler11}
{Morgenthaler}, A., {Petit}, P., {Morin}, J., {et~al.} 2011, Astro. Nachr.,
  332, 866

\bibitem[{{Moss} \& {Brooke}(2000)}]{moss00}
{Moss}, D., \& {Brooke}, J. 2000, \mnras, 315, 521

\bibitem[{{Nelson} {et~al.}(2013{\natexlab{a}}){Nelson}, {Brown}, {Brun},
  {Miesch}, \& {Toomre}}]{nelson13a}
{Nelson}, N.~J., {Brown}, B.~P., {Brun}, A.~S., {Miesch}, M.~S., \& {Toomre},
  J. 2013{\natexlab{a}}, \apj, 762, 73

\bibitem[{{Nelson} {et~al.}(2013{\natexlab{b}}){Nelson}, {Brown}, {Sacha Brun},
  {Miesch}, \& {Toomre}}]{nelson13b}
{Nelson}, N.~J., {Brown}, B.~P., {Sacha Brun}, A., {Miesch}, M.~S., \&
  {Toomre}, J. 2013{\natexlab{b}}, \solphys, 20

\bibitem[{{Nishikawa} \& {Kusano}(2008)}]{nishikawa08}
{Nishikawa}, N., \& {Kusano}, K. 2008, Physics of Plasmas, 15, 082903

\bibitem[{{Parker}(1955)}]{parker55}
{Parker}, E.~N. 1955, \apj, 122, 293

\bibitem[{{Parker}(1977)}]{parker77}
---. 1977, \araa, 15, 45

\bibitem[{{Parker}(1987)}]{parker87}
---. 1987, \apj, 312, 868

\bibitem[{{Passos} \& {Charbonneau}(2014)}]{passos14}
{Passos}, D., \& {Charbonneau}, P. 2014, \aap, 568, A113

\bibitem[{{Ponty} {et~al.}(2005){Ponty}, {Mininni}, {Montgomery}, {Pinton},
  {Politano}, \& {Pouquet}}]{ponty05}
{Ponty}, Y., {Mininni}, P.~D., {Montgomery}, D.~C., {et~al.} 2005, Phys. Rev.
  Lett., 94, 164502

\bibitem[{{Pouquet} {et~al.}(1976){Pouquet}, {Frisch}, \& {Leorat}}]{pouquet76}
{Pouquet}, A., {Frisch}, U., \& {Leorat}, J. 1976, J. Fluid Mech., 77, 321

\bibitem[{{Racine} {et~al.}(2011){Racine}, {Charbonneau}, {Ghizaru}, {Bouchat},
  \& {Smolarkiewicz}}]{racine11}
{Racine}, {\'E}., {Charbonneau}, P., {Ghizaru}, M., {Bouchat}, A., \&
  {Smolarkiewicz}, P.~K. 2011, \apj, 735, 46

\bibitem[{{Rempel}(2012)}]{rempel12}
{Rempel}, M. 2012, \apj, 750, 62

\bibitem[{{Rempel} {et~al.}(2009){Rempel}, {Sch{\"u}ssler}, \&
  {Kn{\"o}lker}}]{rempel09}
{Rempel}, M., {Sch{\"u}ssler}, M., \& {Kn{\"o}lker}, M. 2009, \apj, 691, 640

\bibitem[{{Ribes} \& {Nesme-Ribes}(1993)}]{ribes93}
{Ribes}, J.~C., \& {Nesme-Ribes}, E. 1993, \aap, 276, 549

\bibitem[{{Saar}(2009)}]{saar09}
{Saar}, S.~H. 2009, in Astro. Soc. Pac. Conf. Ser., Vol. 416, Solar-Stellar
  Dynamos as Revealed by Helio- and Asteroseismology: GONG 2008/SOHO 21, 375

\bibitem[{{Schekochihin} {et~al.}(2007){Schekochihin}, {Iskakov}, {Cowley},
  {McWilliams}, {Proctor}, \& {Yousef}}]{schekochihin07}
{Schekochihin}, A.~A., {Iskakov}, A.~B., {Cowley}, S.~C., {et~al.} 2007, New J.
  Phys., 9, 300

\bibitem[{{Simard} {et~al.}(2013){Simard}, {Charbonneau}, \&
  {Bouchat}}]{simard13}
{Simard}, C., {Charbonneau}, P., \& {Bouchat}, A. 2013, \apj, 768, 16

\bibitem[{{Steenbeck} \& {Krause}(1969)}]{steenbeck69}
{Steenbeck}, M., \& {Krause}, F. 1969, Astro. Nachr., 291, 49

\bibitem[{{Steenbeck} {et~al.}(1966){Steenbeck}, {Krause}, \&
  {R{\"a}dler}}]{steenbeck66}
{Steenbeck}, M., {Krause}, F., \& {R{\"a}dler}, K.-H. 1966, Zeit. Natur. Teil
  A, 21, 369

\bibitem[{{Strugarek} {et~al.}(2013){Strugarek}, {Brun}, {Mathis}, \&
  {Sarazin}}]{strugarek13}
{Strugarek}, A., {Brun}, A.~S., {Mathis}, S., \& {Sarazin}, Y. 2013, \apj, 764,
  189

\bibitem[{{Suresh} \& {Huynh}(1997)}]{suresh97}
{Suresh}, A., \& {Huynh}, H.~T. 1997, J. Comp. Phys., 136, 83

\bibitem[{{Tobias}(1997)}]{tobias97}
{Tobias}, S.~M. 1997, \aap, 322, 1007

\bibitem[{{Usoskin}(2013)}]{usoskin13}
{Usoskin}, I.~G. 2013, Liv. Rev. Sol. Phys., 10, 1

\bibitem[{{{\v C}ada} \& {Torrilhon}(2009)}]{cada09}
{{\v C}ada}, M., \& {Torrilhon}, M. 2009, J. Comp. Phys., 228, 4118

\bibitem[{{Warnecke} {et~al.}(2014){Warnecke}, {K{\"a}pyl{\"a}},
  {K{\"a}pyl{\"a}}, \& {Brandenburg}}]{warnecke14}
{Warnecke}, J., {K{\"a}pyl{\"a}}, P.~J., {K{\"a}pyl{\"a}}, M.~J., \&
  {Brandenburg}, A. 2014, \apjl, 796, L12

\bibitem[{{Yoshimura}(1975)}]{yoshimura75}
{Yoshimura}, H. 1975, \apj, 201, 740

\end{thebibliography}

\begin{appendix}
\section{Slope-Limited Diffusion in ASH} \label{sec:sld}

There are currently three methods for stabilizing the numerics in ASH, which like all spectral codes
is inherently unstable without a sufficient level of explicit diffusion. The first is that described
in \S2 with a renormalization of the molecular diffusion coefficients into eddy coefficients, the
second is the spectral extrapolation method as in the dynamic Smagorinski method \citep{nelson13a},
and most recently the numerically stabilizing slope-limited diffusion (SLD) scheme. SLD has many
possible formulations, with the one employed in ASH being similar to that found in \citet{rempel09}
and \citet{fan13}. At its heart, the SLD method attempts to push the solution toward a state that is
monotonically smooth.

In ASH, a smoothed solution is obtained through the following steps: a slope-limited piecewise
linear approximation to the solution is constructed; the difference between the left and right
reconstructed values at a cell edge is computed; this difference is then multiplied by a diffusive
rate to create a diffusive flux at the cell edge; finally, the divergence of this flux is computed
and added to the solution. Since this diffusion is nonlinear and a finite-volume based scheme, it is
computed in physical space and so must be transformed into spectral space before being added to the
other nonlinear terms advanced with an Adams-Bashforth time-stepping scheme. 

Because of the converging longitudinal grid near the poles in physical space configuration of ASH,
the SLD is filtered in longitude. This is accomplished through two processes. First the full
grid-wise SLD is computed; a second SLD is computed with a continuously increasing cell size as the
poles are approached, which incorporates a greater number of points in a given cell. The cell size
increases as $\csc{\theta}$, with the requirement that there be no fewer than eight SLD cells at
high latitudes. The slope is reconstructed over the larger cell such that the individual points
within the cell see the same slope in the cell. These two diffusive fields are weighted with
$\sin^2{\theta}$ for the first field and by $\cos^2{\theta}$ for the second field. This allows for a
more uniform resolution of the SLD operator over the nonuniform griding of the sphere.

\subsection{Constructing a Slope-Limited Diffusion} \label{sec:sldconst}

A piecewise reconstruction of values at the cell edge that are concurrent with the cell center value
takes the general form of a Taylor series as

\vspace{-0.25truein}
\begin{center}
  \begin{align}
    q_{i+1/2} = q_i + \sum_{k=1} \left(x_{i+1/2}-x_{i}\right)^k \left. \widetilde{\frac{\Delta^k q}{\Delta x^k}}\right|_{i},
  \end{align}
\end{center}

\noindent where each order of the numerical derivative is limited in some fashion to maintain
monotonicity as is indicated with the tilde. There are a few techniques for computing these
higher-order derivatives and their limited values, but they become increasingly complex with each
increasing order of polynomial approximation \citep[e.g.,][]{jiang96,suresh97,borges08,cada09,
  mignone10}. The most well studied schemes are linear reconstructions, which will suffice for
creating a diffusive operator.

The piecewise linear reconstruction at the cell edges requires at least three derivative values and
three quantity values at adjacent cell centers of a given quantity $q$ in a particular
direction. Each direction is treated equally, so the following will focus only on one of them; call
it coordinate $x$ with index $i$. First define the cell-edge derivative approximation as

\vspace{-0.25truein}
\begin{center}
  \begin{align}
    \left. \frac{\partial q}{\partial x}\right|_{i+1/2} = \frac{1}{2}\left[\left. \frac{\partial q}{\partial x}\right|_{i+1}+\left. \frac{\partial q}{\partial x}\right|_{i}\right]
  \end{align}
\end{center}

\noindent Next define the the average cell-centered derivative with

\vspace{-0.25truein}
\begin{center}
  \begin{align}
    \left<\left.\frac{\partial q}{\partial x}\right|_{i}\right> &= \frac{1}{2+w_{i+1}+w_{i-1}}\left[w_{i+1}\left(\left.\frac{\partial q}{\partial x}\right|_{i+1}
                                                                 +\left.\frac{\partial q}{\partial x}\right|_{i}\right) \right. \nonumber \\
     &\left. +w_{i-1}\left(\left. \frac{\partial q}{\partial x}\right|_{i}+\left. \frac{\partial q}{\partial x}\right|_{i-1}\right)\right].
  \end{align}
\end{center}

\noindent where the weights $w$ are $w_{i-1} = \mathrm{minmod}\left[q_{i},q_{i-1}\right]$ and
$w_{i+1} = \mathrm{minmod}\left[q_{i},q_{i+1}\right]$, and 

\vspace{-0.25truein}
\begin{center}
  \begin{align}
    &\mathrm{minmod}\left[x,y\right] = \left\{
    \begin{array}{lr}
      \mathrm{min}\left[1,\frac{x}{y}\right]& \mathrm{for}\, xy>0 \\
      0& \mathrm{otherwise}
    \end{array}
    \right. . \label{eqn:minmod}
  \end{align}
\end{center}

\noindent Such a weighting minimizes the overshoot and undershoot of the solution near a
discontinuity, allowing greater monotonicity. Let the slope-limiting function be defined so that

\vspace{-0.25truein}
\begin{center}
  \begin{align}
    \psi_i = \psi\left[\left<\left.\frac{\partial q}{\partial x}\right|_{i}\right>\left(\left. \frac{\partial q}{\partial x}\right|_{i}\right)^{-1}\right]
  \end{align}
\end{center}

\noindent and bounded so that $0\le\psi_i\le1$. The limited slope at cell center is then
constructed as 

\vspace{-0.25truein}
\begin{center}
  \begin{align}
    \left.\widetilde{\frac{\partial q}{\partial x}}\right|_{i} = \left(1-\psi_i\right)\left<\left.\frac{\partial q}{\partial x}\right|_{i}\right> + \psi_i \left.\frac{\partial q}{\partial x}\right|_{i}
  \end{align}
\end{center}

\noindent Thus the cell-edge difference is $\delta q_{i+1/2} = q_{i+1/2}^R-q_{i+1/2}^L$, or 

\vspace{-0.25truein}
\begin{center}
  \begin{align}
     \delta q_{i+1/2} &= \Delta q_{i+1/2} - \frac{1}{2}\left[\Delta x_{i+1} \left.\widetilde{\frac{\partial u}{\partial x}}\right|_{i+1}+\Delta x_{i} \left.\widetilde{\frac{\partial u}{\partial x}}\right|_{i}\right],
  \end{align}
\end{center}

\noindent where the slope-limited, linearly reconstructed values of $q$ on the left and right sides
of a cell edge are $q_{i+1/2}^L$ and $q_{i+1/2}^R$, $\Delta q_{i+1/2} = q_{i+1}-q_i$, and $\Delta
x_i = x_{i}-x_{i-1}$. In this case, the diffusive flux is then controlled with a coefficient that
sets the rate at which the solution achieves a state that is linearly smooth and monotonic. The
diffusive flux in direction $x$ at the cell edge is then $F_{i+1/2}^x = c_{i+1/2} \delta q_{i+1/2}$
with the diffusion speed at the cell edge defined as

\vspace{-0.25truein}
\begin{center}
  \begin{align}
    c_{i+1/2} &= \beta\left(\delta q_{i+1/2},\Delta q_{i+1/2};\alpha\right)\sqrt{\mathrm{v}_{\mathrm{min}}^2 + {\vv'}_{i+1/2}^2}.
  \end{align}
\end{center}

\noindent Here $\mathrm{v}_{\mathrm{min}}$ is a minimum speed that sets the base level of diffusion,
and the fluctuating velocities locally enhance it. The function $\beta$ further isolates the
diffusion to regions of the largest discontinuities and prevents anti-diffusion. It is defined as
$\beta\left(\delta q_{i+1/2},\Delta q_{i+1/2};\alpha\right) = \left(\mathrm{minmod}\left[\delta
  q_{i+1/2},\Delta q_{i+1/2}\right]\right)^{\alpha}$, with $\alpha$ some positive value. Now, let
$r_i$ be the cell center in radius of a grid cell, $\theta_j$ the center in colatitude, and
$\varphi_k$ the center in longitude.  Following the definition of finite volumes, the rate of change
induced in the quantity $q$ through the diffusive flux in spherical coordinates is

\vspace{-0.25truein}
\begin{center}
  \begin{align}
    \frac{\partial q}{\partial t}_{\mathrm{SLD}} &= \frac{3}{2}\frac{r_{i+1/2}^2 - r_{i-1/2}^2}{r_{i+1/2}^3-r_{i-1/2}^3}\left[2\frac{r_{i+1/2}^2 F_{i+1/2}^r - r_{i-1/2}^2 F_{i-1/2}^r}{r_{i+1/2}^2-r_{i-1/2}^2} \right. \nonumber \\
    &+ \frac{\sin{\theta}_{j+1/2}F_{j+1/2}^{\theta}-\sin{\theta}_{j-1/2}F_{j-1/2}^{\theta}}{\cos{\theta}_{j-1/2}-\cos{\theta}_{j+1/2}} \nonumber \\
    &\left. + \frac{{\theta}_{j+1/2}-{\theta}_{j-1/2}}{\cos{\theta}_{j-1/2}-\cos{\theta}_{j+1/2}}\frac{F_{k+1/2}^{\varphi}-F_{k-1/2}^{\varphi}}{\varphi_{k+1/2}-\varphi_{k-1/2}}\right], \label{eqn:spheresld}
  \end{align}
\end{center}

\noindent where unchanging indices have been suppressed.

In order to conserve energy in the domain, the diffusive and dissipative processes must be accounted
for. Since the entropy diffusion is automatically energy conserving, only the dissipative exchange
of energy from the velocity fields to the entropy field need be examined. With this in mind, the
viscous component of the entropy production due to SLD is

\vspace{-0.25truein}
\begin{center}
  \begin{align}
    \orho \overline{T} \left.\frac{\partial s}{\partial t}\right|_{\mathrm{SLD}} = \left.\frac{\partial}{\partial t}\left(\frac{1}{2}\orho\vv^2\right)\right|_{\mathrm{SLD}} - \orho\vv\cdot\left.\frac{\partial \vv}{\partial t}\right|_{\mathrm{SLD}},
  \end{align}
\end{center}

\noindent where $\orho$ and $\overline{T}$ are the background density and temperature in an ASH
simulation. Basically it is the difference between the diffusion of the kinetic energy and the
energy difference arising from the diffusion of the velocity field
(e.g. $\nabla\cdot\mathbf{F}_{\mathrm{ke}} - \orho\vv\cdot\nabla\cdot\mathcal{F}_{\vv}$).

\subsection{Mathematical Implications of SLD} \label{sec:sldimp}

Since this diffusion scheme is derived from finite volumes, it is conservative to within machine
precision, with the adage ``one cell's loss is another's gain'' holding here. It is useful to show
that this diffusion reduces to a Laplacian diffusion under certain assumptions.  Assume first that
the diffusion speed is set to the constant $c=v_{\mathrm{min}}$. Second, let the function $\beta=1$
and further let there be no discontinuities in the domain (so that the weights $w$ are all
unity). For simplicity, also assume a uniform mesh in one dimension, thus the divergence of the
total diffusive flux is

\vspace{-0.25truein}
\begin{center}
  \begin{align}
    \dvg{\mathbf{F}} &= \frac{c}{\Delta x} \left(\delta q_{i+1/2} - \delta q_{i-1/2} \right), \\
                     &= c\Delta x \left(\frac{q_{i+1}-2 q_i + q_{i-1}}{\Delta x^2} 
                        - \frac{1}{2\Delta x}\left[\left.\widetilde{\frac{\partial u}{\partial x}}\right|_{i+1} 
                        - \left.\widetilde{\frac{\partial u}{\partial x}}\right|_{i-1}\right] \right). \nonumber
  \end{align}
\end{center}

\noindent Notice that the first term is just the second-order approximation to the second
derivative, whereas the second is the slope-limited approximation to the second derivative. So, this
operator may be Taylor expanded to see that

\vspace{-0.25truein}
\begin{center}
  \begin{align}
    \dvg{\mathbf{F}} &= c\Delta x \left[\frac{\partial^2 q}{\partial x^2} - \widetilde{\frac{\partial^2 q}{\partial x^2}} + \mathcal{O}\left(\Delta x^2\right)\right]_i. \label{eqn:sldfluxdvg}
  \end{align}
\end{center}

\noindent Equation (\ref{eqn:sldfluxdvg}) elucidates the first-order behavior of the slope-limited
diffusion operator. The difference between the unfiltered and the filtered Laplacian operators
reduces the overall diffusion by subtracting off a smoothed Laplacian, which is by construction
always less oscillatory than the local Laplacian. This leaves only the local fluctuations to be
diffused away with the coefficient $c\Delta x$. The higher-order components yield a filtered
hyper-diffusivity.

\subsection{Estimated Fluid Parameters} \label{sec:parameters}

As a point of reference, and given the values of the eddy diffusivities presented in
\S\ref{sec:methods}, a few standard fluid parameters can be estimated. The first is an effective
fluid Reynolds number $\mathrm{Re}=u_{\mathrm{rms}} d/\nu_{\mathrm{eff}}$. The length-scale $d$ is
taken to be the depth of the convection zone (175~Mm), and the rms velocity near the upper boundary
is measured to be $\mpers{100}$. Given the argument in Appendix A.2, the effective diffusion
coefficient could be estimated to be $\nu_{\mathrm{eff}} \approx c\Delta x \approx 1.75\times
10^{13} \mathrm{cm}^2\, \mathrm{s}^{-1}$ at the upper boundary, where $c$ is the diffusion advection
speed. Yet this is not a complete estimate though, as the strength of the SLD operator's damping
depends upon the local value of the velocity gradient.  This spatial dependence is largely governed
by the morphology of resolved flows. As discussed in A.2, the SLD diffusion operator can become akin
to a diffusion coefficient multiplied by the difference of a grid-level Laplacian and a filtered
Laplacian. Thus, the average value of $|\nabla^2\vv -\widetilde{\nabla}^2\vv|$ needs to be estimated
to better approximate the SLD diffusion coefficient. The radial gradient of the flows is fairly
small when compared to the horizontal gradients as the convective cells tend to stretch across the
convection zone. Further, within the relatively laminar conditions of an upflow, this difference is
again small and so the contribution of the upflows can be neglected. Hence,

\vspace{-0.25truein}
\begin{center}
  \begin{align}
    \avg{|\nabla^2\vv -\widetilde{\nabla}^2\vv|} &\approx \frac{A_{\mathrm{d}}}{A} \left( \frac{u_{\mathrm{rms}}}{\Delta x^2} - \frac{u_{\mathrm{rms}}}{\ell^2}\right)
    \approx \frac{A_{\mathrm{d}}}{A} \left(1 - \frac{\Delta x^2}{\ell^2}\right) \frac{u_{\mathrm{rms}}}{\Delta x^2} \nonumber \\
    &\Rightarrow c \Delta x \left(\nabla^2\vv -\widetilde{\nabla}^2\vv\right) \approx c \Delta x \frac{A_\mathrm{d}}{A} \nabla^2\vv \nonumber
  \end{align}
\end{center}

\noindent where $A$ is the total surface area, $A_\mathrm{d}$ is the area occupied by steep
gradients at the edges of a downflow, $\Delta x$ is an average grid spacing, and $\ell$ is the
filter length scale.

The effective SLD diffusion coefficient can thus be quantified as $\nu_{\mathrm{eff}} = c \Delta x
A_{\mathrm{d}}/A$.  This leaves the task of estimating the effective diffusion profile of the
downflows. In this simulation they are primarily composed of two regions, the downflow core and the
steep transition zones between the core and the surrounding upflows as might be deduced from Figures
\ref{fig4}(b)-(f). Those transition zones are where the velocity gradient is largest and they occupy
approximately one-third of the horizontal area of a downflow, whereas the downflows themselves cover
roughly one-third of the total surface area.  Therefore, the effective diffusion coefficient at the
upper boundary is $2\times10^{12} \mathrm{cm}^2\, \mathrm{s}^{-1}$.  As with the Newtonian diffusion
coefficients for the thermal and magnetic fields, the value $c$ follows a $\orho^{\; -1/2}$
profile. Thus, $\nu_{\mathrm{eff}}$ is reduced by a factor of four at mid-convection zone, yielding
an effective Reynolds number of $\mathrm{Re}_{\mathrm{eff}}\approx 350$ there.

\section{Evolution of Differential Rotation Kinetic Energy}

The derivation of the equation for the time evolution of the volume-integrated energy contained in
the differential rotation, as assessed in \S\ref{sec:dre}, involves some subtlety in the role of the
boundary terms.  First, note that the evolution of the mean specific angular momentum can be written
as the divergence of a flux as $\partial_t\avg{L}=-\dvg{\avg{\mathbf{F}_L}}$, where for the
anelastic system $\avg{L}=\orho\lambda\avg{\vcp}$ and where the angular momentum fluxes
$\avg{\mathbf{F}_L}$ are specified below. Thus, $\avg{\Omega} \partial_t
\avg{L}=1/2\partial_t\orho\avg{\vcp}^2$. Next, multiply the evolution equation by the angular
velocity $\avg{\Omega}$ and apply the chain rule for the divergence operator as
$\avg{\Omega}\dvg{\avg{\mathbf{F}_L}}=\dvg{(\avg{\Omega}\avg{\mathbf{F}_L})}-\avg{\mathbf{F}_L}\cnabla\avg{\Omega}$
to see that

\vspace{-0.25truein}
\begin{center}
   \begin{align}
      \displaystyle \ddtime{}\frac{1}{2}\orho\avg{\vcp}^2 &= \avg{\mathbf{F}_L}\cnabla\avg{\Omega} -\dvg{\left(\avg{\Omega}\avg{\mathbf{F}_L}\right)},
   \end{align}
\end{center}

\noindent with $\avg{\mathbf{F}_L} = \orho\lambda\avg{\vcp\vv}-\lambda\avg{\mathcal{D}}\cdot\pht -
\lambda\avg{\Bcp\vB}/4\pi$, where $\mathcal{D}$ is the viscous diffusion tensor. Since the viscous
eddy diffusivity is very small in this simulation, and the slope-limited diffusion provides an
especially small level of diffusion to the smooth mean fields, this term is neglected. Hence, when
integrated over the volume of the domain and applying Gauss's theorem to the complete divergence
term, the evolution of the total energy contained in the differential rotation can be written as

\vspace{-0.25truein}
\begin{center}
   \begin{align}
      &K_{\varphi}\left(t\right) =\!\!\!\displaystyle \int_V dV \ddtime{}\frac{1}{2}\orho\avg{\vcp}^2 
      = \!\!\!\int_V dV \lambda\left[\orho\avg{\vcp\vv} - \frac{\avg{\Bcp\vB}}{4\pi} \right]\cnabla\avg{\Omega} \nonumber \\
      &-\int_S dS \avg{\Omega}\left[\orho\lambda\avg{\vcr\vcp} - \frac{\avg{\lambda\Bcp\Bcr}}{4\pi}\right]
       = K_{\varphi}^V\left(t\right) + K_{\varphi}^S\left(t\right), \label{eqn:kphi}
   \end{align}
\end{center}

\noindent where the properties of the full spherical geometry are leveraged so that only radial
boundaries can in principle contribute to the evolution of the system. In the above integral, $S$ is
the union of the spherical surfaces at the upper and lower radial boundaries. However, the
impenetrability of the radial boundaries requires that the first term in the surface integral be
zero. Similarly, the perfectly conducting condition at the lower boundary requires that $\Bcr=0$
there, so the second term also vanishes at the lower boundary. In contrast, the upper boundary is a
potential field condition, meaning that the magnetic flux term may not necessarily vanish
there. However, this value can be shown to be zero with the equations utilized in ASH. Given that
$\vB$ is a solenoidal field, one can define $\vB = \curl{\curl{C\rht}} + \curl{A\rht}$, with two
scalar fields $C=C(t,r,\theta,\varphi)$ and $A=A(t,r,\theta,\varphi)$. Therefore at the upper
boundary where $r=r_2$, it can be seen that $\Bcr=-\nabla^2_H C$ and $\lambda\Bcp
=\partial_{\varphi,\mathrm{r}}C - \sint\partial_\theta A$, with $\nabla^2_H$ the Laplacian on
spherical surfaces. Therefore, $\Bcr=\sum_{\ell,\mathrm{m}} \ell(\ell+1)
C_{\ell}^{\mathrm{m}}Y_{\ell}^{\mathrm{m}}/r_2^2$, where
$Y_{\ell}^{\mathrm{m}}=N_{\ell}^{\mathrm{m}}P_{\ell}^{\mathrm{m}}(\theta)\exp{\left(i m
  \varphi\right)}$ are spherical harmonics, the $N_{\ell}^{\mathrm{m}}$ the spherical harmonic
normalization coefficient, the $P_{\ell}^{\mathrm{m}}$ are associated Legendre polynomials, and the
$C_{\ell}^{\mathrm{m}}=C_{\ell}^{\mathrm{m}}(r_2)$ are the spherical harmonic expansion coefficients
of the vector potential component $C$. The potential field boundary condition at $r=r_2$ implies
that $\partial_r C_{\ell}^{\mathrm{m}} = -\ell C_{\ell}^{\mathrm{m}}/r_2$ and $A=0$, so at $r=r_2$
the following holds $\lambda\Bcp = -\sum_{\ell,\mathrm{m}} i \mathrm{m} \ell
C_{\ell}^{\mathrm{m}}Y_{\ell}^{\mathrm{m}}/r_2$. Thus, the surface integral in Equation
(\ref{eqn:kphi}) becomes

\vspace{-0.25truein}
\begin{center}
   \begin{align}
       &K_{\varphi}^S\left(t\right) =\int_S dS \avg{\Omega}\frac{\avg{\lambda\Bcp\Bcr}}{4\pi} = 
       \int_S \frac{dS}{16\pi^3 r_2^3} \sum_{\ell_1, \mathrm{m}_1} \left[\tilde{\Omega}_{\ell_1} \int_0^{2\pi}\!\!\!\!\!\!\!\!d\varphi' Y_{\ell_1}^{\mathrm{m}_1}\right] \nonumber \\ 
       &\times\!\!\! \sum_{\substack{\ell_2, \ell_3, \\ \mathrm{m}_2,\mathrm{m}_3}} \left[ -i \mathrm{m}_2 \ell_2 \ell_3\left(\ell_3+1\right) C_{\ell_2}^{\mathrm{m}_2} C_{\ell_3}^{\mathrm{m}_3} \int_0^{2\pi} \!\!\!\!\!\!\!\!d\varphi' Y_{\ell_2}^{\mathrm{m}_2} Y_{\ell_3}^{\mathrm{m}_3}\right],
   \end{align}
\end{center}

\noindent where $\tilde{\Omega}_{\ell}$ are the coefficients of the spherical harmonics expansion of
the axisymmetric differential rotation. The integrals over $\varphi'$ require that $\mathrm{m}_1=0$
and $\mathrm{m_2+m_3} = 0$, leaving

\vspace{-0.25truein}
\begin{center}
   \begin{align}
     &K_{\varphi}^S\left(t\right) = \frac{-i}{2 r_2}\sum_{\substack{\ell_1, \ell_2 \\ \ell_3, \mathrm{m}_2}} \mathrm{m}_2 \ell_2 \ell_3\left(\ell_3+1\right) N_{\ell_1}^{0} N_{\ell_2}^{\mathrm{m}_2} N_{\ell_3}^{-\mathrm{m}_2} \tilde{\Omega}_{\ell_1} C_{\ell_2}^{\mathrm{m}_2} C_{\ell_3}^{-\mathrm{m}_2} \nonumber \\
     &\times \int_0^\pi d\theta \sint P_{\ell_1}^{0} P_{\ell_2}^{\mathrm{m}_2} P_{\ell_3}^{-\mathrm{m}_2}.
   \end{align}
\end{center}

\noindent Employing the identity $P_{\ell_3}^{-\mathrm{m}_2} =
(-1)^{\mathrm{m}_2}(\ell_3+\mathrm{m}_2)!/(\ell_3-\mathrm{m}_2)! P_{\ell_3}^{\mathrm{m}_2}$, the
integral over the product of three associated Legendre polynomials $P_{\ell}^m$ becomes a Gaunt
integral of the form

\vspace{-0.25truein}
\begin{center}
   \begin{align}
       &\int_0^\pi d\theta \sint P_{\ell_1}^{0} P_{\ell_2}^{\mathrm{m}_2} P_{\ell_3}^{\mathrm{m}_2} = 
       \sqrt{\frac{\left(2\ell_1 +1\right) \left(2\ell_2 +1\right) \left(2\ell_3 +1\right)}{4\pi}} \nonumber \\
       &\times\!\!\!\begin{pmatrix} \ell_1 & \ell_2 & \ell_3 \\ 0 & 0 & 0 \end{pmatrix}\!\!\! \begin{pmatrix} \ell_1 & \ell_2 & \ell_3 \\ \mathrm{m}_2 & \mathrm{m}_2 & 0 \end{pmatrix} = \begin{cases} f\left(\ell_1,\ell_2,\ell_3\right), & \!\!\!\!\!\!m_2 =0 \\ 0, & \!\!\!\!\!\!m_2\ne0 \end{cases},
   \end{align}
\end{center}

\noindent where the two parenthetical terms are Wigner 3-j symbols. The integral must be zero since
$\mathrm{m}_2+\mathrm{m}_2\neq 0$ unless $\mathrm{m}_2=0$, and in the latter case the summand is
zero due to the presence of the factor of $\mathrm{m}_2$. Therefore, the surface terms vanish,
leaving only the following volume integral

\vspace{-0.25truein}
\begin{center}
   \begin{align}
     &K_{\varphi}\left(t\right) = \int_V dV \bigg[\overbrace{\orho\lambda\avg{\vcp}\avg{\vv}\cnabla\avg{\Omega}}^{\mathrm{MC}} 
      + \overbrace{\orho\lambda\avg{\vcp'\vv'}\cnabla\avg{\Omega}}^{\mathrm{RS}} \nonumber \\
     &-\overbrace{\frac{\lambda}{4\pi}\avg{\Bcp}\avg{\vB}\cnabla\avg{\Omega}}^{\mathrm{MM}}
      -\overbrace{\frac{\lambda}{4\pi}\avg{\Bcp'\vB'}\cnabla\avg{\Omega}}^{\mathrm{FM}} \bigg].
   \end{align}
\end{center}

Finally, it can be shown, following the derivation of the surface integral above, that the advection
of the angular velocity by the meridional circulation cannot change the global kinetic energy of the
differential rotation. Physically, this is a result of the fact that this mechanism simply advects
energy from one part of the domain to another. Therefore, the MC term in the integral vanishes as
well, leaving Equation (\ref{eqn:evokdr}) to describe the evolution of the volume-integrated kinetic
energy in the differential rotation.

\end{appendix}

\end{document}